\documentclass[preprint,authoryear,11pt]{elsarticle}

\usepackage{graphics}
\usepackage{graphicx}
\graphicspath{{Figures/}}
\usepackage{epsfig}
\usepackage{subfigure}
\usepackage{amssymb}
\usepackage{amsthm}
\usepackage{bm}
\usepackage{amsmath}
\usepackage{color}
\usepackage{geometry}
\usepackage{setspace}

\allowdisplaybreaks
\usepackage{enumerate}
\usepackage{booktabs}

\usepackage{hyperref}
\hypersetup{
	colorlinks=true,       
}

\biboptions{}

\journal{}

\begin{document}

\begin{frontmatter}



\title{Low-frequency tunable topological interface states
  \\ in soft phononic crystal cylinders}


\author[1,2]{Yingjie Chen}
\author[3]{Bin Wu\corref{cor1}}
\ead{bin.wu@nuigalway.ie}
\author[1]{Jian Li}
\author[2]{Stephan Rudykh}
\author[1,4]{Weiqiu Chen}

\cortext[cor1]{Corresponding author.}
\address[1]{Key Laboratory of Soft Machines and Smart Devices of Zhejiang Province \\ and Department of Engineering Mechanics,\\ Zhejiang University, Hangzhou 310027, P.R. China;\\[6pt]}
\address[2]{Department of Mechanical Engineering,\\ University of Wisconsin Madison, Madison, WI 53706; \\[6pt]}
\address[3]{School of Mathematics, Statistics and Applied Mathematics,\\ NUI Galway, University Road, Galway, Ireland; \\[6pt]}
\address[4]{Soft Matter Research Center, Zhejiang University, Hangzhou 310027, China; }

\begin{abstract}
	
Topological phononic crystals have attracted intensive attention due to their peculiar topologically protected interface or edge states. Their operating frequency, however, is generally fixed once designed and fabricated. Here, we propose to overcome this limitation by utilizing soft topological phononic crystals. In particular, we design a simple one-dimensional periodic system of soft cylindrical waveguides to realize mechanically tunable topological interface states for longitudinal waves. To this end, we employ the nonlinear elasticity theory and its linearized incremental version to fully account for both geometric and material nonlinearities of the system. We derive the dispersion relation for small-amplitude longitudinal motions superimposed on the finitely deformed state. In addition, our analytical results provide information about the corresponding Bloch wave modes, displacement field distributions, and signal transmission coefficients for finite cylindrical waveguides with identical or various unit-cell topologies.  Our numerical results illustrate the low-frequency topological interface state occurring at the interface between two topologically distinct soft phononic cylinders. Moreover, we show that the corresponding frequency in the overlapped band gap can be continuously adjusted by an external force. This analytical result is also validated by the finite element simulations. Finally, we provide the topological phase diagrams to demonstrate the tunable position and existence condition of the topological interface states when tuning the external loading. The low-frequency tunable topological interface states with remarkable field enhancement may find a wide range of potential applications such as tunable energy harvesters, low-pass filters and high-sensitivity  detectors for biomedical applications.

\end{abstract}

\begin{keyword}
Soft phononic cylinder \sep active tunability\sep low-frequency interface state\sep topological phase diagram \sep material nonlinearity



\end{keyword}

\end{frontmatter}




\underline{\underline{}}\section{Introduction}

Phononic crystals (PCs) have attracted intensive attention thanks to their outstanding properties in the manipulation of acoustic/elastic waves. The intrinsically artificial periodic composites can give rise to the wave band gap (BG) -- a special state, where acoustic/elastic waves are prohibited within a certain frequency range. This prominent characteristic of PCs can be attributed to the Bragg scattering \citep{kushwaha1993acoustic}, local resonance \citep{liu2000locally} and inertial amplification \citep{yilmaz2007phononic}. The unique BG character and strong dispersive properties in passbands may produce anomalous wave behaviors such as acoustic/elastic wave filtering, focusing, directional propagation, negative refraction and cloaking \citep{hussein2014dynamics, ma2016acoustic,cummer2016controlling,slesarenko&etal18apl,krushynska&etal18apl}.

Recently, the topic of topological acoustic or mechanical PCs has emerged to offer exciting opportunities for designing materials with broadband one-way transport properties.  These peculiar material systems can generate topologically protected unidirectional, backscattering-immune interface or edge states \citep{zhang2018topological, ma2019topological}. The theoretical development has motivated some experimental efforts in realization of the topological PCs \citep{susstrunk2015observation, xiao2015geometric, peng2016experimental, lu2017observation, yu2018elastic, yin2018band, li2018observation, ding2019experimental}, indicating the feasibility of the concept for potential applications in wave filters, energy harvesters, acoustic rectifiers, vibration isolators, acoustic imaging, and bio-sensors. The topological characteristics of PC band structures can be characterized by their topological invariants such as the Berry phase \citep{zak1989berry} for two-dimensional (2D) systems or Zak phase \citep{atala2013direct} for one-dimensional (1D) media. Since topological PCs possess the global properties of band structures, their nontrivial topological states are extremely robust to the defects and boundary effects \citep{xiao2014surface, wang2015topological}.

Topological PCs can be categorized into the following three classes. The first approach -- an analogue of the quantum Hall effect -- is to break the time-reversal symmetry of the systems and realize the topologically protected edge states through introducing gyroscopic inertial effects \citep{wang2015topological}, external flow fields \citep{chen2016tunable} or time-modulated materials \citep{chen2019mechanical}. This type of topological PCs -- referred to as the \emph{acoustic/mechanical Chern insulators} -- has been experimentally verified by \citet{ding2019experimental}. This method, however, is still challenging to be applied in real-life situations, due to the complicated implementation of external uniform motion in the lattice and the inherent dynamic instabilities and noise in a moving medium. The second strategy exploits the pseudospin-dependent edge states, and breaks the spatial-inversion symmetry of the time-reversal invariant topological PCs, also known as \emph{quantum spin Hall topological insulators} \citep{zhang2017experimental, xia2017topological, chen2018elastic}. These topological PCs are intrinsically based on the spin-orbit coupling mechanism -- an analogue to the quantum spin Hall effect. In principle, the quantum spin Hall topological insulators have a double Dirac degeneracy in band structures and the topological phase inversion appears at the double Dirac point, where the pseudospin state forms the topological edge state \citep{deng2019comparison}. Depending on the appropriate polarization excitation, the quantum spin Hall topological insulators support robust forward or backward edge states due to their time-reversal symmetry \citep{yang2018topological, yu2018elastic}. The last method is based on the quantum valley Hall effect, which provides a pair of valley vortex states with opposite chirality \citep{lu2017observation, pal2017edge, zhu2018design, wang2020tailoring}. The \emph{quantum valley Hall PCs} also possess the topologically protected one-way edge states along the interface of two domains with different valley vortex states. A systematic comparison between acoustic topological states based on valley Hall and quantum spin Hall effects can be found in the recent work by \citet{deng2019comparison}. In the 1D phononic systems, a combination of PCs with different topological properties may result in topological interface states within the overlapped BGs; see, for example, the observations and predictions in discrete spring-mass \citep{chen2018study}, water wave \citep{yang2016topological}, acoustic \citep{xiao2015geometric} and elastic \citep{yin2018band} systems. For more detailed discussion of the recent progress in the topological acoustic/mechanical systems, interested readers are referred to the comprehensive review articles by \citet{zhang2018topological} and \citet{ma2019topological}.

A major limitation of passive topological PCs is that their operating frequency range of topological states is fixed and extremely narrow (usually corresponding to a single frequency of transmission peak in the BG). To realize a wider operating frequency range of topologically protected states, several methods have been proposed to design actively tunable topological PCs. By adjusting the airflow velocity field  and unit-cell geometric size, the time-reversal symmetry of a 2D PC was broken to tune the BG topology and realize a tunable topological acoustic Chern insulator \citep{chen2016tunable}. The intelligent magnetoelastic materials were introduced by \citet{feng2019magnetically} into the topological system to realize magnetically tunable topological interface states for Lamb waves in 1D PC slabs. The periodic electric boundary conditions were exploited by \citet{zhou2020actively} to generate actively tunable topologically protected interface mode in a 1D piezoelectric rod system. \citet{wang2020tunable2} investigated the topological interface mode in a 1D granular PC composed of two sub-lattices, which can be tuned by varying the pre-compression between the spheres. Although all the above-mentioned works demonstrate the tunability of topologically protected edge/interface states, they operate in the high-frequency scenarios.

Soft PCs offer both the low-frequency operating ranges and high tunability by external stimuli such as pre-stretch \citep{bertoldi2017flexible, galich2017elastic, galich&etal18ijes, wu2018propagation, li2019harnessing, parnell2019soft, gao2019harnessing}, electric \citep{galich&rudykh17jam, wu2017guided, wu2018tuning, mao2019electrostatically, chen2020effects, wang2020tunable, Zhu2020} or magnetic field \citep{karami&etal19jam}. These abilities motivate the exploration of low-frequency tunable topological states in soft PCs. By changing the filling ratio and tuning the mechanical load, the dynamically tunable topological interface state was experimentally observed by \citet{li2018observation} in a circular-hole soft PC plate made of two domains with different topological properties. The design of a 2D quantum valley Hall PC was presented by \citet{liu2018tunable}, where the topological states at the domain interface are triggered by geometric nonlinear effects due to the applied strain. These two works, however, neglect changes in material stiffness induced by the pre-stretch. The influence of the essential material nonlinearity for soft matter was considered by \citet{nguyen2019tunable} in the context of soft topological PCs. They designed a 2D quantum valley Hall PC consisting of soft annular cylinders embedded in an elastic matrix, and utilized the pre-stretch and inflation to actively tune the frequencies of topologically protected edge states. \citet{zhou2020voltage} designed a soft membrane-type PC for the voltage-controlled quantum valley Hall effect in a dielectric elastomeric membrane with sprayed metallic particles. Recently, \citet{huang2020flexible} examined a 1D soft periodic system composed of topologically different plates and realized tunable topological interface states by applying deformation. Nevertheless, the geometric structures and loading ways in the aforementioned soft topological PCs are relatively complex, and it is not easy to combine two differently deformed domains while keeping a smooth interface.

{\color{red} It is well known that the Bragg band gaps (BGs) are usually produced by the material periodicity, geometric periodicity, and periodic boundary conditions \citep{hussein2014dynamics}. With appropriate geometric and/or material design of metamaterials, their wave propagation and attenuation behaviors (such as transmission and reflection) could be optimized \citep{derin2019exhibition,abdulkarim2020design,ozturk2020synergetic,ozturk2020mechanical}.} Inspired by previous works \citep{xiao2015geometric, yin2018band}, here we design a 1D soft phononic crystal cylinder (PCC) composed of step-wise sub-cylinders to realize low-frequency tunable topological interface states under the application of mechanical load. We fully account for both the geometric and material nonlinearities in our theoretical and numerical analyses. The proposed soft topological PCC is a \emph{single-phase} material structure allowing to induce different deformation states in its base elements while preserving a smooth interface between the base elements.  {Due to the low stiffness of soft materials, their operating frequency range is much lower than that of hard materials with the same structure.} We focus on (i) tunability of the topological interface states (in the low-frequency range) by an applied axial force, and (ii) influence of the strain-stiffening effect on the tunability and existence of the topological interface states. To this end, we derive analytically the dispersion relations and acoustic characteristics for small-amplitude longitudinal waves propagating in the finitely deformed PCC. This information is complemented by our numerical calculations including finite element (FE) simulations, elucidating the relations between the morphology, applied loading and material nonlinearity effects on the band structures, transmission characteristics and topological phase diagrams.

This paper is organized as follows. The theoretical background on nonlinear elasticity theory and its associated incremental theory \citep{ogden1997non} is summarized in Sec.~\ref{section2}. The nonlinear static response of the proposed soft PCC with alternating cross-sections is analyzed in Sec.~\ref{Sec3}. Section~\ref{Sec4} describes the derivations of the dispersion relation, transmission coefficient and displacement field of a finite PCC waveguide with various periodic unit cells. Numerical calculations are described in Sec.~\ref{section5}. {\color{red} For a mixed finite neo-Hookean PCC waveguide, the frequency of topological interface states is lowered monotonically by the increasing axial force. However, for a Gent PCC waveguide, the axial force affects the topological interface state frequency in a non-monotonous way that an increase in the axial force leads to the continuous decrease of frequency to a minimum value, and then the frequency is increased reversely by a further increase of the axial force.} Section~\ref{section6} concludes the work with a summary and discussion. Some mathematical derivations and FE simulation procedures are provided in Appendices A-C.

\section{Theoretical Background}
\label{section2}



\subsection{Nonlinear elasticity}\label{section2.1}


We consider a deformable continuous body that occupies the \emph{undeformed reference} configuration ${{\mathcal{B}}_{r}}$ in the Euclidian space with the boundary $\partial {{\mathcal{B}}_{r}}$ and the outward unit normal $\mathbf{N}$. An arbitrary material point labelled as $X$ in the undeformed configuration is identified by the position vector $\mathbf{X}$. Subjected to a mechanical loading, the body deforms and moves to the \emph{deformed or current} configuration ${{\mathcal{B}}_{t}}$ with the boundary $\partial {{\mathcal{B}}_{t}}$ and the outward unit normal $\mathbf{n}_t$, such that the point $X$ occupies a new position $\mathbf{x}=\bm{\chi }\text{(}\mathbf{X},t\text{)}$ at time $t$ in ${{\mathcal{B}}_{t}}$, where an invertible vector function $\bm{\chi }$ is defined for all points in ${{\mathcal{B}}_{r}}$. The deformation gradient tensor is defined as $\mathbf{F}=\partial\mathbf{x}/\partial\mathbf{X} =\text{Grad}\bm{\chi }$, where `Grad' is the gradient operator with respect to ${{\mathcal{B}}_{r}}$. The components of the deformation gradient tensor are ${{{F}}_{i\alpha }}=\partial {{x}_{i}}/\partial {{X}_{\alpha }}$, where Roman and Greek indices are associated with ${{\mathcal{B}}_{t}}$ and ${{\mathcal{B}}_{r}}$, respectively. The local measure of the volume change is denoted by $J=\det\mathbf{F}>0$.

In the absence of body force, the equilibrium equations can be written in Eulerian and Lagrangian forms, respectively, as
\begin{equation} \label{1}
	{\text{div}\mathbf{\bm{\tau }}=\mathbf{\bm{0}}\quad\text{and}\quad {\text{Div}\mathbf{\bm{T }}=\mathbf{\bm{0}}}},
\end{equation}
where $\mathbf{\bm{\tau }}={{J}^{-1}}\mathbf{FT}$ is the Cauchy stress tensor and $\mathbf{T}$ is the nominal stress tensor. Here `div' and `Div' denote the divergence operators relative to ${{\mathcal{B}}_{t}}$ and ${{\mathcal{B}}_{r}}$, respectively. Note that the nominal stress tensor is the transpose of the first Piola-Kirchhoff stress tensor and that both of them are non-symmetric two-point tensors like the deformation gradient tensor. In index notation, the equilibrium equations \eqref{1} read
\begin{equation} \label{1_index}
	\tau_{ij,i}=0 \quad\text{and}\quad T_{\alpha i,\alpha}=0,
\end{equation}
where the Einstein summation convention is used.

Consider a compressible hyperelastic material described in terms of its strain energy density function $\Omega \left(\mathbf{F} \right)$ (per unit reference volume) such that
\begin{equation} \label{3}
{\bf{T}} = \frac{{\partial \Omega}}{{\partial {\bf{F}}}}\quad\text{and}\quad {\bm{\tau }} = {{J}^{-1}}\mathbf{F}\frac{{\partial \Omega}}{{\partial {\bf{F}}}},
\end{equation}
or in index notation
\begin{equation} \label{3_index}
{{T}_{\alpha i}} = \frac{{\partial \Omega}}{{\partial {{F}_{i \alpha}}}} \quad\text{and}\quad {{\tau }_{ij}} = {{J}^{-1}}{F}_{j \alpha}\frac{{\partial \Omega}}{{\partial {{F}_{i \alpha}}}},
\end{equation}
with the  angular momentum  conservation ${{\tau }_{ij}}={{\tau }_{ji}}$.

Alternatively, the strain energy density function $\Omega$ can be expressed in terms of  the principal stretches, i.e., $\Omega=\Omega({\lambda }_{1},{\lambda }_{2},{\lambda }_{3})$ with $J={\lambda }_{1}{\lambda }_{2}{\lambda }_{3}$  \citep{ogden1997non}. Thus, referring to the principal axes of $\bm{\tau }$, the corresponding principal Cauchy stresses $\tau_i$ $(i=1,2,3)$ are expressed as
\begin{equation} \label{4}
{\tau _i} = {J^{ - 1}}{\lambda _i}\frac{{\partial \Omega \left( {{\lambda _1},{\lambda _2},{\lambda _3}} \right)}}{{\partial {\lambda _i}}}, \quad (\textrm{no summation over } i).
\end{equation}

The mechanical boundary conditions on $\partial {{\mathcal{B}}_{t}}$ can be written in Eulerian form as
\begin{equation} \label{2}
{\mathbf{\bm{\tau n}}_t={{\mathbf{t}}^{a}}}, 
\end{equation}
where ${{\mathbf{t}}^{a}}$ is the applied mechanical traction vector per unit area of $\partial {{\mathcal{B}}_{t}}$.

\subsection{Incremental motions superimposed on finitely deformed state}\label{Sec2-2}

A time-dependent infinitesimal incremental motion $\mathbf{\dot{x}}\left( \mathbf{X},t \right)$ is superimposed on a finitely deformed configuration ${{\mathcal{B}}_{0}}$ (with the boundary $\partial{{\mathcal{B}}_{0}}$ and the outward unit normal $\mathbf{n}$). Here, the incremental quantities are represented by a superposed dot. The incremental equation of motion in the updated Lagrangian form is
\begin{equation} \label{5}
{\rm{div}}{{\bf{\dot T}}_0} = \rho {{\bf{u}}_{,tt}},
\end{equation}
where ${{\mathbf{\dot{T}}}_{0}}={{J}^{-1}}\mathbf{F\dot{T}}$ is the push-forward counterpart of the Lagrangian incremental stress tensor $\mathbf{\dot{T}}$, $\mathbf{u}=\mathbf{\dot{x}}\left( \mathbf{X},t \right)$ is the incremental displacement vector, and $\rho ={{\rho }_{0}}{{J}^{-1}}$ is the current mass density in ${{\mathcal{B}}_{t}}$, with ${{\rho }_{0}}$ denoting the mass density in the reference configuration  ${{\mathcal{B}}_{r}}$. The subscript 0 indicates the resulting push-forward quantities and the subscript $t$ following a comma represents the material time derivative. 

The linearized incremental constitutive law for a compressible hyperelastic material is
\begin{equation} \label{6}
{{\bf{\dot T}}_0} = {\bm{{\cal A}}_0}{\bf{H}},
\end{equation}
where $\mathbf{H}=\text{grad}\mathbf{u}$ is the incremental displacement gradient tensor; `grad' is the gradient operator with respect to ${{\mathcal{B}}_{0}}$. The fourth-order \emph{instantaneous} elasticity tensor ${\bm{{\mathcal{A}}}_{0}}$ is represented in component form by
\begin{equation} \label{7}
{{\cal A}_{0piqj}} = {J^{ - 1}}{F_{p\alpha }}{F_{q\beta }}{{\cal A}_{\alpha i\beta j}} = {{\cal A}_{0qjpi}},
\end{equation}
in which $\bm{\mathcal{A}}$ indicates the \emph{referential} elasticity tensor with its components given by
\begin{equation} \label{8}
{{\cal A}_{\alpha i\beta j}} = \frac{{{\partial ^2} \Omega}}{{\partial {F_{i\alpha }}\partial {F_{j\beta }}}},
\end{equation}
Following \citet{ogden1997non} and referring to the principal axes of $\bm{\tau }$, the non-zero components of ${\bm{{\mathcal{A}}}_{0}}$ for compressible isotropic hyperelastic materials can be expressed, in terms of the three principal stretches ${\lambda }_{i}$, as
\begin{equation} \label{9}
\begin{array}{l}
{{\cal A}_{0iijj}} = {{\cal A}_{0jjii}} = {J^{ - 1}}{\lambda _i}{\lambda _j}{\Omega_{ij}}, \vspace{2ex}\\
{{\cal A}_{0ijij}} = {J^{ - 1}}\left\{ \begin{array}{l}
\frac{{{\lambda _i}{\Omega_i} - {\lambda _j}{\Omega_j}}}{{\lambda _i^2 - \lambda _j^2}}\lambda _i^2,\;\;(i \ne j,\;{\lambda _i} \ne {\lambda _j}), \vspace{1ex}\\
\frac{1}{2}\left( {J{{\cal A}_{0iiii}} - J{{\cal A}_{0iijj}} + {\lambda _i}{\Omega_i}} \right),\;\;(i \ne j,\;{\lambda _i} = {\lambda _j}),
\end{array} \right. \vspace{2ex}\\
{{\cal A}_{0ijji}} = {{\cal A}_{0jiij}} = {{\cal A}_{0ijij}} - {J^{ - 1}}{\lambda _i}{\Omega_i} = {{\cal A}_{0ijij}} - {\tau _i},\;\;(i \ne j),
\end{array}
\end{equation}
where ${{\Omega}_{i}}=\partial \Omega/\partial {{\lambda }}_{i}$  and ${{\Omega}_{ij}}={{\partial }^{2}}\Omega/\partial {{\lambda }_{i}}\partial {{\lambda }_{j}}$.

In the updated Lagrangian form, the incremental mechanical boundary conditions, which are to be satisfied on $\partial {{\mathcal{B}}_{0}}$, are written as
\begin{equation} \label{10}
{\bf{\dot T}}_0^{\rm{T}}{\bf{n}} = {\bf{\dot t}}_0^A,
\end{equation}
where the superscript ${(\cdot)}^{\text{T}}$  signifies the usual transpose of a tensor and $\mathbf{\dot{t}}_{0}^{A}$ is the updated Lagrangian incremental mechanical traction vector per unit area of $\partial {{\mathcal{B}}_{0}}$.


\section{Nonlinear deformation of a soft PCC}\label{Sec3}


\begin{figure}[htbp]
	\centering	
	\includegraphics[width=0.83\textwidth]{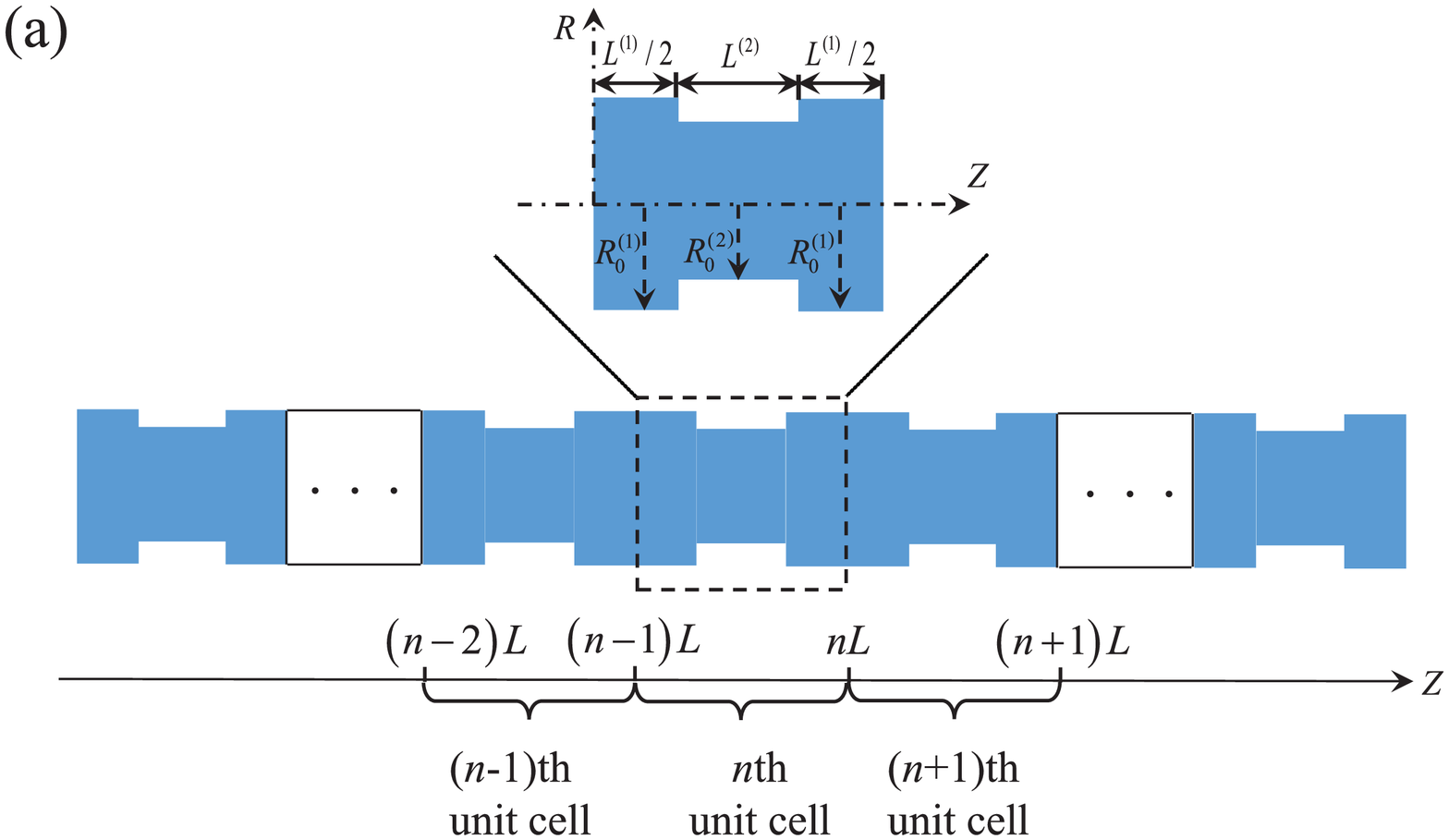}
	\includegraphics[width=0.9\textwidth]{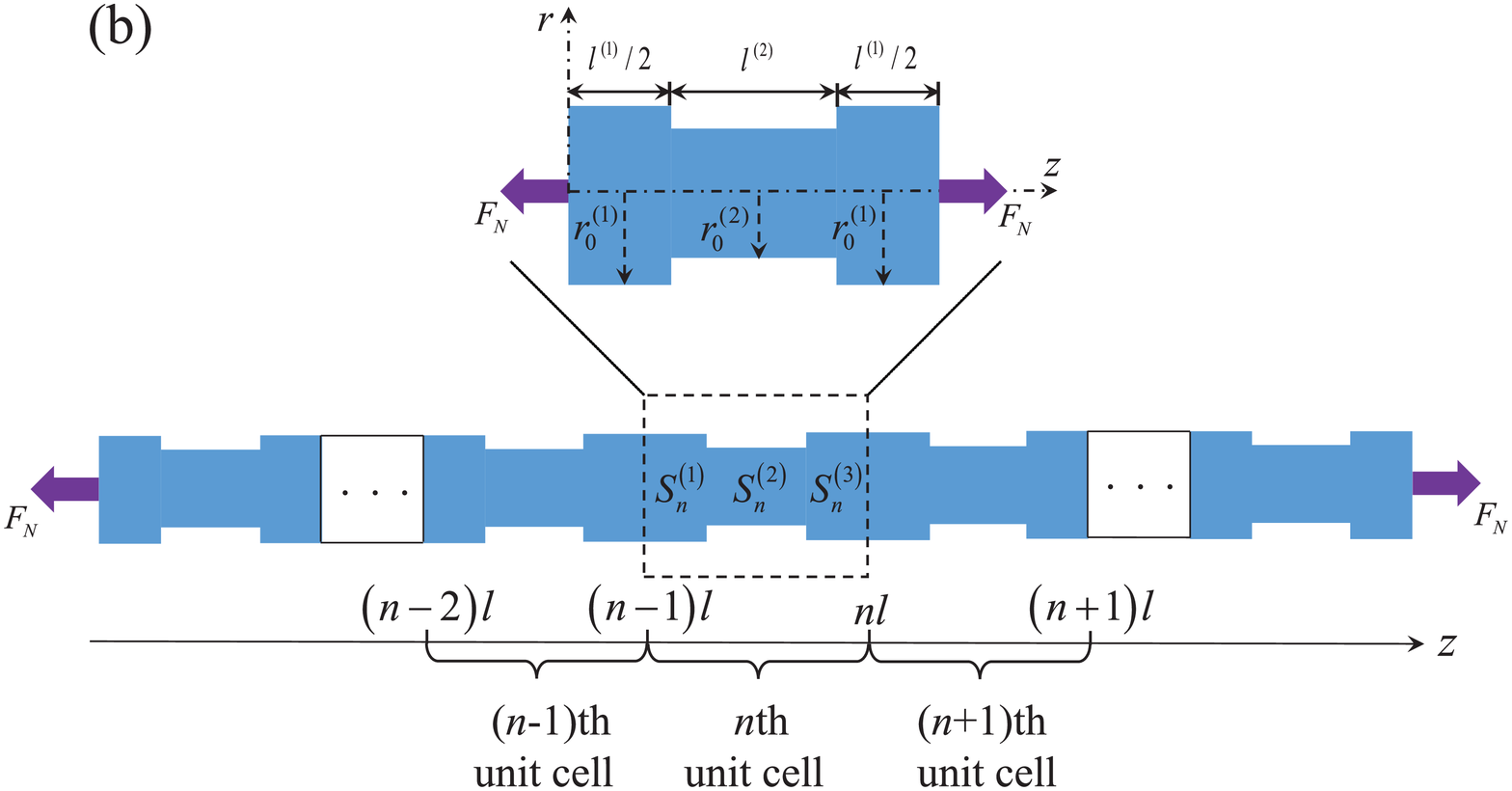}
	\caption{Schematic diagram of an infinite soft PCC composed of step-wise sub-cylinders and its unit cell along with related geometric size and cylindrical coordinates: (a) undeformed configuration and (b) deformed configuration induced by an axial force ${{F}_{N}}$.}
	\label{Fig1}
\end{figure}
 
Consider a \emph{single phase} hyperelastic PCC structure with periodically varying cross-sectional areas as shown in Fig.~\ref{Fig1}(a). Each unit cell has two wider sub-cylinders 1 and 3 of length ${{L}^{\left( 1 \right)}}/2$ and inner radius $R_{\text{0}}^{\left( \text{1} \right)}$, sandwiching a narrower sub-cylinder 2 of length ${{L}^{\left( 2 \right)}}$ and inner radius $R_{\text{0}}^{\left( \text{2} \right)}$. Here and thereafter, the superscript ${{\left( \cdot  \right)}^{\left( p \right)}}$ denotes the physical quantities of the sub-cylinder $p$ ($p=1,2,3$).
The wider sub-cylinders $p=1$ and 3 have identical geometric sizes (i.e., $R_{\text{0}}^{\left( \text{1} \right)}=R_{\text{0}}^{\left( \text{3} \right)}$ and ${{L}^{\left( 1 \right)}}={{L}^{\left( 3 \right)}}$).  In the undeformed configuration, the unit-cell length is $L={{L}^{\left( 1 \right)}}+{{L}^{\left( 2 \right)}}$ along the $Z$ direction. Note that the 1D PCC with inversion symmetry has two inversion centers and without loss of generality, we assign the origin to be at the center of the wider sub-cylinder (see Fig.~\ref{Fig1}). {\color{red} By varying the initial length fraction of sub-cylinder 1 or 2, the Bragg BG could exhibit the evolutionary process of open, close and reopen. The topological transition point where the BG closes is mechanically tunable, which is our main goal and is shown in Sec.~\ref{section5}.}

{\color{red} As shown in Fig.~\ref{Fig1}(b), under the application of tensile axial force, the length of the deformed PCC becomes longer and its lateral size becomes smaller, respectively, than those of the undeformed PCC. The length fraction of the deformed sub-cylinder is different from that of the undeformed one. We note that, due to the geometric inhomogeneity, complex local deformations can develop near the interfaces between the wider and narrower sub-cylinders when subjected to an axial force ${{F}_{N}}$ \citep{parnell2019soft}}. These local deformations, however, are only confined in small regions in the vicinity of the interfaces and barely affect the response of topological interface states in the low-frequency regime of interest here. Therefore, the nonlinear deformation can be approximately assumed uniform  in the theoretical model. As we shall show, this assumption  has been validated by the FE simulations (see Sec.~\ref{Sec5-2}). The deformed configurations of the soft PCC and its unit cell are shown in Fig.~\ref{Fig1}(b). The uniform axisymmetric deformations can be described in two cylindrical coordinate systems $\left( R,\Theta,Z \right)$ and $\left( r,\theta, z \right)$ as follows:
\begin{equation} \label{11}
r = {\lambda _1}R,\quad \theta  = \Theta ,\quad z = {\lambda _3}Z,
\end{equation}
where ${{\lambda }_{1}}$ and ${{\lambda }_{3}}$ are the principal stretches along the radial and axial directions, respectively. Thus, the geometric sizes of each sub-cylinder become 
\begin{equation} \label{12}
{\rm{  }}r_0^{\left( p \right)} = \lambda _1^{\left(p  \right)}R_0^{\left( p \right)}, \quad {l^{\left( p \right)}} = \lambda _3^{\left( p \right)}{L^{\left( p \right)}},
\end{equation}
where $\lambda _{1}^{\left( p \right)}$ and $\lambda _{3}^{\left( p \right)}$ represent the principal stretches of sub-cylinder $p$, and $r_{0}^{\left( p \right)}$, ${{l}^{\left( 1 \right)}}/2={{l}^{\left( 3 \right)}}/2$ and ${{l}^{\left( 2 \right)}}$ are the radius of sub-cylinder $p$ and the lengths of the wider and narrower sub-cylinders in the deformed state, respectively. In addition, $l={{l}^{\left( 1 \right)}}+{{l}^{\left( 2 \right)}}$ is the length of the deformed unit cell. Therefore, the deformation gradient tensor of sub-cylinder $p$ is expressed as ${{\mathbf{F}}^{\left( p \right)}}=\text{diag}[\lambda _{1}^{\left( p \right)},\lambda _{1}^{\left( p \right)},\lambda _{3}^{\left( p \right)}]$, which will be determined by the initial boundary conditions on the lateral surface. 

In order to analyze the longitudinal wave propagation in the soft PCC, the compressible Gent material model \citep{gent1996new} is adopted to characterize the hyperelastic cylinder, which is described as
\begin{equation} \label{13}
{\Omega^{\left( p \right)}} =  - \frac{{{\mu }{J_{m}}}}{2}\ln \left( {1 - \frac{{I_1^{\left( p \right)} - 3}}{{J_{m}}}} \right) - {\mu}\ln {J^{\left( p \right)}} + \left( {\frac{{{\Lambda}}}{2} - \frac{{{\mu}}}{{J_{m}}}} \right){\left( {{J^{\left( p \right)}} - 1} \right)^2},
\end{equation}
where ${{\mu }}$ and ${\Lambda }$ are the shear modulus and the first Lam$\acute{\text{e}}$'s parameter in the undeformed configuration, $I_{1}^{\left( p \right)}=2{{( \lambda _{1}^{\left( p \right)})}^{2}}+{{(\lambda _{3}^{\left( p \right)})}^{2}}$ denotes the first strain invariant and ${{J}^{\left( p \right)}}={{( \lambda _{1}^{\left( p \right)} )}^{2}}\lambda _{3}^{\left( p \right)}$. The bulk modulus is then calculated as ${{K}}={{\Lambda }}+2{{\mu }}/3$. The parameter ${J_{m}}$ is the dimensionless Gent constant used to characterize the strain-stiffening effect of the PCC. Recall that the compressible neo-Hookean model is recovered from Eq.~\eqref{13} when ${J_{m}}\to \infty $. It should be emphasized that the material properties (i.e., $\mu$, $\Lambda$, $K$ and ${J_{m}}$) are the same for the three sub-cylinders, but $I_{1}^{\left( p \right)}$ and ${{J}^{\left( p \right)}}$ are different because of various cross-sections.

In virtue of Eqs.~\eqref{4} and \eqref{13}, we obtain the principal Cauchy stress components for sub-cylinder $p$ as 
\begin{equation} \label{14}
\begin{array}{l}
\tau _1^{\left( p \right)} = \tau _2^{\left( p \right)} = \dfrac{\mu }{{{J^{\left( p \right)}}}}\left[ {\frac{J_{m}}{{J_{m} - I_1^{\left( p \right)} + 3}}{{\left( {\lambda _1^{\left( p \right)}} \right)}^2} - 1} \right] + \left( {\Lambda  - \dfrac{{2\mu }}{J_{m}}} \right)\left( {{J^{\left( p \right)}} - 1} \right), \vspace{1ex}\\
\tau _3^{\left( p \right)} = \dfrac{\mu }{{{J^{\left( p \right)}}}}\left[ {\frac{J_{m}}{{J_{m} - I_1^{\left( p \right)} + 3}}{{\left( {\lambda _3^{\left( p \right)}} \right)}^2} - 1} \right] + \left( {\Lambda  - \dfrac{{2\mu }}{J_{m}}} \right)\left( {{J^{\left( p \right)}} - 1} \right),
\end{array}
\end{equation}
Considering the axial force ${{F}_{N}}$ applied along the $z$ direction as well as the traction-free boundary condition on the lateral surface $r=r_{0}^{\left( p \right)}$, we have
\begin{equation} \label{15}
\tau _1^{\left( p \right)} = \tau _2^{\left( p \right)} = 0,\quad \tau _3^{\left( p \right)} = \frac{{{F}_{N}}}{{{s^{\left( p \right)}}}},
\end{equation}
where ${{s}^{\left( p \right)}}=\pi {{( r_{0}^{\left( p \right)})}^{2}}$ is the area of the deformed cross-section of sub-cylinder $p$. Therefore, the nonlinear algebraic equations \eqref{14} and \eqref{15} can be utilized to completely determine the principal stretch ratios $\lambda _{1}^{\left( p \right)}$ and $\lambda _{3}^{\left( p \right)}$ ($\lambda _{1}^{\left( 1 \right)}=\lambda _{1}^{\left( 3 \right)}$ and $\lambda _{3}^{\left( 1 \right)}=\lambda _{3}^{\left( 3 \right)}$) once the axial force ${{F}_{N}}$ is prescribed.


\section{Analysis of incremental longitudinal wave propagation}\label{Sec4}

 
After obtaining the nonlinear axisymmetric deformations in Sec.~\ref{Sec3}, the solutions of the superimposed incremental longitudinal waves in an initially deformed PCC are derived in Sec.~\ref{Sec4-1}. The transfer matrix method \citep{yeh1977electromagnetic} in conjunction with the Bloch-Floquet theorem \citep{kittel1996introduction} is then employed in Sec.~\ref{Sec4-2} to derive the dispersion relation of incremental wave motions in an infinite PCC, which in turn determines the displacement mode shape of unit cell in Sec.~\ref{Sec4-3}. Furthermore, the transmission coefficient of a finite PCC with \emph{identical} unit cells is provided in Sec.~\ref{Sec4-4}. For a finite cylindrical waveguide consisting of two types of PCCs with \emph{different} unit cells, we derive its transmission coefficient and displacement distribution in Secs.~\ref{Sec4-5} and \ref{Sec4-6}, respectively.


\subsection{Wave solutions of incremental motions} \label{Sec4-1}

For each sub-cylinder $p$, the incremental constitutive law \eqref{6} for the superimposed longitudinal waves can be expressed in component form as
\begin{equation} \label{16}
\begin{array}{l}
\dot T_{011}^{\left( p \right)} = {\cal A}_{01111}^{\left( p \right)}H_{11}^{\left( p \right)} + {\cal A}_{01122}^{\left( p \right)}H_{22}^{\left( p \right)} + {\cal A}_{01133}^{\left( p \right)}H_{33}^{\left( p \right)},\vspace{1ex}\\
\dot T_{022}^{\left( p \right)} = {\cal A}_{01122}^{\left( p \right)}H_{11}^{\left( p \right)} + {\cal A}_{02222}^{\left( p \right)}H_{22}^{\left( p \right)} + {\cal A}_{02233}^{\left( p \right)}H_{33}^{\left( p \right)},\vspace{1ex}\\
\dot T_{033}^{\left( p \right)} = {\cal A}_{01133}^{\left( p \right)}H_{11}^{\left( p \right)} + {\cal A}_{02233}^{\left( p \right)}H_{22}^{\left( p \right)} + {\cal A}_{03333}^{\left( p \right)}H_{33}^{\left( p \right)},
\end{array}
\end{equation}
where the non-zero components of the instantaneous elasticity tensor ${\bm{{\mathcal{A}}}_{0}}$ for the compressible Gent model characterized by Eq.~\eqref{13} may be derived from Eq.~\eqref{9} as
\begin{equation} \label{17}
\begin{array}{l}
{\cal A}_{01111}^{\left( p \right)} = {\cal A}_{02222}^{\left( p \right)} = \dfrac{1}{{{J^{\left( p \right)}}}}\left[ {\frac{{\mu J_{m}{{\left( {\lambda _1^{\left( p \right)}} \right)}^2}}}{{J_{m} - I_1^{\left( p \right)} + 3}}\left( {1 + \frac{{2{{\left( {\lambda _1^{\left( p \right)}} \right)}^2}}}{{J_{m} - I_1^{\left( p \right)} + 3}}} \right) + \mu  + {{\left( {{J^{\left( p \right)}}} \right)}^2}\left( {\Lambda  - \dfrac{{2\mu }}{J_{m}}} \right)} \right], \vspace{1ex}\\
{\cal A}_{01122}^{\left( p \right)} = \frac{{2\mu J_{m}{{\left( {\lambda _1^{\left( p \right)}} \right)}^2}}}{{\lambda _3^{\left( p \right)}{{\left( {J_{m} - I_{\rm{1}}^{\left( p \right)} + 3} \right)}^2}}} + \left( {\Lambda  - \dfrac{{2\mu }}{J_{m}}} \right)\left( {2{J^{\left( p \right)}} - 1} \right), \vspace{1ex}\\
{\cal A}_{01133}^{\left( p \right)} = {\cal A}_{02233}^{\left( p \right)} = \frac{{2\mu J_{m}\lambda _3^{\left( p \right)}}}{{{{\left( {J_{m} - I_{\rm{1}}^{\left( p \right)} + 3} \right)}^2}}} + \left( {\Lambda  - \dfrac{{2\mu }}{J_{m}}} \right)\left( {2{J^{\left( p \right)}} - 1} \right), \vspace{1ex}\\
{\cal A}_{03333}^{\left( p \right)} = \dfrac{1}{{{J^{\left( p \right)}}}}\left[ {\frac{{\mu J_{m}{{\left( {\lambda _3^{\left( p \right)}} \right)}^2}}}{{J_{m} - I_{\rm{1}}^{\left( p \right)} + 3}}\left( {1 + \frac{{2{{\left( {\lambda _3^{\left( p \right)}} \right)}^2}}}{{J_{m} - I_{\rm{1}}^{\left( p \right)} + 3}}} \right) + \mu  + {{\left( {{J^{\left( p \right)}}} \right)}^2}\left( {\Lambda  - \dfrac{{2\mu }}{J_{m}}} \right)} \right].
\end{array}
\end{equation}

As is well-accepted in the classical rod theory, the assumption of 1D stress state \citep{wu2018tuning} is made in the following derivation, which results in $\dot{T}_{011}^{\left( p \right)}=\dot{T}_{022}^{\left( p \right)}=0$. As a result, utilizing Eq.~\eqref{16}$_{1,2}$, we derive $H_{11}^{\left( p \right)}$ and $H_{22}^{\left( p \right)}$ in terms of $H_{33}^{\left( p \right)}$ as
\begin{equation} \label{18}
\left[ {\begin{array}{*{20}{c}}
	{H_{11}^{\left( p \right)}}\\
	{H_{22}^{\left( p \right)}}
	\end{array}} \right] =  - \left[ {\begin{array}{*{20}{c}}
	{{\cal P}_1^{\left( p \right)}}\\
	{{\cal P}_2^{\left( p \right)}}
	\end{array}} \right]H_{33}^{\left( p \right)},
\end{equation}
where
\begin{equation} \label{19}
{\cal P}_1^{\left( p \right)} = \dfrac{{{\cal A}_{02222}^{\left( p \right)}{\cal A}_{01133}^{\left( p \right)} - {\cal A}_{01122}^{\left( p \right)}{\cal A}_{02233}^{\left( p \right)}}}{{{\cal A}_{01111}^{\left( p \right)}{\cal A}_{02222}^{\left( p \right)} - {{( {{\cal A}_{01122}^{\left( p \right)}} )}^2}}}, \quad 
{\cal P}_2^{\left( p \right)} = \dfrac{{ - {\cal A}_{01122}^{\left( p \right)}{\cal A}_{01133}^{\left( p \right)} + {\cal A}_{01111}^{\left( p \right)}{\cal A}_{02233}^{\left( p \right)}}}{{{\cal A}_{01111}^{\left( p \right)}{\cal A}_{02222}^{\left( p \right)} - {{( {{\cal A}_{01122}^{\left( p \right)}} )}^2}}}.
\end{equation}
Inserting Eq.~\eqref{18} into Eq.~\eqref{16}$_{3}$ yields
\begin{equation} \label{20}
\dot T_{033}^{\left( p \right)} = {\cal A}_0^{e\left( p \right)}H_{33}^{\left( p \right)},
\end{equation}
where  $\mathcal{A}_{0}^{e\left( p \right)}=\mathcal{A}_{03333}^{\left( p \right)}-\mathcal{A}_{01133}^{\left( p \right)}\mathcal{P}_{1}^{\left( p \right)}-\mathcal{A}_{02233}^{\left( p \right)}\mathcal{P}_{2}^{\left( p \right)}$ is the effective elastic stiffness. Thus, Eq.~\eqref{20} is the \emph{reduced} incremental constitutive relation in the updated Lagrangian form.

Here, by introducing the incremental axial displacement $w$, we can rewrite $H_{33}^{\left( p \right)}$ as $H_{33}^{\left( p \right)}=\textrm{d}{{w}^{\left( p \right)}}/\textrm{d}z$. Due to the postulation of 1D stress state as well as the applied axial force, the incremental equation \eqref{5} of motion is simplified, if ignoring the lateral inertial effect, as
\begin{equation} \label{21}
\dot T_{033,z}^{\left( p \right)} = {\rho ^{\left( p \right)}}w_{,tt}^{\left( p \right)}.
\end{equation}
Note that taking the lateral inertial effect into account will lead to the Love rod theory \citep{graff1991wave}, which exceeds the scope of the present study. Substituting Eq.~\eqref{20} into Eq.~\eqref{21} and considering that $\mathcal{A}_{0}^{e\left( p \right)}$ is constant in each sub-cylinder $p$, we have
\begin{equation} \label{22}
{\cal A}_0^{e\left( p \right)}w_{,zz}^{\left( p \right)} = {\rho ^{\left( p \right)}}w_{,tt}^{\left( p \right)},
\end{equation}
which is the incremental wave equation for the superimposed longitudinal motions, where all physical fields depend on $z$ and $t$ only.

Consequently, for the harmonic time-dependency ${{{\text{e}}}^{ - {\rm{i}}\omega t}}$ with $\omega$ being the angular frequency, the incremental axial displacement in sub-cylinder $p$ of the $n$th unit cell (see Fig.~\ref{Fig1}(b)) can be written as
\begin{equation} \label{23}
w_n^{\left( p \right)}\left( z,t \right)= \overline w_n^{\left( p \right)}\left( z \right){{\text{e}}^{ - {\rm{i}}\omega t}} = \left(a_n^{\left( p \right)}{\text{e}^{{\rm{i}}{k^{\left( p \right)}}\left( {z - nl} \right)}} + b_n^{\left( p \right)}{\text{e}^{ - {\rm{i}}{k^{\left( p \right)}}\left( {z - nl} \right)}}\right){{{\text{e}}}^{ - {\rm{i}}\omega t}},
\end{equation}
where $\overline{w}_{n}^{\left( p \right)}\left( z \right)$ is the displacement amplitude, $a_{n}^{\left( p \right)}$ and $b_{n}^{\left( p \right)}$ are the undetermined complex coefficients denoting the amplitudes of incident and reflected waves, respectively, and ${{k}^{\left( p \right)}}=\omega/c^{\left( p \right)}$ (with $c^{\left( p \right)}=\sqrt{\mathcal{A}_{0}^{e\left( p \right)}/{{\rho }^{\left( p \right)}}}$) represents the axial wave number in sub-cylinder $p$.

Owing to the difference in cross-sections for various sub-cylinders, it is appropriate to choose the incremental axial force $Q_n^{\left( p \right)}$ rather than the incremental stress  $\dot T_{033n}^{\left( p \right)}$ to be continuous at the interfaces delimiting the sub-cylinders. Inserting Eq.~\eqref{23} into Eq.~\eqref{20} and multiplying the resultant expression by the deformed cross-sectional area leads to the incremental axial force as
\begin{equation} \label{24}
Q_n^{\left( p \right)}\left( z,t \right) = {s^{\left( p \right)}}\dot T_{033n}^{\left( p \right)}\left( z \right) =\overline Q_n^{\left( p \right)}\left( z \right){{{\text{e}}}^{ - {\rm{i}}\omega t}},
\end{equation}
where
\begin{equation} \label{Qnp}
\overline Q_n^{\left( p \right)}\left( z \right)={\rm{i}}{s^{\left( p \right)}k^{\left( p \right)}} {\cal A}_0^{e\left( p \right)}\left[ {a_n^{\left( p \right)}{\text{e}^{{\rm{i}}{k^{\left( p \right)}}\left( {z - nl} \right)}} - b_n^{\left( p \right)}{\text{e}^{ - {\rm{i}}{k^{\left( p \right)}}\left( {z - nl} \right)}}} \right],
\end{equation}
is the axial force amplitude.


\subsection{Dispersion relation for an infinite PCC} \label{Sec4-2}

For simplicity, the state vector in sub-cylinder $p$ of the $n$th unit cell (see Fig.~\ref{Fig1}(b)) is defined as
\begin{equation} \label{25}
{\bf{S}}_n^{\left( p \right)} = \left[ {a_n^{\left( p \right)},b_n^{\left( p \right)}} \right]^{\textrm{T}}.
\end{equation}
The state vectors are not independent of each other and can be connected through the interfacial continuity conditions between different sub-cylinders, which are expressed as
\begin{align} \label{26}
\overline w_{n - 1}^{\left( 3 \right)} &= \overline w_{n}^{\left( 1 \right)},\quad \overline Q_{n - 1}^{\left( 3 \right)} = \overline Q_{n}^{\left( 1 \right)},\quad {\rm{  at  }} \text{ } z = \left( {n - 1} \right)l,\vspace{1ex} \notag \\
\overline w_{n}^{\left( 1 \right)} &= \overline w_{n}^{\left( 2 \right)},\quad \overline Q_{n}^{\left( 1 \right)} = \overline Q_{n}^{\left( 2 \right)},\quad{\rm{  at  }}\text{ } z = \left( {n - 1} \right)l + {l^{\left( 1 \right)}}/2,\vspace{1ex}\\
\overline w_{n}^{\left( 2 \right)} &= \overline w_{n}^{\left( 3 \right)},\quad \overline Q_{n}^{\left( 2 \right)} = \overline Q_{n}^{\left( 3 \right)}, \quad {\rm{  at  }} \text{ } z = nl - {l^{\left( 1 \right)}}/2. \notag
\end{align}
Substituting Eqs.~\eqref{23}-\eqref{25} into Eq.~\eqref{26} and noting $s^{\left( 1 \right)} = s^{\left( 3 \right)}$, $k^{\left( 1 \right)}=k^{\left( 3 \right)}$ and ${\cal A}_0^{e\left( 1 \right)}={\cal A}_0^{e\left( 3 \right)}$, we rewrite the displacement and force continuity conditions \eqref{26} as
\begin{equation} \label{28}
\begin{array}{l}
{\left[ {\begin{array}{*{20}{c}}1&1\\1&{ - 1} \end{array}} \right]{\bf{S}}_{n-1}^{\left( 3 \right)}} = \left[ {\begin{array}{*{20}{c}}{{\text{e}^{ - {\rm{i}}{k^{\left( 1 \right)}}l}}}&{{\text{e}^{{\rm{i}}{k^{\left( 1 \right)}}l}}}\vspace{0.5ex}\\
{{\text{e}^{ - {\rm{i}}{k^{\left( 1 \right)}}l}}}&{ - {\text{e}^{{\rm{i}}{k^{\left( 1 \right)}}l}}}
\end{array}} \right]{\bf{S}}_{n}^{\left( 1 \right)}, \vspace{2ex}\\

{\left[ {\begin{array}{*{20}{c}}
		{{\text{e}^{ - {\rm{i}}{k^{\left( 1 \right)}}\left( {{l^{\left( 1 \right)}}/2 + {l^{\left( 2 \right)}}} \right)}}}&{{\text{e}^{{\rm{i}}{k^{\left( 1 \right)}}\left( {{l^{\left( 1 \right)}}/2 + {l^{\left( 2 \right)}}} \right)}}}\vspace{0.5ex}\\
		{{\text{e}^{ - {\rm{i}}{k^{\left( 1 \right)}}\left( {{l^{\left( 1 \right)}}/2 + {l^{\left( 2 \right)}}} \right)}}}&{ - {\text{e}^{{\rm{i}}{k^{\left( 1 \right)}}\left( {{l^{\left( 1 \right)}}/2 + {l^{\left( 2 \right)}}} \right)}}}
		\end{array}} \right]{\bf{S}}_{n}^{\left( 1 \right)}}\vspace{1ex}\\
	\qquad\qquad= \left[ {\begin{array}{*{20}{c}}
	{{\text{e}^{ - {\rm{i}}{k^{\left( 2 \right)}}\left( {{l^{\left( 1 \right)}}/2 + {l^{\left( 2 \right)}}} \right)}}}&{{\text{e}^{{\rm{i}}{k^{\left( 2 \right)}}\left( {{l^{\left( 1 \right)}}/2 + {l^{\left( 2 \right)}}} \right)}}}\vspace{0.5ex}\\
	{\frac{{{Z^{\left( 2 \right)}}}}{{{Z^{\left( 1 \right)}}}}{\text{e}^{ - {\rm{i}}{k^{\left( 2 \right)}}\left( {{l^{\left( 1 \right)}}/2 + {l^{\left( 2 \right)}}} \right)}}}&{ - \frac{{{Z^{\left( 2 \right)}}}}{{{Z^{\left( 1 \right)}}}}{\text{e}^{{\rm{i}}{k^{\left( 2 \right)}}\left( {{l^{\left( 1 \right)}}/2 + {l^{\left( 2 \right)}}} \right)}}}
	\end{array}} \right]{\bf{S}}_{n}^{\left( 2 \right)}, \vspace{2ex}\\

{\left[ {\begin{array}{*{20}{c}}
		{{\text{e}^{ - {\rm{i}}{k^{\left( 2 \right)}}{l^{\left( 1 \right)}}/2}}}&{{\text{e}^{{\rm{i}}{k^{\left( 2 \right)}}{l^{\left( 1 \right)}}/2}}}\vspace{0.5ex}\\
		{{\text{e}^{ - {\rm{i}}{k^{\left( 2 \right)}}{l^{\left( 1 \right)}}/2}}}&{ - {\text{e}^{{\rm{i}}{k^{\left( 2 \right)}}{l^{\left( 1 \right)}}/2}}}
		\end{array}} \right]{\bf{S}}_{n}^{\left( 2 \right)}} = \left[ {\begin{array}{*{20}{c}}
	{{\text{e}^{ - {\rm{i}}{k^{\left( 1 \right)}}{l^{\left( 1 \right)}}/2}}}&{{\text{e}^{{\rm{i}}{k^{\left( 1 \right)}}{l^{\left( 1 \right)}}/2}}}\vspace{0.5ex}\\
	{\frac{{{Z^{\left( 1 \right)}}}}{{{Z^{\left( 2 \right)}}}}{\text{e}^{ - {\rm{i}}{k^{\left( 1 \right)}}{l^{\left( 1 \right)}}/2}}}&{ - \frac{{{Z^{\left( 1 \right)}}}}{{{Z^{\left( 2 \right)}}}}{\text{e}^{{\rm{i}}{k^{\left( 1 \right)}}{l^{\left( 1 \right)}}/2}}}
	\end{array}} \right]{\bf{S}}_{n}^{\left( 3 \right)},
\end{array}
\end{equation}
where ${{Z}^{\left( p \right)}}={{s}^{\left( p \right)}}{{k}^{\left( p \right)}}\mathcal{A}_{0}^{e\left( p \right)}$ with ${{Z}^{\left( 1 \right)}}={{Z}^{\left( 3 \right)}}$. Through some mathematical manipulations, the transfer relation in Eq.~\eqref{28} can be expressed as
\begin{equation} \label{29}
{\bf{S}}_{n-1}^{\left( 3 \right)} = \left[ {\begin{array}{*{20}{c}}
	{{f_1}}&{{f_2}}\\
	{{f_3}}&{{f_4}}
	\end{array}} \right]{\bf{S}}_{n}^{\left( 3 \right)} \equiv {\bf{M}}{\bf{S}}_{n}^{\left( 3 \right)},
\end{equation}
where $\mathbf{M}$ is the $2\times2$ unit-cell transfer matrix that relates the state vector in one sub-cylinder of a unit cell to that in the same sub-cylinder of the adjacent unit cell, and its components are 
\begin{equation} \label{30}
\begin{array}{l}
{f_1} = {\text{e}^{ - \textrm{i}{k^{\left( 1 \right)}}{l^{\left( 1 \right)}}}}\left[ {\cos {k^{\left( 2 \right)}}{l^{\left( 2 \right)}} - \dfrac{1}{2}\textrm{i}\left( {\dfrac{{{Z^{\left( 2 \right)}}}}{{{Z^{\left( 1 \right)}}}} + \dfrac{{{Z^{\left( 1 \right)}}}}{{{Z^{\left( 2 \right)}}}}} \right)\sin {k^{\left( 2 \right)}}{l^{\left( 2 \right)}}} \right],\vspace{1ex}\\
{f_2} =-{f_3} = \dfrac{1}{2}\textrm{i}\left( {\dfrac{{{Z^{\left( 1 \right)}}}}{{{Z^{\left( 2 \right)}}}} - \dfrac{{{Z^{\left( 2 \right)}}}}{{{Z^{\left( 1 \right)}}}}} \right)\sin {k^{\left( 2 \right)}}{l^{\left( 2 \right)}},\vspace{1ex}\\
{f_4} = {\text{e}^{\textrm{i}{k^{\left( 1 \right)}}{l^{\left( 1 \right)}}}}\left[ {\cos {k^{\left( 2 \right)}}{l^{\left( 2 \right)}} + \dfrac{1}{2}\textrm{i}\left( {\dfrac{{{Z^{\left( 2 \right)}}}}{{{Z^{\left( 1 \right)}}}} + \dfrac{{{Z^{\left( 1 \right)}}}}{{{Z^{\left( 2 \right)}}}}} \right)\sin {k^{\left( 2 \right)}}{l^{\left( 2 \right)}}} \right].
\end{array}
\end{equation}
Note that $\mathbf{M}$ is a unimodular matrix \citep{yeh1977electromagnetic}, and we have
\begin{equation} \label{31}
{f_1}{f_4} - {f_2}{f_3} = 1.
\end{equation}

The deformed PCC is still periodic along the axial direction. Based on the Bloch-Floquet theorem \citep{kittel1996introduction}, the relation of state vectors of the same sub-cylinder in adjacent unit cells takes the form as follows:
\begin{equation} \label{32}
{\bf{S}}_n^{\left( 3 \right)} = {\text{e}^{{\rm{i}}ql}}{\bf{S}}_{n - 1}^{\left( 3 \right)},
\end{equation}
where $q$ is the Bloch wave number. It follows from Eqs.~\eqref{29} and \eqref{32} that the state vector of the Bloch wave satisfies the following eigenvalue problem:
\begin{equation} \label{33}
\left( {{\bf{M}} - {\text{e}^{ - {\rm{i}} ql}}{\bf{I}}} \right) {\bf{S}}_{n}^{\left( 3 \right)}=0,
\end{equation}
As a necessary condition for nontrivial solutions, the determinant of the coefficient matrix of Eq.~\eqref{33} vanishes, which yields the following dispersion relation as
\begin{equation} \label{34}
\cos\left( {ql} \right) = \cos \left( {{k^{\left( 1 \right)}}{l^{\left( 1 \right)}}} \right)\cos \left( {{k^{\left( 2 \right)}}{l^{\left( 2 \right)}}} \right) - \frac{1}{2}\left( {\frac{{{Z^{\left( 1 \right)}}}}{{{Z^{\left( 2 \right)}}}} + \frac{{{Z^{\left( 2 \right)}}}}{{{Z^{\left( 1 \right)}}}}} \right)\sin \left( {{k^{\left( 1 \right)}}{l^{\left( 1 \right)}}} \right)\sin \left( {{k^{\left( 2 \right)}}{l^{\left( 2 \right)}}} \right).
\end{equation}
Thus, Eq.~\eqref{34} determines the relation between $q$ and $\omega$ (i.e., the band structure) for incremental longitudinal waves.
%


\subsection{Displacement mode shape of the deformed unit cell for Bloch waves}\label{Sec4-3}

To interpret the topological characteristics of band structures, it is feasible to examine the symmetry properties of mode shapes of passband/BG edge states \citep{xiao2014surface}. Thus, we will provide the derivation of displacement mode shapes in the deformed unit cell in this subsection. Corresponding to the eigenvalue $ {\text{e}^{ - {\rm{i}} ql}}$, the eigenvector of the transfer matrix for sub-cylinder 3 of the first unit cell is obtained from Eq.~\eqref{33} as
\begin{equation} \label{35}
{\bf{S}}_1^{\left( 3 \right)} = \left[ {\begin{array}{*{20}{c}}
	{{f_2}} \vspace{0.5ex} \\
	{{\text{e}^{ - {\rm{i}}ql}} - {f_1}}
	\end{array}} \right],
\end{equation}
where ${\bf{S}}_1^{\left( 3 \right)}= [{a_1^{\left( 3 \right)}},{b_1^{\left( 3 \right)}}]^\textrm{T}$ denotes the state vector made of the coefficients of incident and reflected waves in sub-cylinder 3. Making use of the interfacial continuity conditions \eqref{26}$_{2,3}$ for the displacement in the first unit cell, we have 
\begin{equation} \label{36}
\begin{array}{l}
{\left[ {\begin{array}{*{20}{c}}
		{{\text{e}^{ - {\rm{i}}{k^{\left( 1 \right)}}\left( {{l^{\left( 1 \right)}}/2 + {l^{\left( 2 \right)}}} \right)}}}&{{\text{e}^{ {\rm{i}}{k^{\left( 1 \right)}}\left( {{l^{\left( 1 \right)}}/2 + {l^{\left( 2 \right)}}} \right)}}}\vspace{0.5ex}\\
		{{\text{e}^{ -  {\rm{i}}{k^{\left( 1 \right)}}\left( {{l^{\left( 1 \right)}}/2 + {l^{\left( 2 \right)}}} \right)}}}&{ - {\text{e}^{ {\rm{i}}{k^{\left( 1 \right)}}\left( {{l^{\left( 1 \right)}}/2 + {l^{\left( 2 \right)}}} \right)}}}
		\end{array}} \right]{\bf{S}}_{1}^{\left( 1 \right)}}\vspace{1ex} \\
\qquad\qquad = \left[ {\begin{array}{*{20}{c}}
	{{\text{e}^{ -  {\rm{i}}{k^{\left( 2 \right)}}\left( {{l^{\left( 1 \right)}}/2 + {l^{\left( 2 \right)}}} \right)}}}&{{\text{e}^{ {\rm{i}}{k^{\left( 2 \right)}}\left( {{l^{\left( 1 \right)}}/2 + {l^{\left( 2 \right)}}} \right)}}}\vspace{0.5ex}\\
	{\frac{{{Z^{\left( 2 \right)}}}}{{{Z^{\left( 1 \right)}}}}{\text{e}^{ -  {\rm{i}}{k^{\left( 2 \right)}}\left( {{l^{\left( 1 \right)}}/2 + {l^{\left( 2 \right)}}} \right)}}}&{ - \frac{{{Z^{\left( 2 \right)}}}}{{{Z^{\left( 1 \right)}}}}{\text{e}^{ {\rm{i}}{k^{\left( 2 \right)}}\left( {{l^{\left( 1 \right)}}/2 + {l^{\left( 2 \right)}}} \right)}}}
	\end{array}} \right]{\bf{S}}_{1}^{\left( 2 \right)},\vspace{2ex}\\

{\left[ {\begin{array}{*{20}{c}}
		{{\text{e}^{ - {\rm{i}}{k^{\left( 2 \right)}}{l^{\left( 1 \right)}}/2}}}&{{\text{e}^{{\rm{i}}{k^{\left( 2 \right)}}{l^{\left( 1 \right)}}/2}}}\vspace{0.5ex}\\
		{{\text{e}^{ - {\rm{i}}{k^{\left( 2 \right)}}{l^{\left( 1 \right)}}/2}}}&{ - {\text{e}^{{\rm{i}}{k^{\left( 2 \right)}}{l^{\left( 1 \right)}}/2}}}
		\end{array}} \right]{\bf{S}}_{1}^{\left( 2 \right)}} = 
\left[ {\begin{array}{*{20}{c}} {{\text{e}^{ - {\rm{i}}{k^{\left( 1 \right)}}{l^{\left( 1 \right)}}/2}}}&{{\text{e}^{{\rm{i}}{k^{\left( 1 \right)}}{l^{\left( 1 \right)}}/2}}}\vspace{0.5ex}\\
	{\frac{{{Z^{\left( 1 \right)}}}}{{{Z^{\left( 2 \right)}}}}{\text{e}^{ - {\rm{i}}{k^{\left( 1 \right)}}{l^{\left( 1 \right)}}/2}}}&{ - \frac{{{Z^{\left( 1 \right)}}}}{{{Z^{\left( 2 \right)}}}}{\text{e}^{{\rm{i}}{k^{\left( 1 \right)}}{l^{\left( 1 \right)}}/2}}}
	\end{array}} \right] {\bf{S}}_{1}^{\left( 3 \right)}.
\end{array}
\end{equation}
where ${\bf{S}}_1^{\left( 1 \right)}= [{a_1^{\left( 1 \right)}},{b_1^{\left( 1 \right)}}]^\textrm{T}$ and ${\bf{S}}_1^{\left( 2 \right)}= [{a_1^{\left( 2 \right)}},{b_1^{\left( 2 \right)}}]^\textrm{T}$ are the state vectors in sub-cylinders 1 and 2 of the first unit cell.

Once the state vectors ${\bf{S}}_1^{\left( p \right)}$ are determined from Eqs.~\eqref{35} and \eqref{36}, we obtain the displacement distributions in the deformed unit cell for incremental Bloch waves as
\begin{equation} \label{37}
\begin{array}{l}
{{\overline w}_1^{\left( 1 \right)}}\left( z \right) = {a_1^{\left( 1 \right)}}{\text{e}^{ {\rm{i}}{k^{\left( 1 \right)}}\left( {z - l} \right)}} + {b_1^{\left( 1 \right)}}{\text{e}^{ -  {\rm{i}}{k^{\left( 1 \right)}}\left( {z - l} \right)}},\quad{\rm{        }}0 \le z \le {l^{\left( 1 \right)}}/2, \vspace{1.5ex}\\
{{\overline w}_1^{\left( 2 \right)}}\left( z \right) = {a_1^{\left(2 \right)}}{\text{e}^{ {\rm{i}}{k^{\left( 2 \right)}}\left( {z - l} \right)}} + {b_1^{\left( 2 \right)}}{\text{e}^{ -  {\rm{i}}{k^{\left( 2 \right)}}\left( {z - l} \right)}},\quad{\rm{       }}{l^{\left( 1 \right)}}/2 \le z \le {l^{\left( 1 \right)}}/2 + {l^{\left( 2 \right)}}, \vspace{1.5ex}\\
{{\overline w}_1^{\left( 3 \right)}}\left( z \right) = {a_1^{\left( 3 \right)}}{\text{e}^{ {\rm{i}}{k^{\left( 1 \right)}}\left( {z - l} \right)}} + {b_1^{\left(3 \right)}}{\text{e}^{ - {\rm{i}}{k^{\left( 1 \right)}}\left( {z - l} \right)}},\quad{\rm{        }}{l^{\left( 1 \right)}}/2 + {l^{\left( 2 \right)}} \le z \le l.
\end{array}
\end{equation}
%


\subsection{Transmission coefficient of a finite PCC with identical unit cells} \label{Sec4-4}


\begin{figure}[htbp]
	\centering
	\setlength{\abovecaptionskip}{5pt}		
	\includegraphics[width=0.86\textwidth]{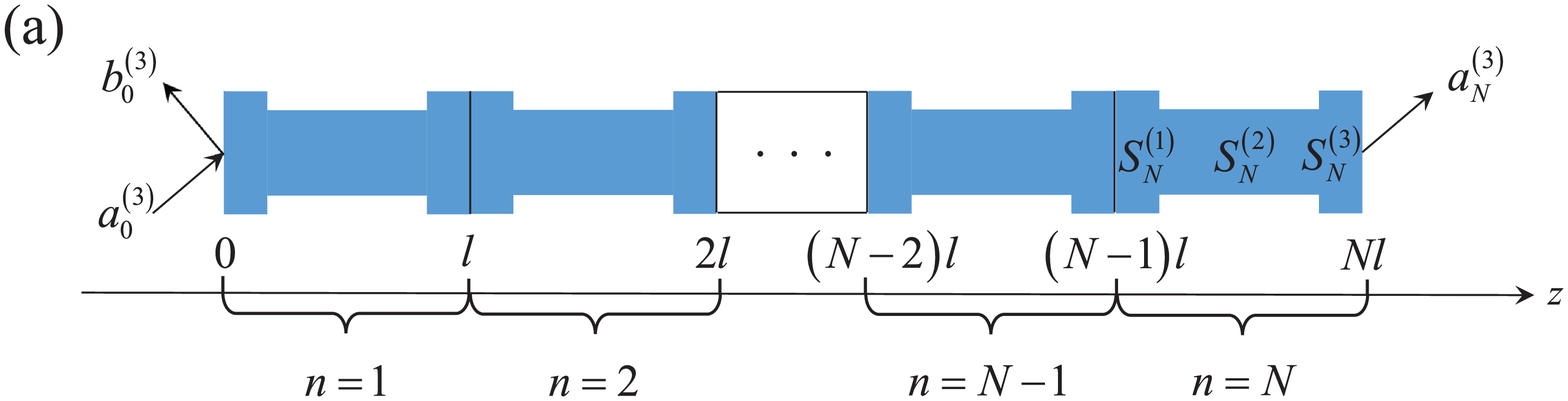}
	\includegraphics[width=0.92\textwidth]{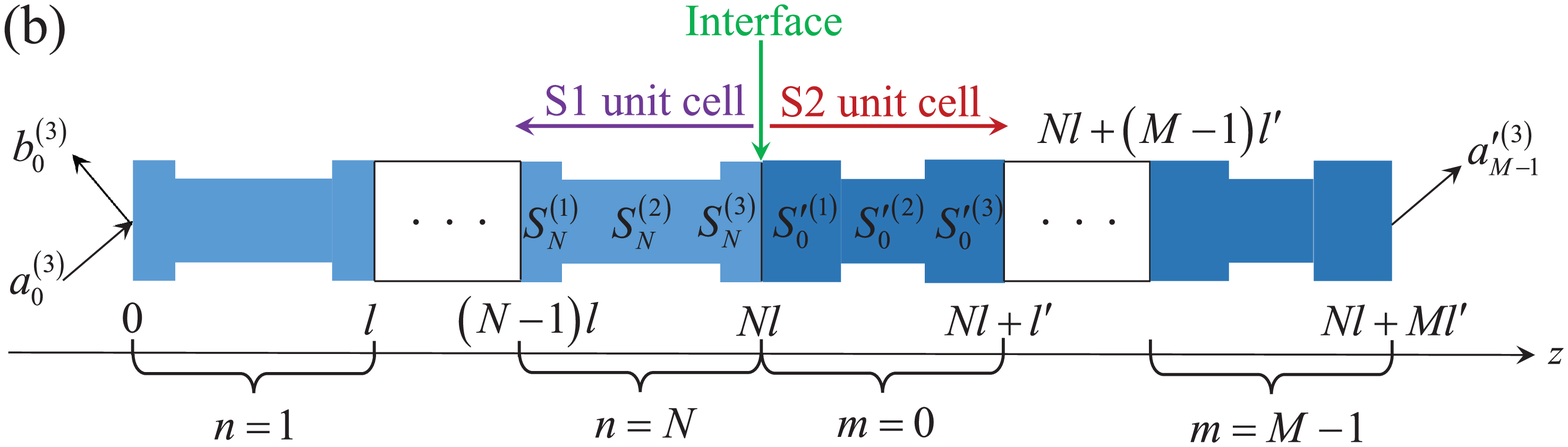}
	\caption{Schematic diagrams of (a) a finite soft PCC consisting of $N$ identical deformed unit cells and (b) a finite cylindrical waveguide made of two types of soft PCCs with different deformed unit cells.}
	\label{Fig2}
\end{figure}

Now consider a finite PCC with $N$ identical \emph{deformed} unit cells arranged in the axial direction (see Fig.~\ref{Fig2}(a)). Inserting Eq.~\eqref{25} into Eq.~\eqref{29}, using Eq.~\eqref{31} and performing matrix transfer $N$ times, we have
\begin{equation} \label{38}
\left[ {\begin{array}{*{20}{c}}
	{a_N^{\left( 3 \right)}}\vspace{0.5ex} \\
	{b_N^{\left( 3 \right)}}
	\end{array}} \right] = \left[ {\begin{array}{*{20}{c}}
	{{f_4}}&{ - {f_2}}\vspace{0.5ex} \\
	{ - {f_3}}&{{f_1}}
	\end{array}} \right]^{ N} \left[ {\begin{array}{*{20}{c}}
	{a_{0}^{\left( 3 \right)}}\vspace{0.5ex} \\
	{b_{0}^{\left( 3 \right)}}
	\end{array}} \right] \equiv {{\bf{M}}_t}\left[ {\begin{array}{*{20}{c}}
	{a_0^{\left( 3 \right)}}\vspace{0.5ex} \\
	{b_0^{\left( 3 \right)}}
	\end{array}} \right],
\end{equation}
where $a_{0}^{\left( 3 \right)}$ and  $b_{0}^{\left( 3 \right)}$ are the amplitude coefficients of incident and reflected waves at the incident side, respectively, and ${{\mathbf{M}}_{t}}$ is the global transfer matrix.

To calculate the transmission spectrum in the finite hyperelastic PCC, we set the reflection coefficient at the output side to be zero (i.e., $b_N^{\left( 3 \right)}=0$). As a result,  the wave coefficient ratios $b_{0}^{\left( 3 \right)}/{a_0^{\left( 3 \right)}}$ and $a_{N}^{\left( 3 \right)}/{a_0^{\left( 3 \right)}}$ are determined from Eq.~\eqref{38} as
\begin{equation} \label{ratio1}
\begin{array}{l}
{b_0^{\left( 3 \right)}}/{a_0^{\left( 3 \right)}} =  - \dfrac{{{M_{t21}}}}{{{M_{t22}}}}, \vspace{1ex}\\
{a_N^{\left( 3 \right)}}/{a_0^{\left( 3 \right)}} = {{M_{t11}} - \dfrac{{{M_{t12}}{M_{t21}}}}{{{M_{t22}}}}},
\end{array}
\end{equation}
where ${{{M}}_{tij}}$ are the components of the global transfer matrix. The transmission coefficient ${t_N}$ defined as the absolute square of ${{a_N^{\left( 3 \right)}}}/{{a_0^{\left( 3 \right)}}}$ is then calculated as
\begin{equation} \label{trans1}
{t_N} = \left| {{a_N^{\left( 3 \right)}}}/{{a_0^{\left( 3 \right)}}} \right|^2 = \left| {M_{t11}} - \frac{{{M_{t12}}{M_{t21}}}}{{{M_{t22}}}} \right|^2.
\end{equation}
%


\subsection{Transmission coefficient of a finite waveguide with two types of different unit cells} \label{Sec4-5}

In order to investigate the existence of topological interface states, the transmission behaviors of a finite cylindrical waveguide composed of two types of different \emph{deformed} unit cells (i.e., $N$ unit cells of S1-type and $M$ unit cells of S2-type arranged consecutively in the axial direction) are considered in this subsection, and its schematic diagram is shown in Fig.~\ref{Fig2}(b). Based on Eq.~\eqref{38}, the transfer relation of $N$ unit cells of S1-type is expressed as
\begin{equation} \label{41}
{\bf{S}}_N^{\left( 3 \right)} = {\left[ {\begin{array}{*{20}{c}}
		{{f_4}}&{ - {f_2}}\\
		{ - {f_3}}&{{f_1}}
		\end{array}} \right]^N}{\bf{S}}_0^{\left( 3 \right)} \equiv {\bf{M}}_t^{\left( 1 \right)} {\bf{S}}_0^{\left( 3 \right)},
\end{equation}
where ${\bf{M}}_t^{\left( 1 \right)}$ is the transfer matrix of $N$ unit cells of S1-type.

Analogous to Eqs.~\eqref{23} and \eqref{Qnp}, the axial displacement and force amplitudes in the $m$th unit cell for the S2-type PCC are
\begin{equation} \label{42}
\begin{array}{l}
\overline w_m^{\prime \left( p \right)}\left( z \right) = a_m^{\prime \left( p \right)}{\text{e}^{{\rm{i}}{k^{\prime \left( p \right)}}\left( {z' - ml'} \right)}} + b_m^{\prime \left( p \right)}{\text{e}^{ - {\rm{i}}{k^{\prime \left( p \right)}}\left( {z' - ml'} \right)}}, \vspace{1ex}\\
\overline Q_m^{\prime\left( p \right)}\left( z \right)={\rm{i}} {{Z}^{\prime \left( p \right)}} \left[ {a_m^{\prime \left( p \right)}{\text{e}^{{\rm{i}}{k^{\prime \left( p \right)}}\left( {z' - ml'} \right)}} - b_m^{\prime \left( p \right)}{\text{e}^{ - {\rm{i}}{k^{ \prime \left( p \right)}}\left( {z' - ml'} \right)}}} \right],
\end{array}
\end{equation}
where $z'=z-Nl-l'$, ${{\left( \cdot  \right)}^\prime }$ stands for the related parameters and physical quantities of the S2-type unit cell, and $m$ is chosen to vary from 0 to $M-1$ for the purpose of illustration (see Fig.~\ref{Fig2}(b)). Referring to Eqs.~\eqref{29} and \eqref{31}, we can obtain the transfer relation between the state vectors for sub-cylinder 3 in two adjacent S2-type unit cells as
\begin{equation} \label{43}
{\bf{S}}_m^{\prime \left( 3 \right)} = \left[ {\begin{array}{*{20}{c}}
	{f}_{4}^{\prime }&-{f}_{2}^{\prime } \vspace{0.5ex}\\
	-{f}_{3}^{\prime } &{f}_{1}^{\prime }
	\end{array}} \right] {\bf{S}}_{m-1}^{\prime \left( 3 \right)}.
\end{equation}
Similar to Eq.~\eqref{41}, the transfer relation of the last $M-1$ unit cells of S2-type is
\begin{equation} \label{44}
{\bf{S}}_{M-1}^{\prime \left( 3 \right)} = {\left[ {\begin{array}{*{20}{c}}
		{f}_{4}^{\prime } & -{f}_{2}^{\prime }  \vspace{0.5ex} \\
		-{f}_{3}^{\prime } & {f}_{1}^{\prime }  \\
		\end{array}} \right]^{M - 1}}{\bf{S}}_0^{\prime \left( 3 \right)} \equiv {\bf{M}}_t^{\left( 2 \right)}{\bf{S}}_0^{\prime \left( 3 \right)},
\end{equation}
where ${\bf{M}}_t^{\left( 2 \right)}$ indicates the transfer matrix for the last $M-1$ S2-type unit cells.

Furthermore, the interfacial continuity condition between the two different PCCs and those in the first S2-type unit cell ($m=0$) are written as 
\begin{equation} \label{45}
\begin{array}{l}
\overline w_{N}^{\left( 3 \right)} =\overline w_0^{\prime \left( 1 \right)},\quad \overline Q_{N}^{\left( 3 \right)} = \overline Q_0^{\prime \left( 1 \right)},\quad {\rm{  at  }} \text{ } z = Nl, \vspace{1ex}\\
\overline w_0^{\prime \left( 1 \right)} =\overline w_0^{\prime \left(2 \right)},\quad \overline Q_0^{\prime \left( 1 \right)} = \overline Q_0^{\prime \left( 2 \right)},\quad{\rm{  at  }} \text{ } z = Nl+{{{l}'}^{\left( 1 \right)}}/2, \vspace{1ex}\\
\overline w_0^{\prime \left( 2 \right)} =\overline w_0^{\prime \left(3 \right)},\quad \overline Q_0^{\prime \left( 2 \right)} = \overline Q_0^{\prime \left( 3 \right)}, \quad {\rm{  at  }} \text{ } z = Nl+{l}'-{{{l}'}^{\left( 1 \right)}}/2.
\end{array}
\end{equation}
Using Eqs.~\eqref{23}, \eqref{Qnp} and \eqref{42}, Eq.~\eqref{45} becomes
\begin{equation} \label{46}
\begin{array}{l}
\left[ {\begin{array}{*{20}{c}}
	1&1\\
	1&{ - 1}
	\end{array}} \right]{\bf{S}}_N^{\left( 3 \right)}=\left[ {\begin{array}{*{20}{c}}
	{{\text{e}^{ - {\rm{i}}{{k'}^{\left( 1 \right)}}l'}}}&{{\text{e}^{{\rm{i}}{{k'}^{\left( 1 \right)}}l'}}} \vspace{0.5ex} \\
	{\frac{{{{Z'}^{\left( 1 \right)}}}}{{{Z^{\left( 1 \right)}}}}{\text{e}^{ - {\rm{i}}{{k'}^{\left( 1 \right)}}l'}}}&{ - \frac{{{{Z'}^{\left( 1 \right)}}}}{{{Z^{\left( 1 \right)}}}}{\text{e}^{{\rm{i}}{{k'}^{\left( 1 \right)}}l'}}}
	\end{array}} \right]{\bf{S}}_0^{\prime \left( 1 \right)},
\vspace{2ex}\\

\left[ {\begin{array}{*{20}{c}}
	{{\text{e}^{ - {\rm{i}}{{k'}^{\left( 1 \right)}}\left( {{{l'}^{\left( 1 \right)}}/2 + {{l'}^{\left( 2 \right)}}} \right)}}}&{{\text{e}^{{\rm{i}}{{k'}^{\left( 1 \right)}}\left( {{{l'}^{\left( 1 \right)}}/2 + {{l'}^{\left( 2 \right)}}} \right)}}} \vspace{0.5ex}\\
	{{\text{e}^{ - {\rm{i}}{{k'}^{\left( 1 \right)}}\left( {{{l'}^{\left( 1 \right)}}/2 + {{l'}^{\left( 2 \right)}}} \right)}}}&{ - {\text{e}^{{\rm{i}}{{k'}^{\left( 1 \right)}}\left( {{{l'}^{\left( 1 \right)}}/2 + {{l'}^{\left( 2 \right)}}} \right)}}}
	\end{array}} \right]{\bf{S}}_0^{\prime \left( 1 \right)}
\vspace{1ex}\\
    \qquad \qquad=\left[ {\begin{array}{*{20}{c}}
	{{\text{e}^{ - {\rm{i}}{{k'}^{\left( 2 \right)}}\left( {{{l'}^{\left( 1 \right)}}/2 + {{l'}^{\left( 2 \right)}}} \right)}}}&{{\text{e}^{{\rm{i}}{{k'}^{\left( 2 \right)}}\left( {{{l'}^{\left( 1 \right)}}/2 + {{l'}^{\left( 2 \right)}}} \right)}}} \vspace{0.5ex}\\
	{\frac{{{{Z'}^{\left( 2 \right)}}}}{{{{Z'}^{\left( 1 \right)}}}}{\text{e}^{ - {\rm{i}}{{k'}^{\left( 2 \right)}}\left( {{{l'}^{\left( 1 \right)}}/2 + {{l'}^{\left( 2 \right)}}} \right)}}}&{ - \frac{{{{Z'}^{\left( 2 \right)}}}}{{{{Z'}^{\left( 1 \right)}}}}{\text{e}^{{\rm{i}}{{k'}^{\left( 2 \right)}}\left( {{{l'}^{\left( 1 \right)}}/2 + {{l'}^{\left( 2 \right)}}} \right)}}}
	\end{array}} \right]{\bf{S}}_0^{\prime \left( 2 \right)},
\vspace{2ex}\\

\left[ {\begin{array}{*{20}{c}}
	{{\text{e}^{ - {\rm{i}}{{k'}^{\left( 2 \right)}}{{l'}^{\left( 1 \right)}}/2}}}&{{\text{e}^{{\rm{i}}{{k'}^{\left( 2 \right)}}{{l'}^{\left( 1 \right)}}/2}}} \vspace{0.5ex} \\
	{{\text{e}^{ - {\rm{i}}{{k'}^{\left( 2 \right)}}{{l'}^{\left( 1 \right)}}/2}}}&{ - {\text{e}^{{\rm{i}}{{k'}^{\left( 2 \right)}}{{l'}^{\left( 1 \right)}}/2}}}
	\end{array}} \right]{\bf{S}}_0^{\prime \left( 2 \right)}=\left[ {\begin{array}{*{20}{c}}
	{{\text{e}^{ - {\rm{i}}{{k'}^{\left( 1 \right)}}{{l'}^{\left( 1 \right)}}/2}}}&{{\text{e}^{{\rm{i}}{{k'}^{\left( 1 \right)}}{{l'}^{\left( 1 \right)}}/2}}}\vspace{0.5ex} \\
	{\frac{{{{Z'}^{\left( 1 \right)}}}}{{{{Z'}^{\left( 2 \right)}}}}{\text{e}^{ - {\rm{i}}{{k'}^{\left( 1 \right)}}{{l'}^{\left( 1 \right)}}/2}}}&{ - \frac{{{{Z'}^{\left( 1 \right)}}}}{{{{Z'}^{\left( 2 \right)}}}}{\text{e}^{{\rm{i}}{{k'}^{\left( 1 \right)}}{{l'}^{\left( 1 \right)}}/2}}}
	\end{array}} \right]{\bf{S}}_0^{\prime \left( 3 \right)}.
\end{array}
\end{equation}
Through some mathematical manipulations, Eq.~\eqref{46} is rewritten as
\begin{equation} \label{47}
{\bf{S}}_0^{\prime \left( 3 \right)}={{\mathbf{M}}_{\operatorname{int}}} {\bf{S}}_N^{\left( 3 \right)},
\end{equation}
where ${\bf{M}}_{\textrm{int}}$ is the $2\times2$ interfacial transfer matrix between the two different PCCs and its components are omitted here due to their redundancy.

Combining Eqs.~\eqref{41}, \eqref{44} and \eqref{47}, we obtain the final global transfer relation of the finite cylindrical waveguide as
\begin{equation} \label{48}
{\bf{S}}_{M-1}^{\prime \left( 3 \right)} = {\bf{K}} {\bf{S}}_0^{\left( 3 \right)}.
\end{equation}
where ${\bf{K}} = {\bf{M}}_t^{\left( 2 \right)}{{\bf{M}}_{{\mathop{\rm int}} }}{\bf{M}}_t^{\left( 1 \right)}$ is the $2\times2$ global transfer matrix. Provided that the reflection coefficient at the output side is equal to zero (i.e., $b_{M-1}^{\prime \left( 3 \right)}=0$), the transmission coefficient ${t_{N+M}}$ is calculated as
\begin{equation} \label{49}
{t_{N+M}} = \left| {{a_{M-1}^{\prime \left( 3 \right)}}}/{{a_0^{\left( 3 \right)}}} \right|^2 = \left| {K_{11}} - \frac{{{K_{12}}{K_{21}}}}{{{K_{22}}}} \right|^2,
\end{equation}
where ${{{K}}_{ij}}$ are the components of ${\bf{K}}$. Note that Eq.~\eqref{49} has the same form as Eq.~\eqref{trans1}.


\subsection{Displacement field of a finite waveguide with two types of different unit cells}\label{Sec4-6}

In this subsection, we will derive the incremental displacement distribution of a finite \emph{deformed} waveguide consisting of $N$ unit cells of S1-type and $M$ unit cells of S2-type. In view of the second formula of Eq.~\eqref{48} with $b_{M-1}^{\prime \left( 3 \right)}=0$, we obtain the relation ${b_0^{\left( 3 \right)}}/{a_0^{\left( 3 \right)}} =  -{{{K_{21}}}}/{{{K_{22}}}}$. Without loss of generality, we set ${a_0^{\left( 3 \right)}}=1$ and thus the state vector composed of the complex amplitude coefficients at the input side (see Fig.~\ref{Fig2}(b)) is
\begin{equation} \label{50}
{\bf{S}}_0^{\left( 3 \right)} = \left[1, -{{{K_{21}}}}/{{{K_{22}}}} \right]^{\textrm{T}},
\end{equation}
which, combined with the transfer relation \eqref{29} between two adjacent unit cells, yields the state vector of sub-cylinder 3 in the $n$th S1-type unit cell as
\begin{equation} \label{51}
{\bf{S}}_n^{\left( 3 \right)} = {\left[ {\begin{array}{*{20}{c}}
		{{f_4}}&{{-f_2}}\\
		{{-f_3}}&{{f_1}}
		\end{array}} \right]^{  n}}{\bf{S}}_0^{\left( 3 \right)}, \quad \left(n = 1,2,...,N\right).
\end{equation}
Utilizing the interfacial continuity conditions \eqref{26}$_{2,3}$ for the displacement in the $n$th S1-type unit cell, we obtain the state vectors ${\bf{S}}_n^{\left( 1 \right)}$ and ${\bf{S}}_n^{\left( 2 \right)}$ of sub-cylinders 1 and 2 in terms of ${\bf{S}}_n^{\left( 3 \right)}$.

Similarly, the state vector of sub-cylinder 3 in the $m$th S2-type unit cell is written as
\begin{equation} \label{52}
{\bf{S}}_m^{\prime \left( 3 \right)} = {\left[ {\begin{array}{*{20}{c}}
		{f}_{4}^{\prime } & -{f}_{2}^{\prime }\vspace{0.5ex} \\
		-{f}_{3}^{\prime } & {f}_{1}^{\prime } 
		\end{array}} \right]^{m}}{\bf{S}}_0^{\prime \left( 3 \right)}, \quad \left(m = 0,1,...,M-1\right),
\end{equation}
where ${\bf{S}}_0^{\prime \left( 3 \right)}$ can be achieved from Eqs.~\eqref{47} and \eqref{51}. Analogous to the S1-type unit cell, the state vectors ${\bf{S}}_m^{\prime \left( 1 \right)}$ and ${\bf{S}}_m^{\prime \left( 2 \right)}$ of sub-cylinders 1 and 2 can be expressed by ${\bf{S}}_m^{\prime \left( 3 \right)}$ when using the interfacial continuity conditions for the S2-type unit cell.

After the state vectors ${\bf{S}}_n^{\left( p \right)}$ and ${\bf{S}}_m^{\prime \left( p \right)}$ are determined, the incremental axial displacements for the two different PCCs are calculated, respectively, by Eqs.~\eqref{23} and \eqref{42}$_1$, which then determines the mode distributions $\overline w_n^{\left( p \right)}\left( z \right)$ and $\overline w_m^{\prime \left( p \right)}\left( z \right)$ of the longitudinal waves propagating in the finite waveguide consisting of two types of different deformed unit cells.


\section{Results and discussion}\label{section5}


This section will elucidate the tunable effects of mechanical load on the topological interface state of longitudinal waves propagating in the hyperelastic PCC characterized by the neo-Hookean and Gent models, respectively. As described in Sec.~\ref{Sec3}, the developed theoretical model assumes a uniform nonlinear deformation when subjected to an axial force and neglects the locally nonuniform deformation near the interfaces of different sub-cylinders. Therefore,  the effectiveness of this hypothesis will be validated  in Subsec.~\ref{Sec5-2-2} by performing the FE simulations based on the commercial software package ABAQUS.

In the following numerical calculations, the undeformed unit cell shown in Fig.~\ref{Fig1}(a) has the radius $R_{0}^{\left( 1 \right)}=0.5$~cm and the length ${{L}^{\left( 1 \right)}}={{\phi }_{0}}L$ for sub-cylinder 1, along with the corresponding parameters $R_{0}^{\left( 2 \right)}=0.4$~cm and ${{L}^{\left( 2 \right)}}=\left( 1-{{\phi }_{0}} \right)L$ for sub-cylinder 2, where the total length is $L=10$~cm and $\phi_{0}$ is the initial length fraction of sub-cylinder 1. Note that the radii and the total lengths of the undeformed unit cells are the same for the two base elements of the PCC waveguide, which ensures a smooth interface between the two PCC elements after deformation. In addition, the hyperelastic PCC is characterized by the commercial product Zhermarck Elite Double 32 made of silicon rubber \citep{galich2017elastic} with its initial density, shear modulus and first Lam$\acute{\text{e}}$'s parameter given as ${{\rho}_{0}}=1040$~kg/m$^3$, $\mu =0.444$~MPa and $\Lambda =22.2$~MPa, respectively. The dimensionless axial force is defined as $ {{\overline{F}}_{N}}={{{F}_{N}}}/{\mu {{S}^{\left( \text{2} \right)}}}$, where ${{S}^{\left( \text{2} \right)}}$ is the initial cross-sectional area of sub-cylinder 2. We define the normalized Bloch wave number as $\overline{q}={ql}/{2\pi}$ ranging from $-0.5$ to $0.5$ within the first Brillouin zone \citep{kittel1996introduction}. The ordinary frequency $f$ measured in Hz is given by $f=\omega/(2\pi)$.


\subsection{Nonlinear static deformation} \label{Sec5-1}

\begin{figure}[htbp]
	\centering
	\setlength{\abovecaptionskip}{5pt}
	\setlength{\belowcaptionskip}{0pt}
	\includegraphics[width=0.39\textwidth]{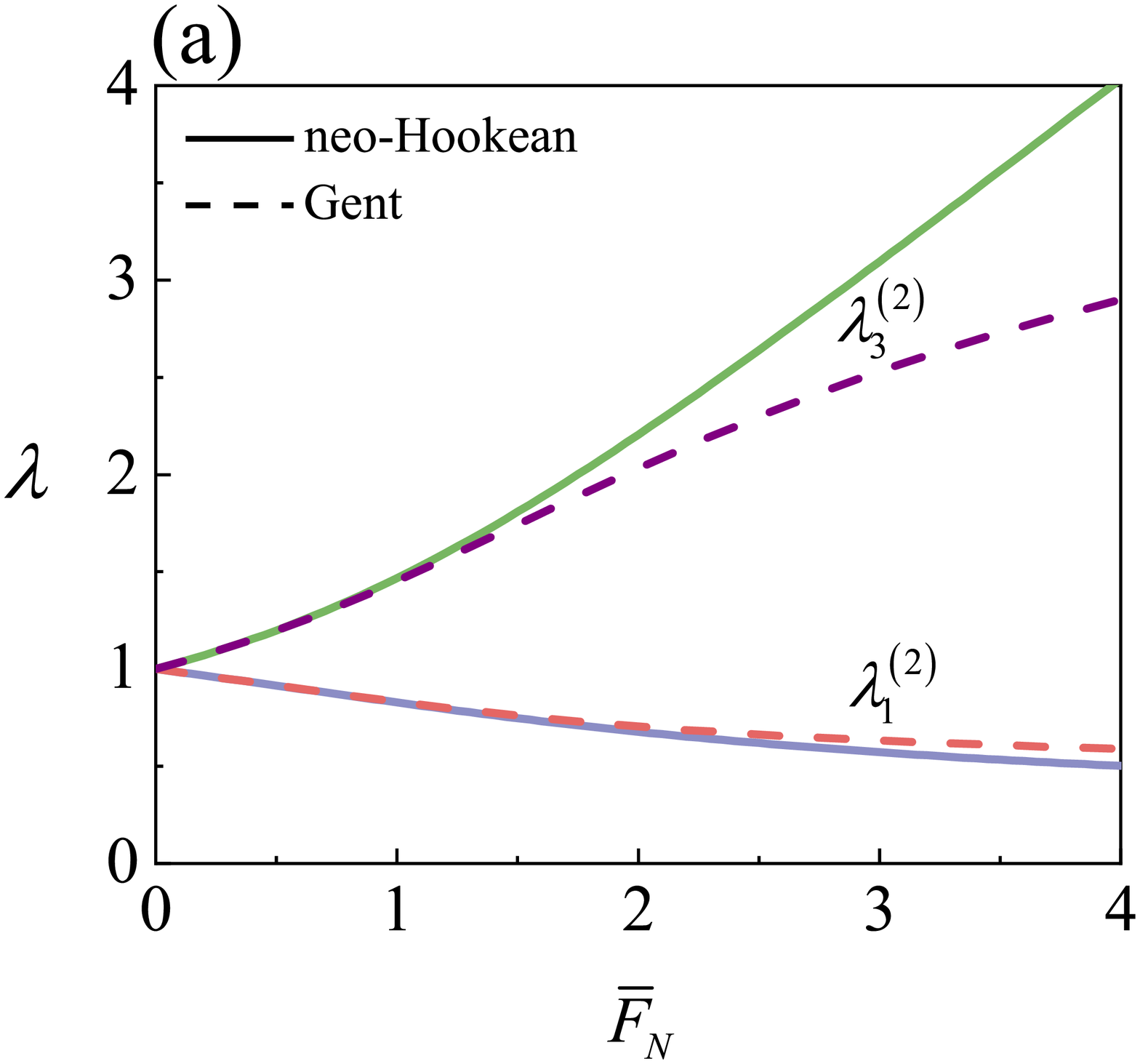}
	\hspace{0.005\textwidth}
	\includegraphics[width=0.41\textwidth]{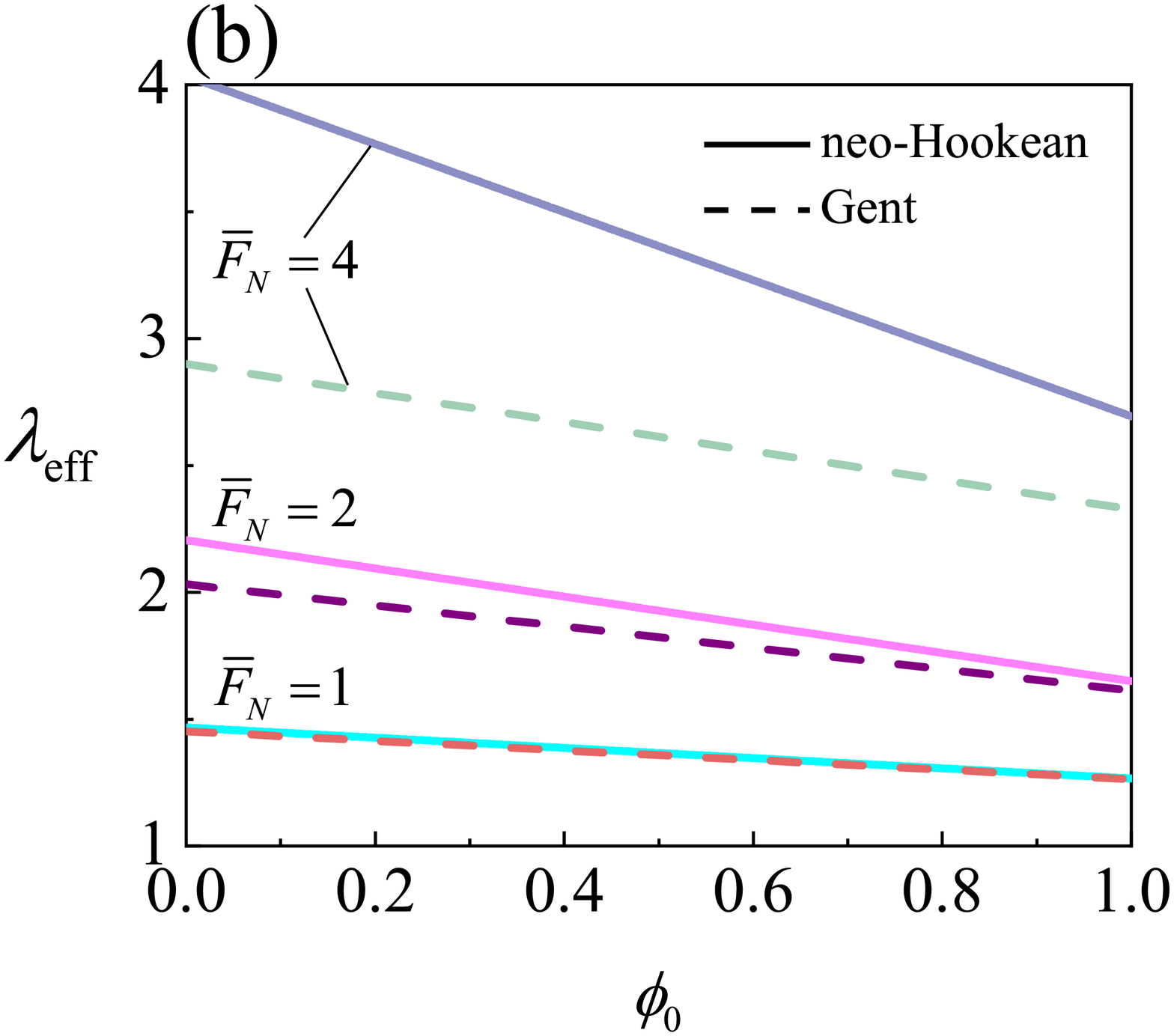}
	\caption{Stretch ratios $\lambda _{1}^{\left( 2 \right)}$ and $\lambda _{3}^{\left( 2 \right)}$ as functions of the normalized axial force ${{\overline{F}}_{N}}$ (a) and overall stretch ratio $\lambda _{\rm eff}$ versus the initial length fraction ${\phi }_{0}$ for ${{\overline{F}}_{N}}=1$, $2$ and $4$ (b) in the neo-Hookean and Gent ($J_{m}=20$) PCC with $R_{0}^{\left( 1 \right)}/R_{0}^{\left( 2 \right)}=1.25$ and $\Lambda/\mu =50$.}
	\label{Fig3}
\end{figure}

First, we examine the nonlinear axisymmetric deformation of the soft PCC under the action of axial force. The analytical results are calculated from the nonlinear algebraic equations \eqref{14} and \eqref{15}. Fig.~\ref{Fig3}(a) shows the variations of axial ($\lambda _{3}^{\left( 2 \right)}$) and radial ($\lambda _{1}^{\left( 2 \right)}$) stretch ratios of sub-cylinder 2 with the dimensionless axial force ${{\overline{F}}_{N}}$ for both the neo-Hookean and Gent ($J_{m}=20$) models.  Indeed, the results for the neo-Hookean model are also recovered by those for the Gent model with a large enough value of $J_{m}$ (e.g., $J_{m}=2000$). Clearly, when the axial tensile force is applied, the length of sub-cylinder 2 increases and its radius shrinks owing to the Poisson's effect. The induced stretch variations of the two material models overlap for ${{\overline{F}}_{N}}\le 1.25$. Nevertheless, the difference between the axial stretches of the neo-Hookean and Gent models becomes more evident with an increase in the axial force when ${{\overline{F}}_{N}}>1.25$. Specifically, at the same axial force level, the Gent PCC with the strain-stiffening effect experiences a smaller deformation as compared to the neo-Hookean PCC. Similar results can be obtained for sub-cylinder 1 in the unit cell.

Intuitively, the sub-cylinder with a smaller cross-section will be elongated more than the counterpart with a larger one in the unit cell. For completeness, we plot in Fig.~\ref{Fig3}(b) the dependence of the overall stretch ratio, defined as
\begin{equation} \label{lmEff}
\lambda _{\rm eff}= {\phi _0}\lambda _3^{\left( 1 \right)} + \left( {1 - {\phi _0}} \right)\lambda _3^{\left( 2 \right)},
\end{equation}
on the initial length fraction ${\phi }_{0}$ for sub-cylinder 1 with a larger cross-section. The results are shown for the neo-Hookean and Gent PCCs subjected to different axial force levels, namely, ${{\overline{F}}_{N}}= 1$, $2$ and $4$. As expected, the PCC with a smaller ${\phi }_{0}$ (i.e., a smaller length fraction of sub-cylinder 1 with a larger radius) develops larger deformations when subjected to an identical axial force. As the axial force level is increased, this variation trend strengthens irrespective of the material model. However, this variation trend of the Gent PCC weakens for a large axial force activating the strain-stiffening effect, compared to that of the neo-Hookean PCC (such as ${{\overline{F}}_{N}}= 4$ in Fig.~\ref{Fig3}(b)). Note that we have selected the sub-cylinders with a relatively small difference in their initial cross-sectional areas to diminish the influence of the inhomogeneous deformation at the interface, especially for a larger axial force ${{\overline{F}}_{N}}$.

Next we will examine the tunable topological interface states propagating in the neo-Hookean and Gent PCC waveguides separately, to illustrate the effects of geometric and material nonlinearities.


\subsection{Tunable topological interface states for the neo-Hookean model} \label{Sec5-2}

\subsubsection{Analysis of band structures and topological characteristics} \label{Sec5-2-1}


%
\begin{figure}[htbp]
	\centering
	\setlength{\abovecaptionskip}{5pt}
	\includegraphics[width=0.3\textwidth]{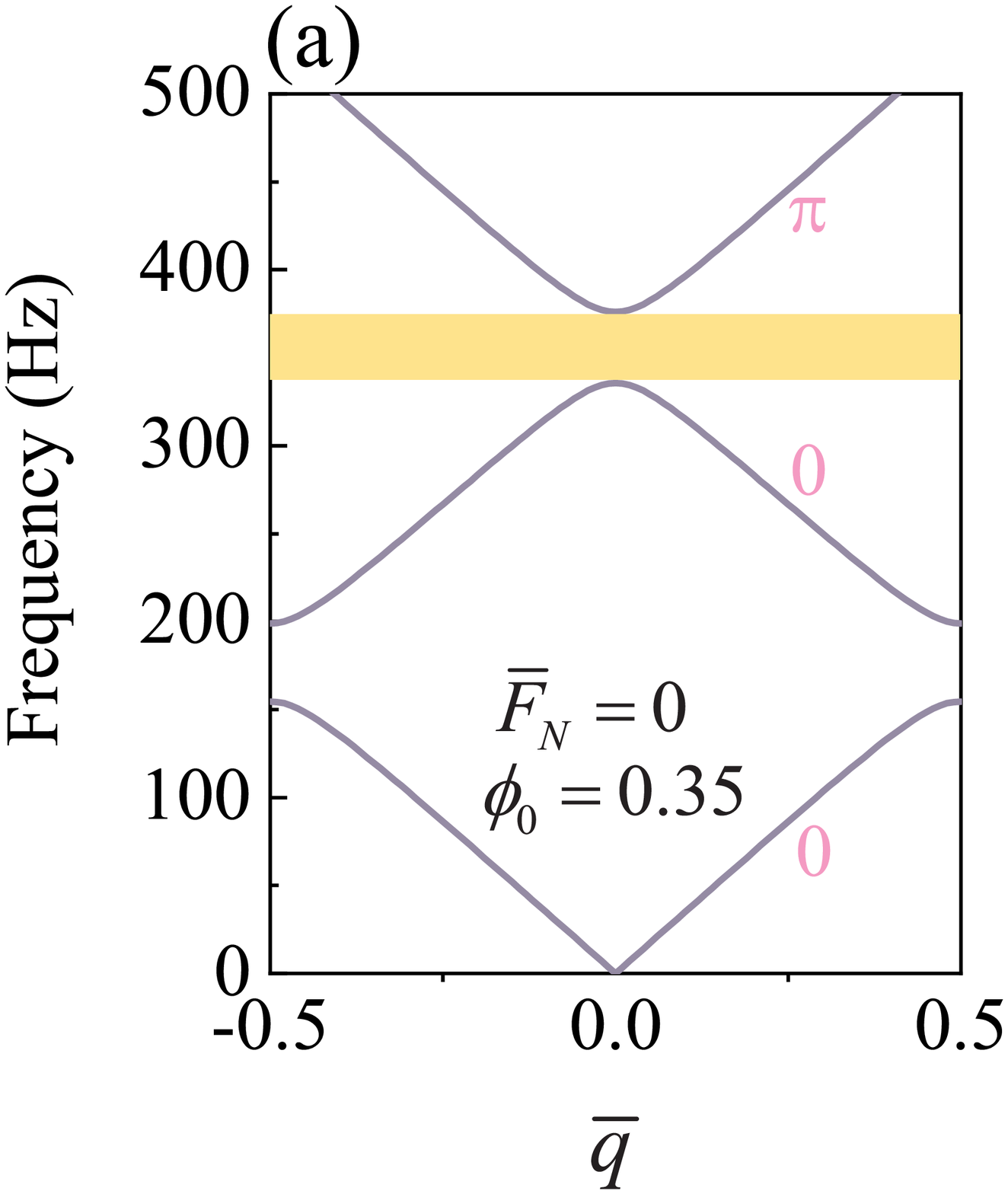}
	\includegraphics[width=0.3\textwidth]{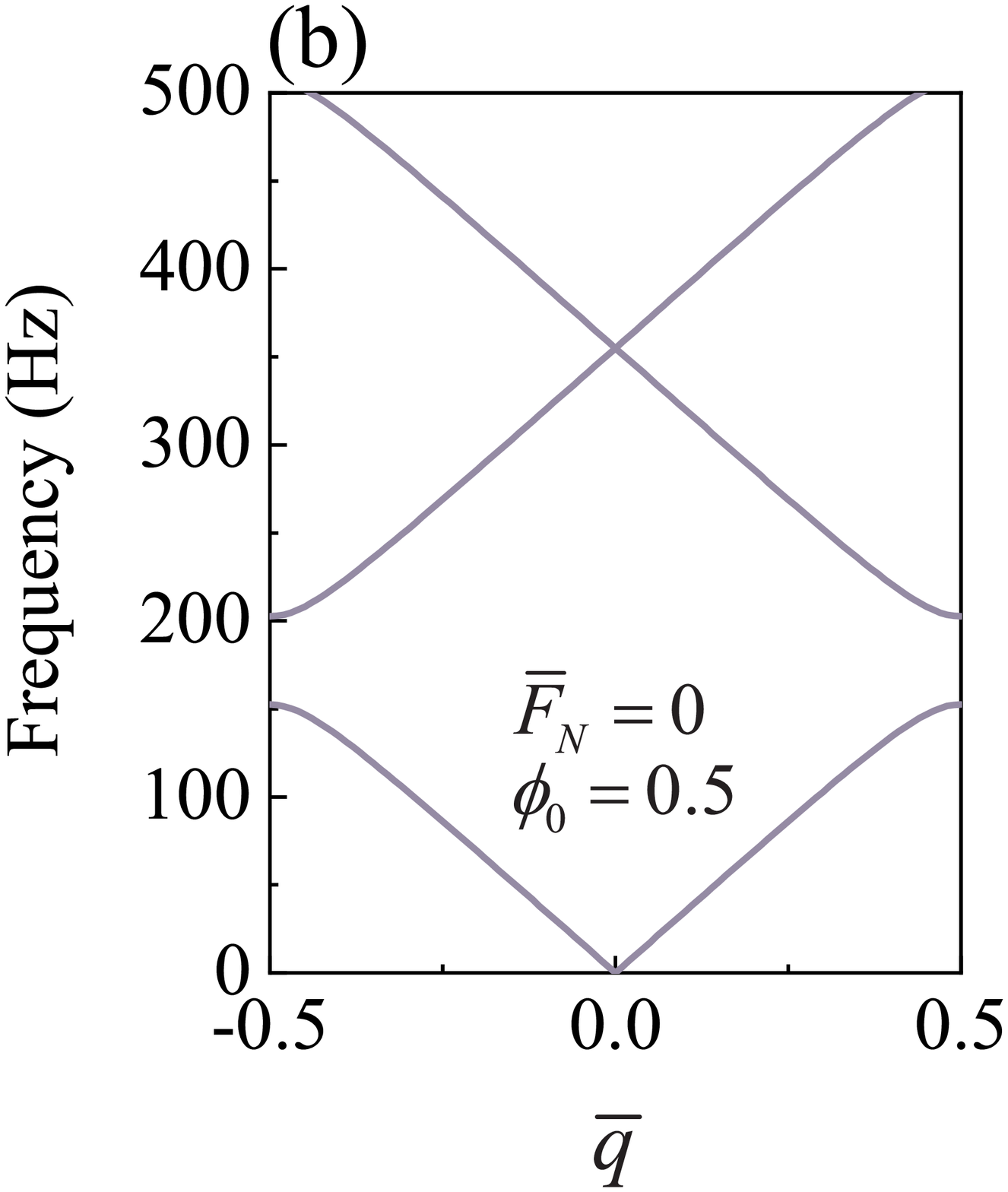}
	\includegraphics[width=0.3\textwidth]{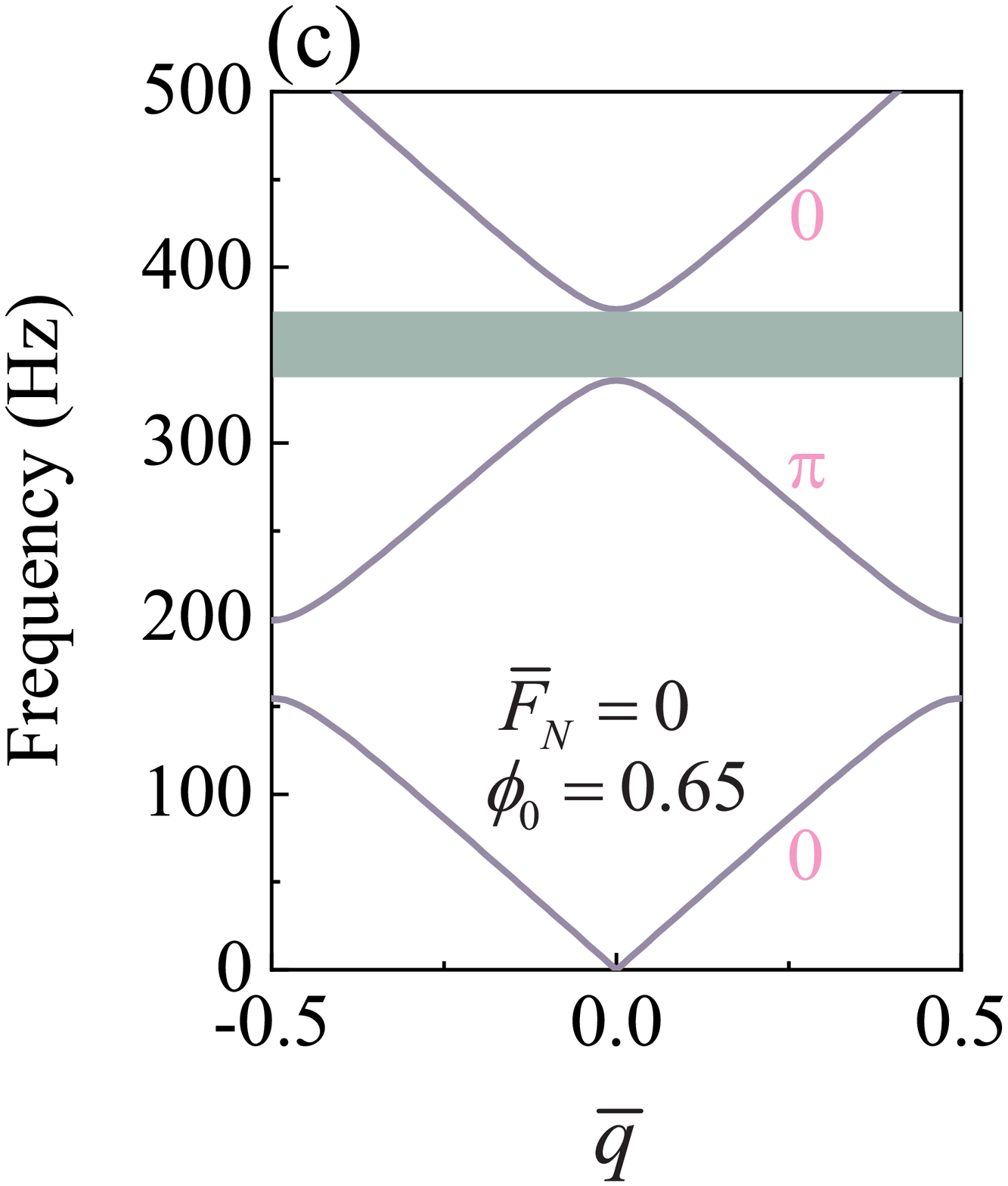}
	\includegraphics[width=0.3\textwidth]{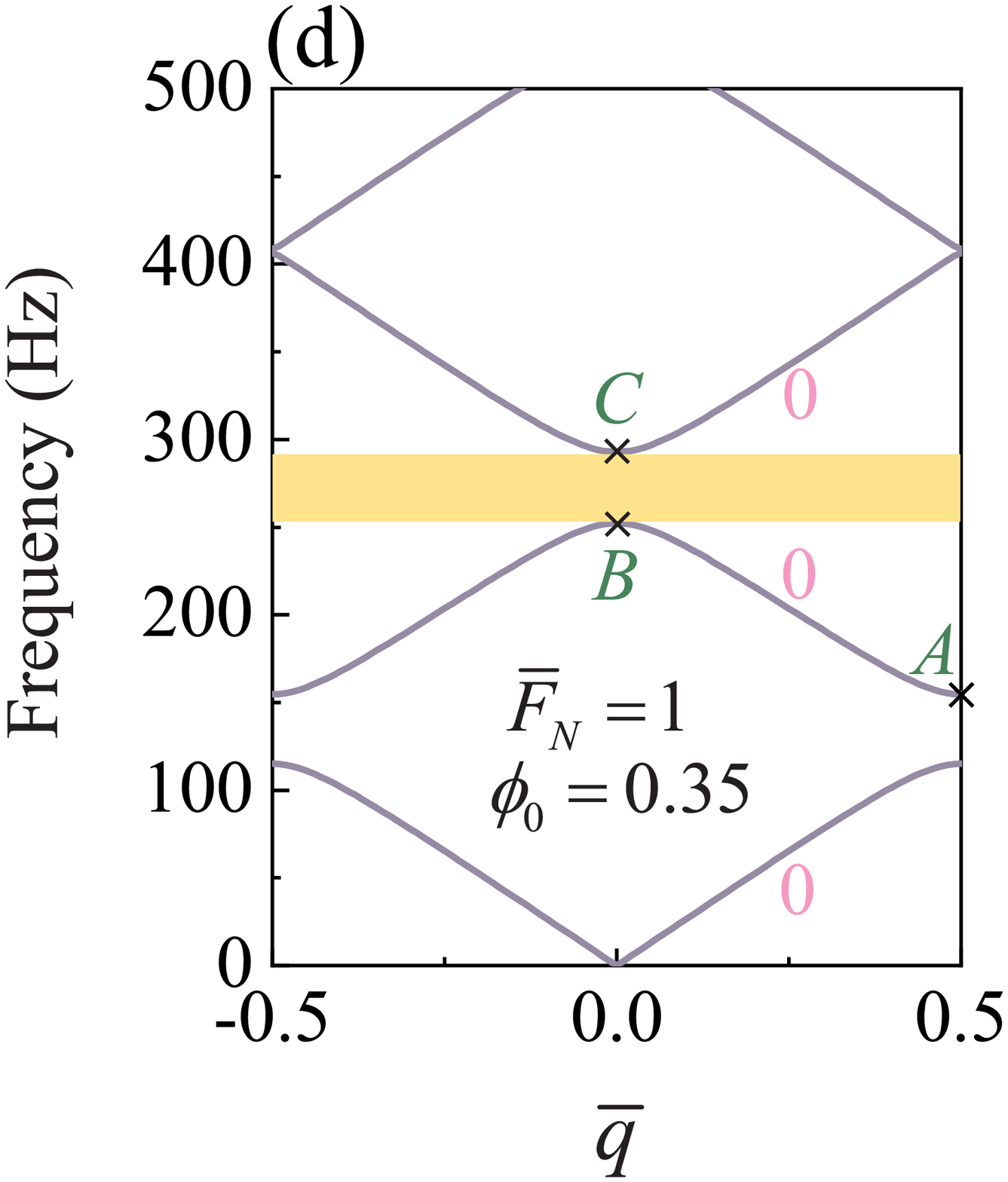}
	\includegraphics[width=0.3\textwidth]{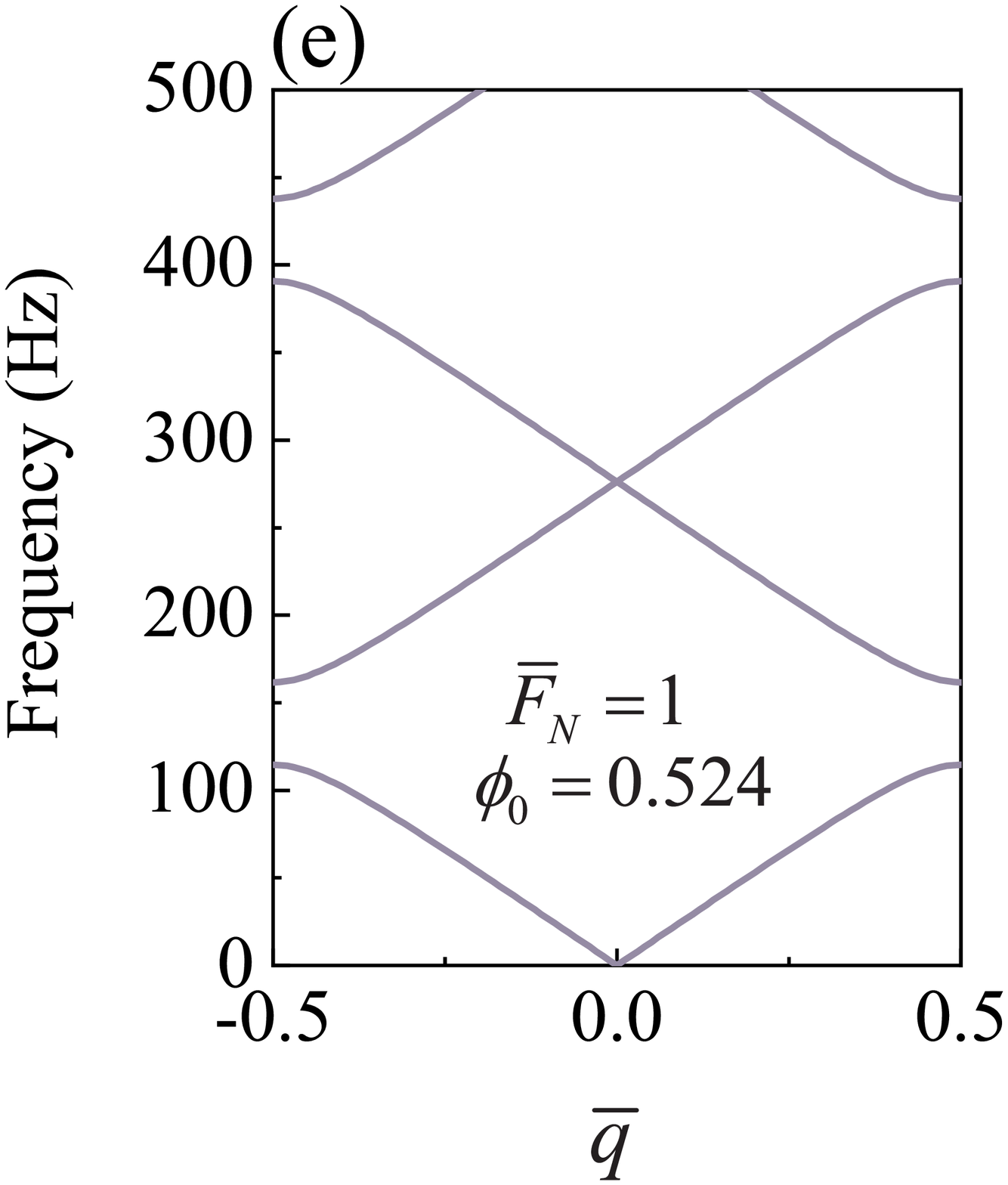} 
	\includegraphics[width=0.3\textwidth]{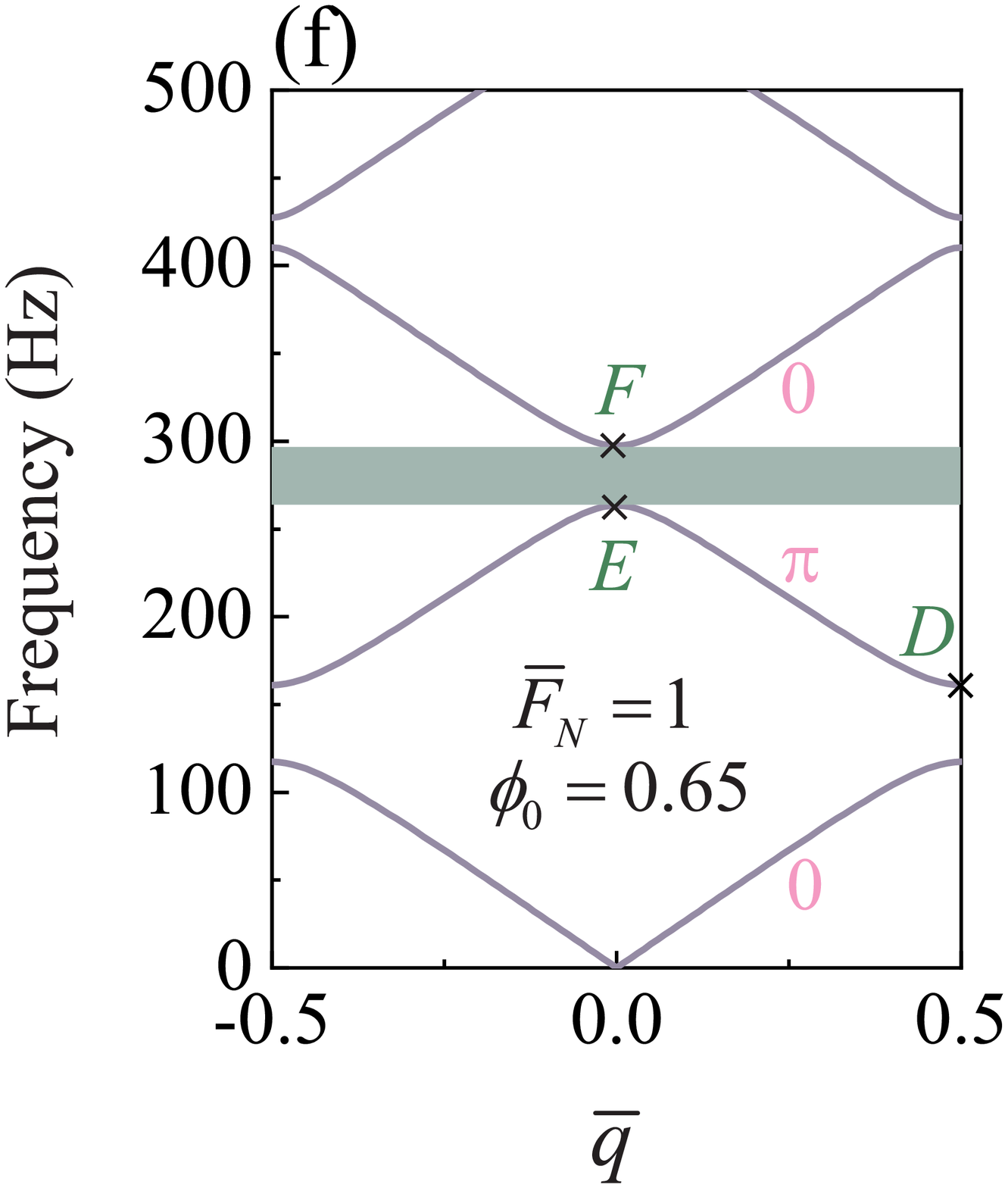}
	\caption{The band structures of longitudinal waves in the neo-Hookean PCC for different axial forces ${{\overline{F}}_{N}}$ and initial length fractions ${\phi }_{0}$ of sub-cylinder 1: (a)-(c) Band inversion process in the absence of axial force (${{\overline{F}}_{N}}=0$) for three values of ${\phi }_{0}=0.35$, 0.5 and 0.65, respectively; (d)-(f) Band inversion process for ${{\overline{F}}_{N}}=1$ and three values of ${\phi }_{0}=0.35$, 0.524 and 0.65, respectively. The Zak phase is marked in magenta on the corresponding bulk band. The yellow and green stripes stand for the second BG signs with $\varsigma>0$ and $\varsigma<0$, respectively.}
	\label{Fig4}
\end{figure}

Fig.~\ref{Fig4} illustrates the band structures, described by Eq.~\eqref{34}, for longitudinal waves in the neo-Hookean PCC for different axial forces ${{\overline{F}}_{N}}$ and geometric parameters ${\phi }_{0}$. The results corresponding to ${{\overline{F}}_{N}}=0$ are shown in Figs.~\ref{Fig4}(a)-(c) while those for ${{\overline{F}}_{N}}=1$ are displayed in Figs.~\ref{Fig4}(d)-(f). The unit cells with various ${\phi }_{0}$ represent different PCC configurations and have the same undeformed length $L$. By varying the initial length fraction ${\phi }_{0}$ of sub-cylinder 1, the second BG for ${{\overline{F}}_{N}}=0$ closes at the center ($\overline{q}=0$) of the first Brillouin zone with a Dirac cone formed by accidental degeneracy (see Fig.~\ref{Fig4}(b)). The Dirac cone (where the two bulk bands have linear dispersion) occurs at ${\phi }_{0}=0.5$ in the absence of an axial force. The degeneracy is broken for any configuration such that ${\phi }_{0}\ne0.5$, and the Dirac cone will be opened to form the second BG. This is illustrated, if we decrease ${\phi}_{0}$ from 0.5 to 0.35 and increase ${\phi }_{0}$ from 0.5 to 0.65, in Figs.~\ref{Fig4}(a) and (c) for the unloaded case. Therefore, the second BG of the soft PCC without axial force exhibits the evolutionary process of open, close and reopen when adjusting the geometric parameter ${\phi }_{0}$. Remarkably, as we shall show below, this transition corresponds to the switching of topological characteristics. Similar BG inversion process is observed for the axially loaded PCC as demonstrated in Figs.~\ref{Fig4}(d)-(f). A topological transition point (i.e., the point for two bands to cross) is also obtained in Fig.~\ref{Fig4}(e) for the case of ${{\overline{F}}_{N}}=1$, where the Dirac cone appears at a different length fraction ${\phi }_{0}\simeq0.524$ due to the nonlinear deformation of unit cell.
  
It should be pointed out that the topological characteristics of a BG is completely determined by the summation of the Zak phases of all the bulk bands below this gap \citep{zak1985symmetry,xiao2014surface,xiao2015geometric}. Originating from electronic systems \citep{kohn1959analytic, zak1989berry}, the so-called Zak phase is a special type of Berry phase, which is a topological invariant characterizing the topological property of bulk bands in 1D periodic systems. The Zak phase for the $j$th isolated band of the 1D PCC system is defined as \citep{xiao2015geometric,yin2018band}
\begin{equation} \label{53}
\theta _{j}^{\textrm{Zak}}=\int_{-\pi /l}^{\pi /l}{\left[ \operatorname{i}\int\limits_{\text{unit cell}}{\dfrac{1}{2\rho c^2}\textrm{d}\mathbf{r}\operatorname{dz}W_{j,q}^{*}\left( z,\mathbf{r} \right){{\partial }_{q}}{{W}_{j,q}}\left( z,\mathbf{r} \right)} \right]}\textrm{d}q,
\end{equation}
where $z$ is the deformed axial coordinate, $\mathbf{r}$ denotes the position in the cross-section plane, $l$ is the length of the deformed unit cell, $\rho$ and $c$ are the current mass density and longitudinal wave velocity in the deformed configuration, respectively, and ${{W}_{j,q}}\left( z,\mathbf{r} \right) = {{{w}}_{j,q}}\left( z,\mathbf{r}\right){{{\text{e}}}^{-\text{i}qz}}$ represents the periodic in-cell part of the normalized Bloch displacement eigenfunction ${{{w}}_{j,q}}\left( z,\mathbf{r}\right)$ in the $j$th band with Bloch wave number $q$. The factor $1/(2\rho c^2)$ is the weight function for the elastic system.

Thus, to distinguish the topological properties of different PCC configurations, it is necessary to obtain the Zak phase value. Here, we make use of a discretized form \eqref{A3} to numerically calculate the Zak phase (the detailed derivation of the numerical procedure is presented in~\ref{AppeA}). Given the inversion symmetry of the 1D PCC with respect to its central cross-sectional plane, the calculated Zak phase value is quantized at either 0 or $\pi $ \citep{zak1989berry}. This quantization also holds true for the PCC under the action of an external mechanical load. The calculated Zak phase values are marked in magenta on the corresponding bands in Fig.~\ref{Fig4}. Moreover, the BG topological property can be characterized by the BG sign $\varsigma$, which is related to the Zak phase and given by \citep{xiao2014surface}
\begin{equation} \label{54}
 {\mathop{\rm sgn}} \left[ {{\varsigma ^{\left( j \right)}}} \right] = {\left( { - 1} \right)^j}{\left( { - 1} \right)^h}{\rm{exp}}\left( {{\rm{i}}\sum\limits_{\beta  = 1}^{j} {\theta _\beta ^{\text{Zak}}} } \right),
\end{equation}
where the integer $h$ is the number of band crossing points below the $j$th BG. The second BGs with $\varsigma>0$ and $\varsigma<0$ are represented by the yellow and green stripes in Fig.~\ref{Fig4}, respectively, to demonstrate different BG topological properties. For simplicity, the two PCC configurations with ${\phi }_{0}=0.35$ and ${\phi }_{0}=0.65$ will be referred to as S1-configuration and S2-configuration, respectively.

While the Zak phase of the first band is the same for S1- and S2-configurations in the absence of an axial force (see Figs.~\ref{Fig4}(a) and (c)), the Zak phase of the second bulk band varies from 0 to $\pi$, which exhibits a topological phase transition. According to Eq.~\eqref{54}, the topological properties of the second BG for S1- and S2-configurations are thus different in spite of the overlapped BG frequencies (see Figs.~\ref{Fig4}(a) and (c)). This indicates the existence of a topological state in this BG at the interface separating S1- and S2-configurations. This topological interface state will be discussed in the next Sec.~\ref{Sec5-2-2}. The phenomenon is also observed in PCC subjected to an axial loading (see Figs.~\ref{Fig4}(d)-(f) for ${{\overline{F}}_{N}}=1$). In this case, however, the BG frequencies for the loaded PCC are lower than those for the unloaded PCC (compare Figs.~\ref{Fig4}(d) and (f) with Figs.~\ref{Fig4}(a) and (c)). For example, the second BG frequency range of S2-configuration is from $336$~Hz to $376$~Hz for ${{\overline{F}}_{N}}=0$, while for ${{\overline{F}}_{N}}=1$ the corresponding BG frequency varies from $263$~Hz to $297$~Hz. Thus, the applied axial force can tune the BG position and the frequency of the topological transition point.
 
\begin{figure}[htbp]
	\centering
	\setlength{\abovecaptionskip}{5pt}
	\includegraphics[width=0.3\textwidth]{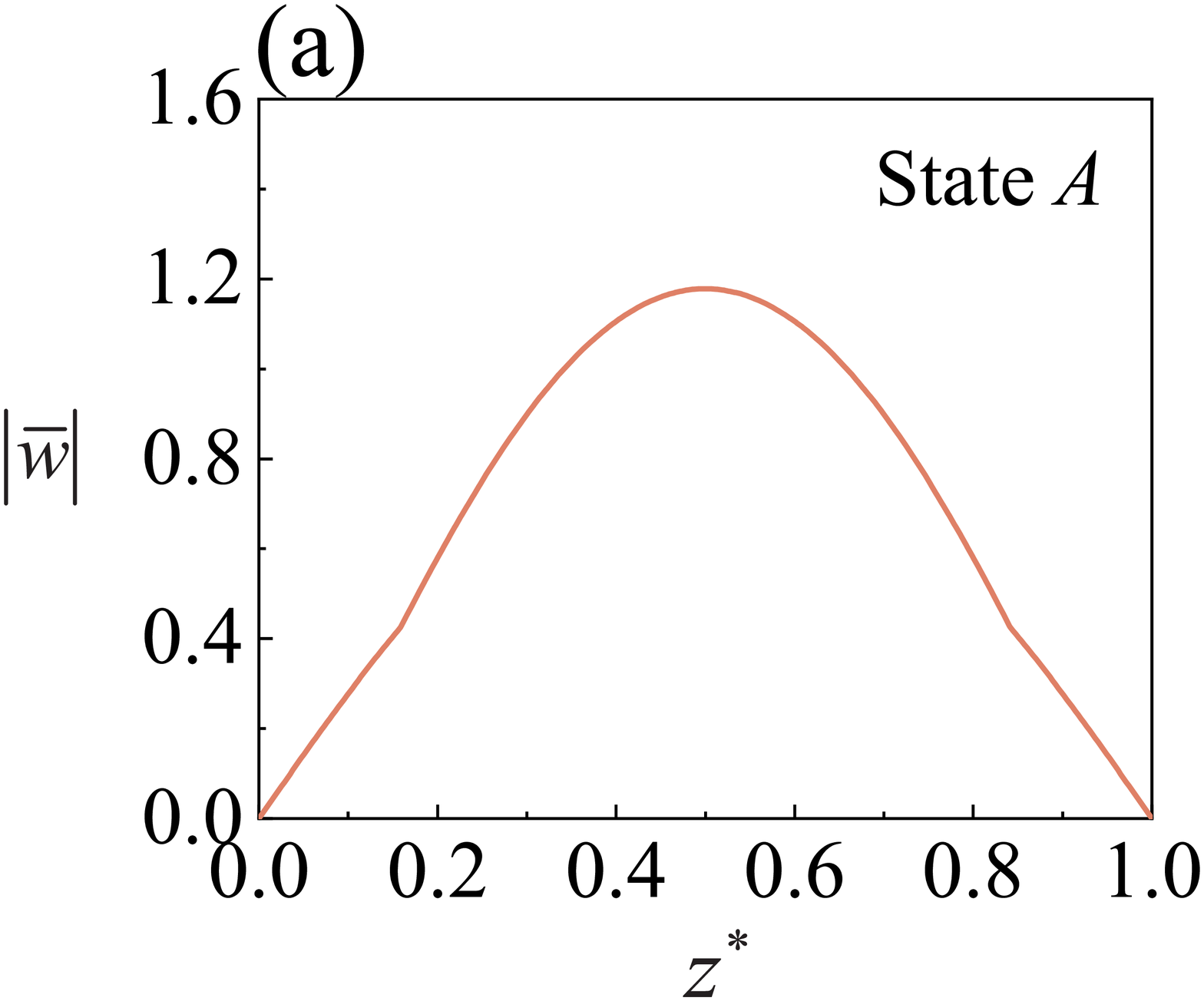}
	\hspace{0.002\textwidth}
	\includegraphics[width=0.3\textwidth]{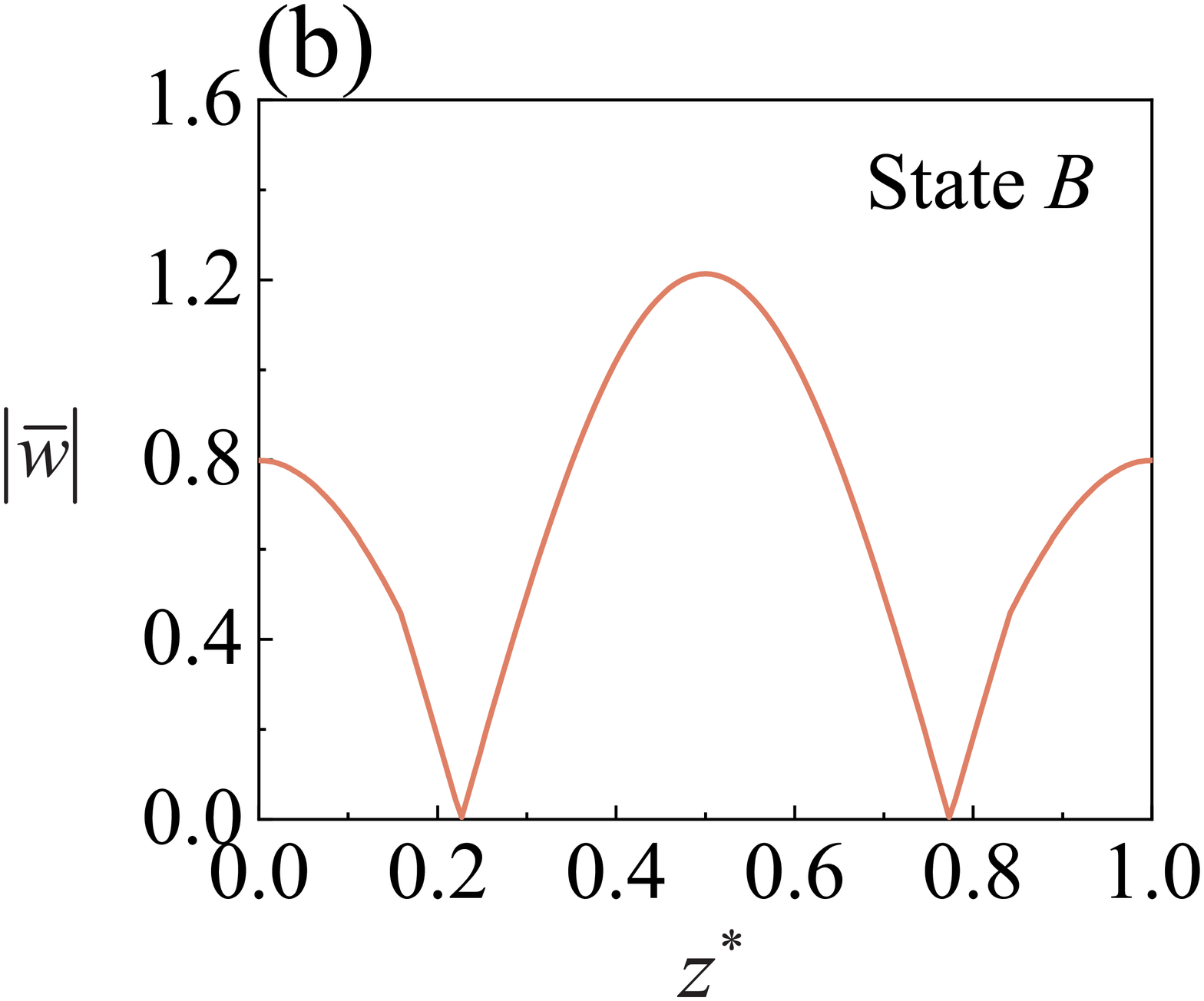}
	\hspace{0.002\textwidth}
	\includegraphics[width=0.3\textwidth]{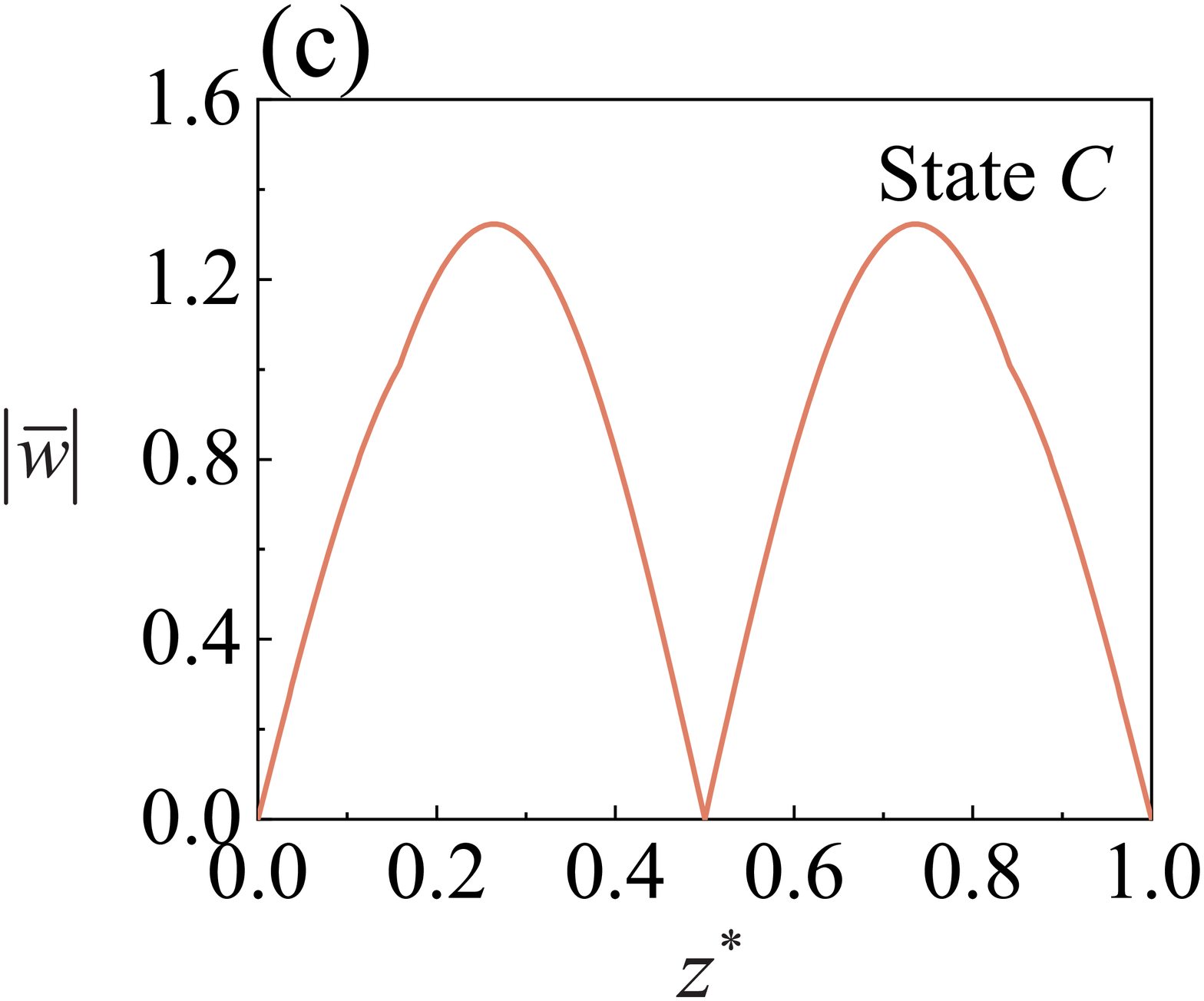}
	\includegraphics[width=0.3\textwidth]{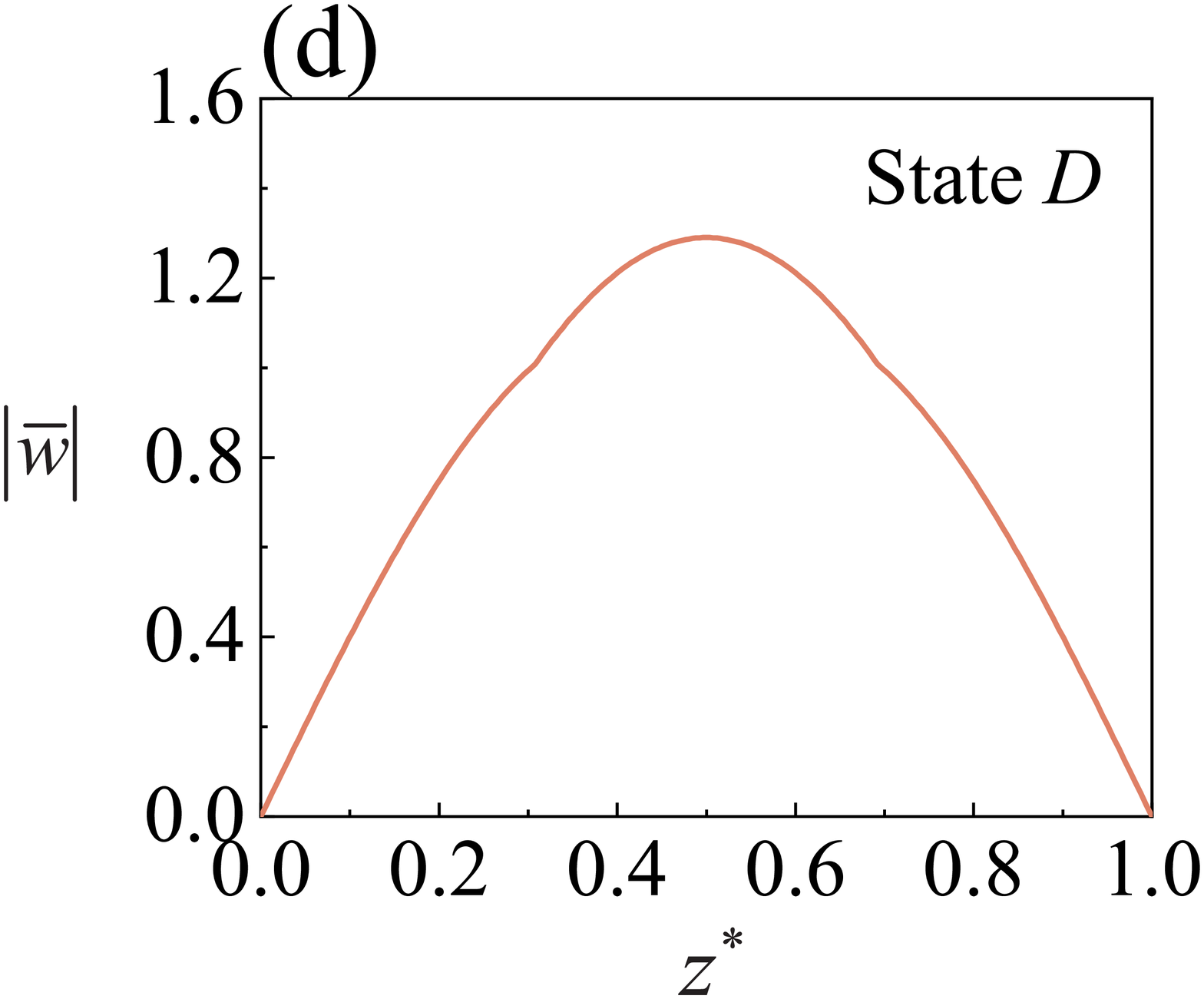}
	\hspace{0.002\textwidth}
	\includegraphics[width=0.3\textwidth]{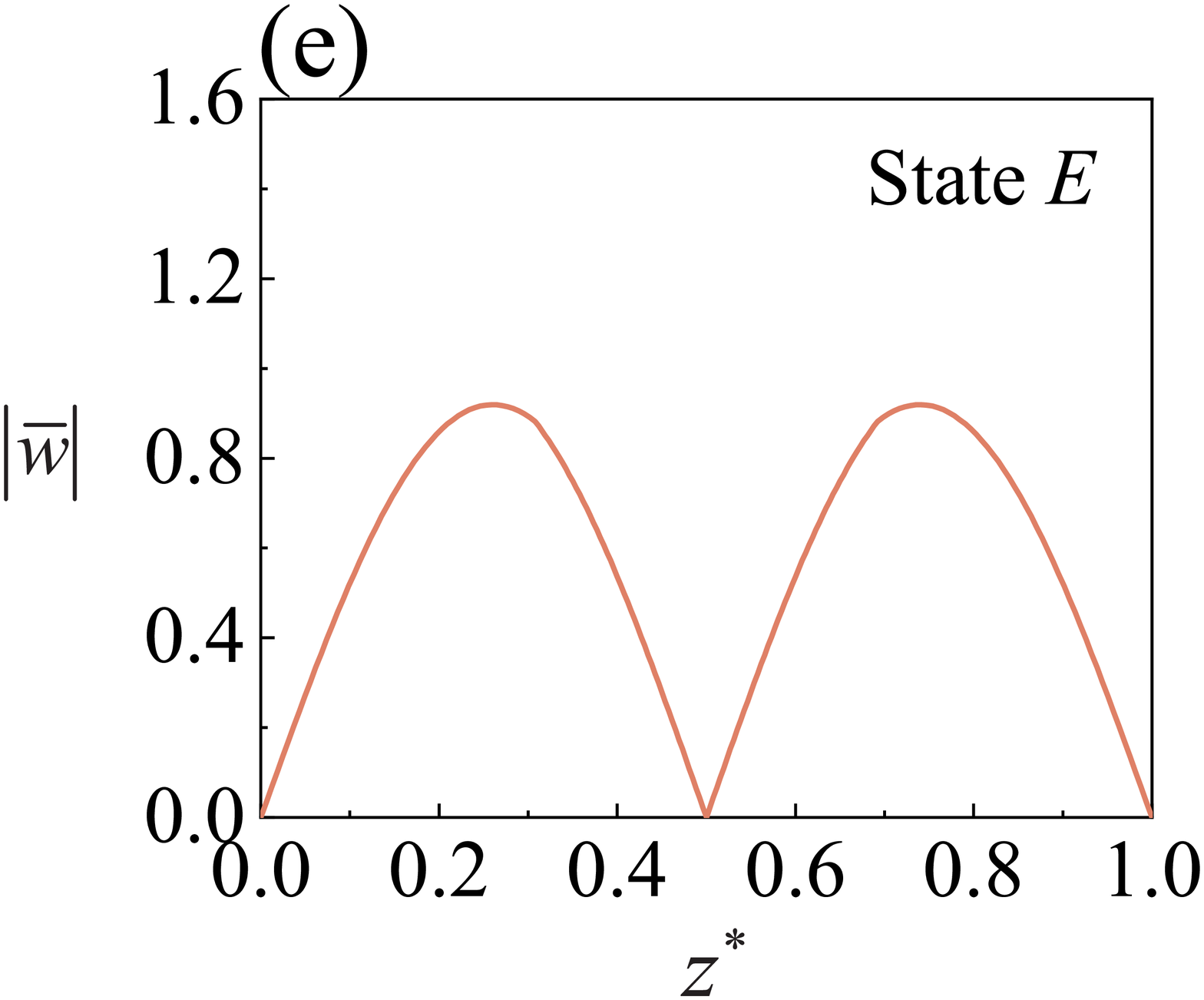}
	\hspace{0.002\textwidth}
	\includegraphics[width=0.3\textwidth]{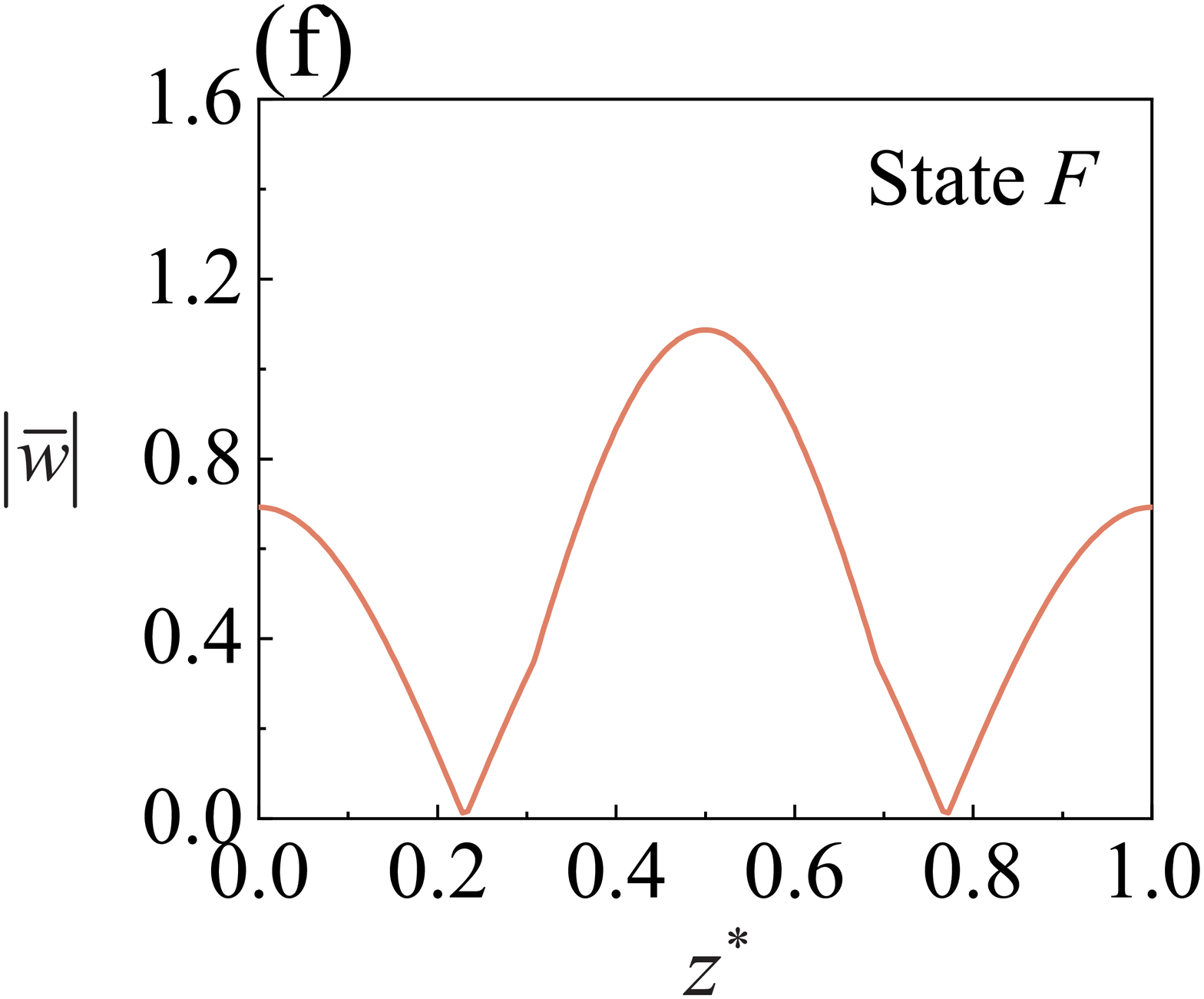}
	\caption{The absolute value of the displacement field $\overline{w}(z)$ of the six band-edge states at the Brillouin zone center and boundary as a function of the normalized axial coordinate ${{z}^{*}=z/l}$ in the deformed neo-Hookean unit cell with ${{\overline{F}}_{N}}=1$. The six band-edge states $A-F$, indicated with cross symbols and marked by capital Roman letters  in Figs.~\ref{Fig4}(d) and (f), are displayed in Figs.~\ref{Fig5}(a)-(f), respectively.}
	\label{Fig5}
\end{figure}

In addition to the direct calculation of Zak phase and BG sign from Eqs.~\eqref{53} and \eqref{54}, the symmetry analysis method of the edge states at the two Brillouin zone symmetry points can also be employed to verify the topological phase transition and to identify the BG topological property \citep{kohn1959analytic, zak1989berry,xiao2014surface}. To perform the symmetry analysis, we make use of Eq.~\eqref{37} and calculate the absolute value of the displacement field $\overline{w}(z)$ for the six band-edge states $A-F$ (indicated with cross symbols in Figs.~\ref{Fig4}(d) and (f)). Fig.~\ref{Fig5} shows the dependence of $|\overline{w}(z)|$ on the normalized axial coordinate ${{z}^{*}=z/l}$ in the deformed neo-Hookean unit cell with ${{\overline{F}}_{N}}=1$.

For a bulk band, the Zak phase is $\theta _{j}^{\textrm{Zak}}=0$ if the two edge states at the Brillouin zone center and boundary possess the same symmetric property. Otherwise, the Zak phase value should be $\theta _{j}^{\textrm{Zak}}=\pi$.  For S1-configuration, the edge states $A$ and $B$ of the second band exhibit the symmetric distributions with respect to the unit-cell center (i.e., even eigenmodes associated with a \emph{nonzero} displacement amplitude at the unit-cell center) (see Figs.~\ref{Fig5}(a) and (b)), and hence the corresponding Zak phase is 0. For S2-configuration, however, the edge states $D$ and $E$ of the second band are symmetric and antisymmetric (odd eigenmode related to the \emph{zero} displacement amplitude at the unit-cell center) (see Figs.~\ref{Fig5}(d) and (e)), and its Zak phase is $\pi$. Therefore, the Zak phase of the second band is altered after the band crossing, which is in full agreement with the topological phase transition shown in Figs.~\ref{Fig4}(d) and (f). In addition, Figs.~\ref{Fig5}(b) and (f) show the symmetric displacement fields with respect to the center of unit cell for edge states $B$ and $F$, while the displacement fields in Figs.~\ref{Fig5}(c) and (e) are antisymmetric for edge states $C$ and $E$. Thus, we can observe the eigenmode switching of the two edge states across the second BG, which characterizes the topological band inversion. Furthermore, if two states at the lower or upper edges of the overlapped BG possess different symmetries, the sign $\varsigma$ of this BG is opposite and thus an interface state exists inside the BG \citep{xiao2014surface}. Here, the lower edge states $B$ and $E$ in Figs.~\ref{Fig5}(b) and (e) for S1- and S2-configurations are symmetric and antisymmetric, respectively. This indicates the different signs $\varsigma$ of the second BG and various topological properties, as shown in Figs.~\ref{Fig4}(d) and (f).


\subsubsection{Transmission spectra and displacement distributions} \label{Sec5-2-2}

\begin{figure}[htb]
	\centering	
	\setlength{\abovecaptionskip}{8pt}
	\includegraphics[width=0.4\textwidth]{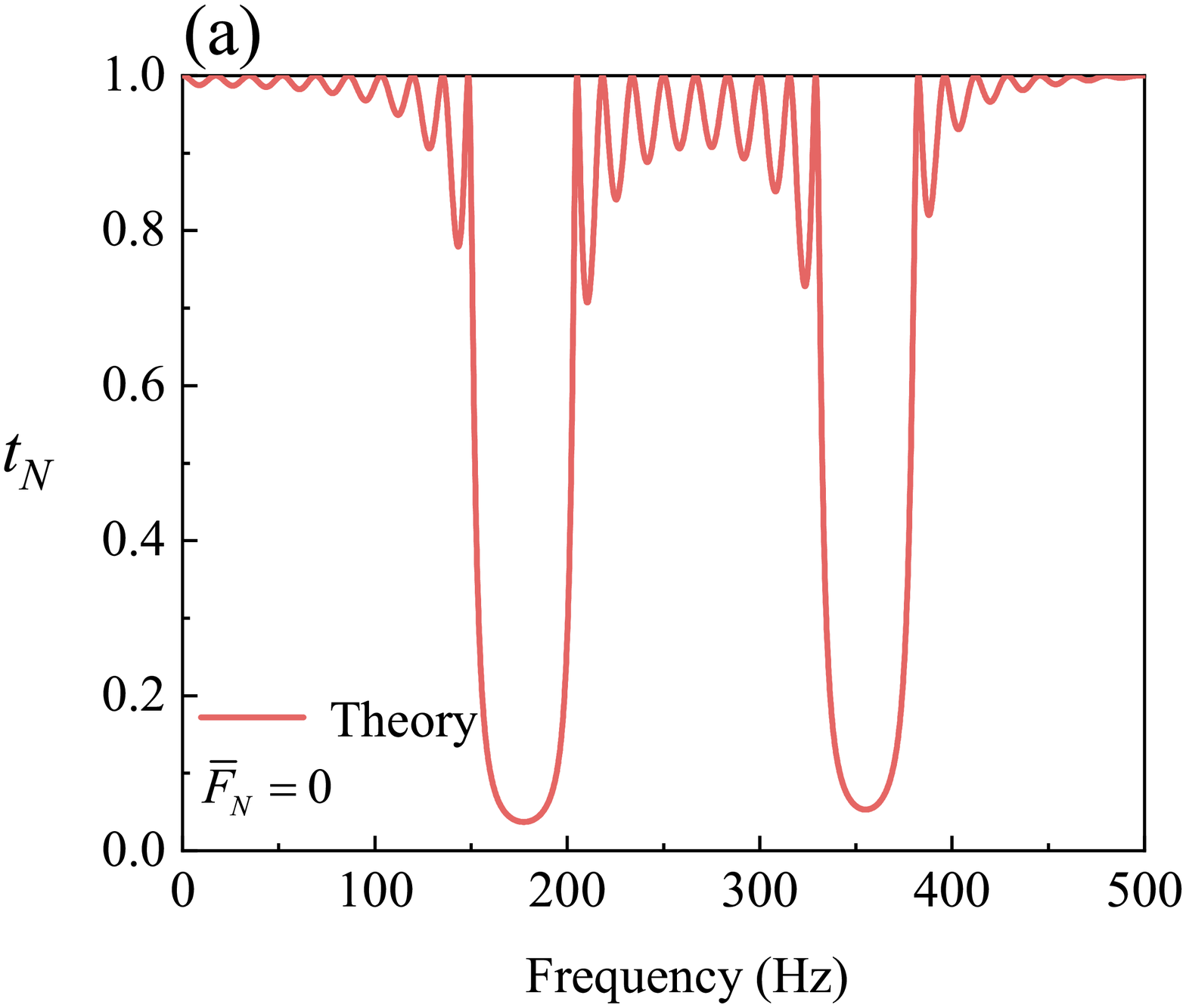}
	\hspace{0.01\textwidth}
	\includegraphics[width=0.4\textwidth]{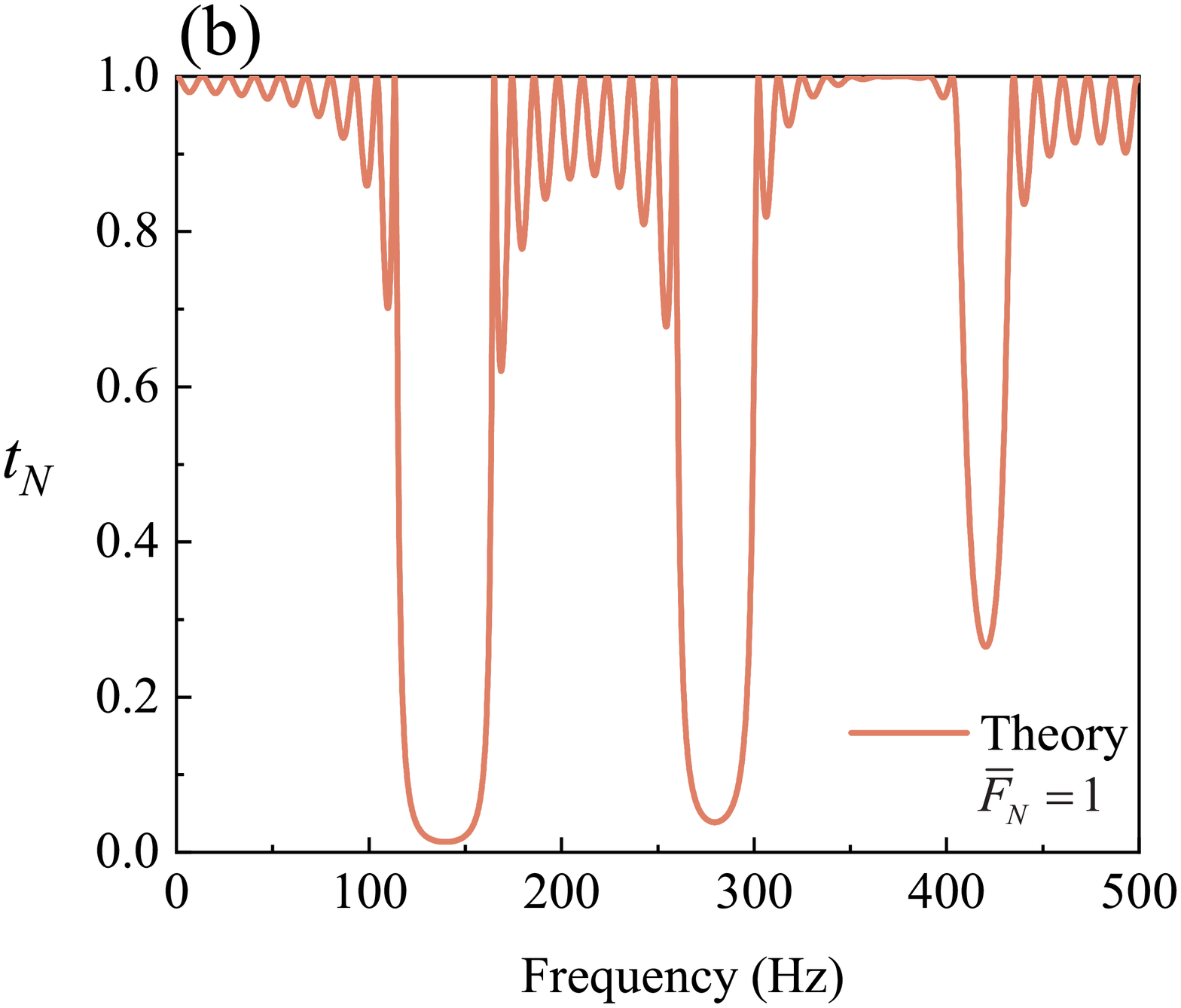}
	\includegraphics[width=0.412\textwidth]{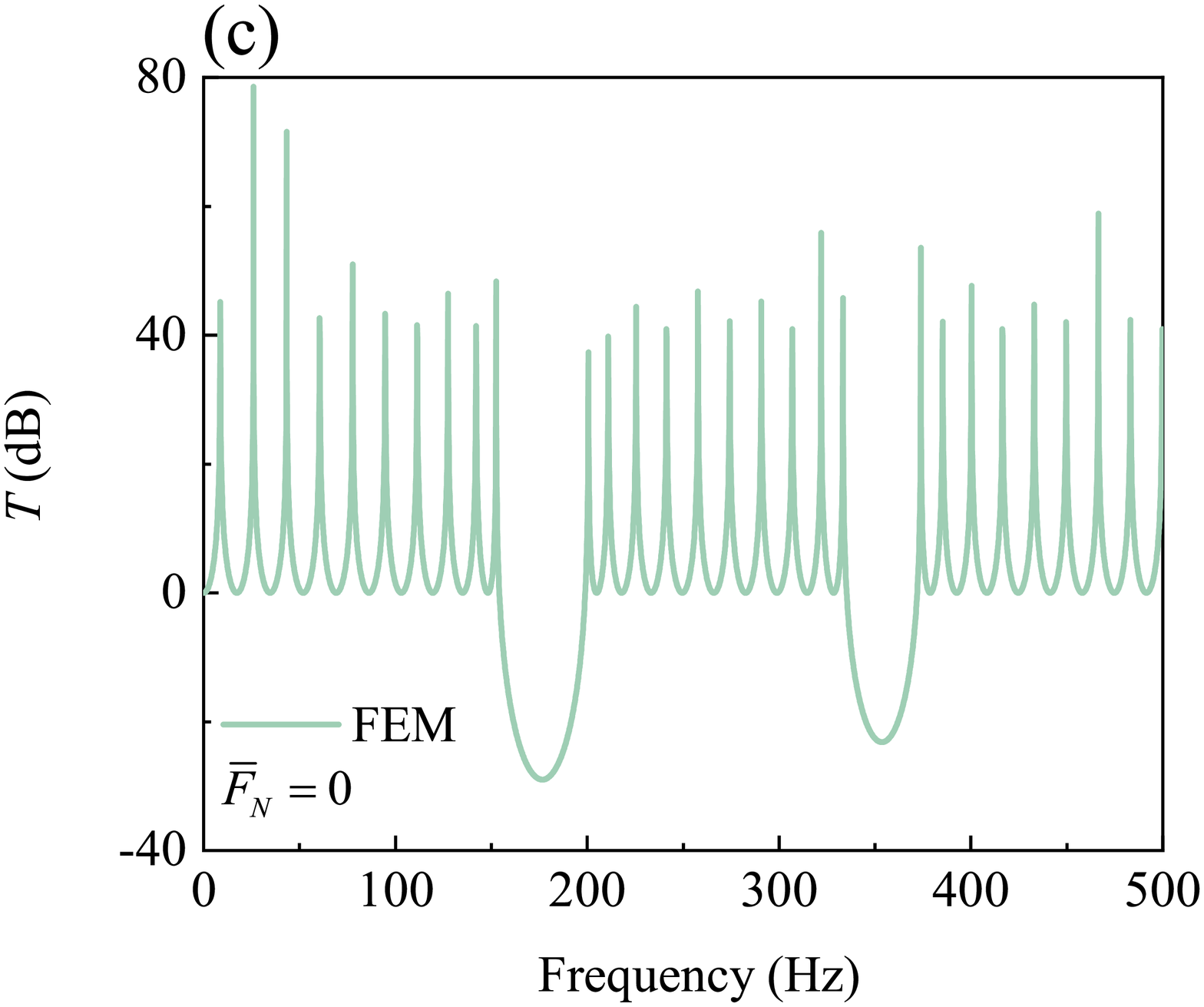}
	\hspace{0.0005\textwidth}
	\includegraphics[width=0.418\textwidth]{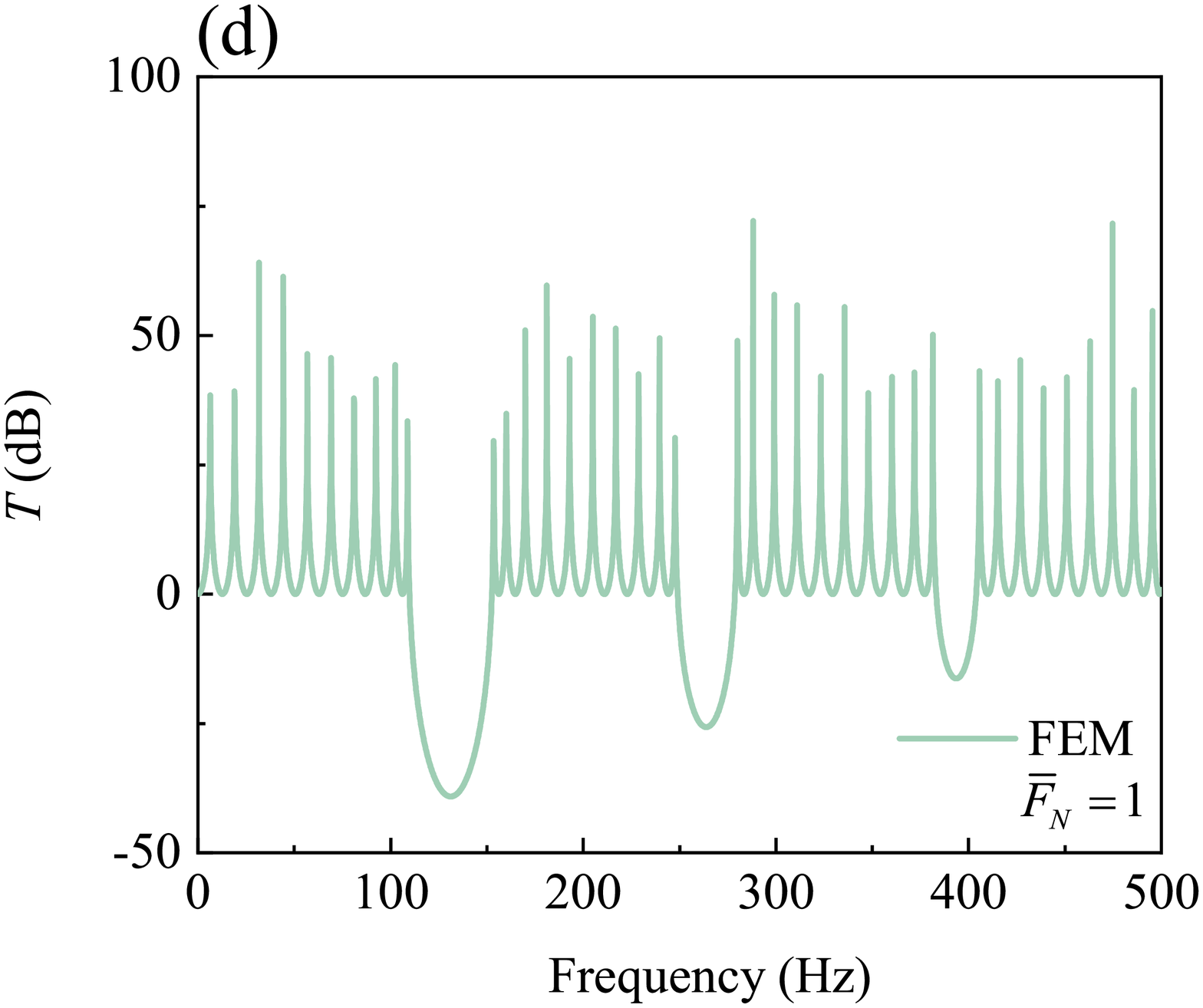}
	\caption{The transmission spectra of a finite neo-Hookean PCC consisting of 10 identical S2 unit cells (${\phi }_{0}=0.65$) calculated by the theoretical solutions (a, b) and the FE method (c, d): (a, c) ${{\overline{F}}_{N}}=0$; (b, d) ${{\overline{F}}_{N}}=1$.}
	\label{Fig6}
\end{figure}

Here, we make use of Eq.~\eqref{trans1} to calculate the transmission spectra (i.e., the transmission coefficient $t_N$ versus the frequency $f$) of a finite-size neo-Hookean PCC. Figs.~\ref{Fig6}(a) and (b) show the results for the finite PCC composed of 10 identical S2-type unit cells subjected to the axial forces ${{\overline{F}}_{N}}=0$ and $1$, respectively. The transmission spectra based on our theoretical analysis agree well with the corresponding band structures (compare Figs.~\ref{Fig6}(a) and (b) with Figs.~\ref{Fig4}(c) and (f)). {\color{red} The transmission coefficient in the BG range approaches zero, indicating that the wave propagation is prohibited; but in the passing bands, the elastic wave is allowed to propagate in the finite PCC as the transmission ratio almost equals one.} The spectrum does not show any peaks in the BGs for the finite-size PCC with identical unit cells.

In addition, the transmission spectra have been calculated independently by means of the finite element (FE) simulations. The details about the FE simulation procedures are described in \ref{AppeB}. Figs.~\ref{Fig6}(c) and (d) shows the FE results for the finite-size PCC under the action of axial forces ${{\overline{F}}_{N}}=0$ and $1$, respectively. Here, the attenuation intensity $T (\text{dB})$ is defined as $T=20\log(A_{\text{output}}/A_{\text{input}})$, where $A_{\text{input}}$ and $A_{\text{output}}$ are the average displacement amplitudes of the input and output signals. The comparison of the theoretical and numerical results demonstrates a good agreement between these intendant methods. There are, however, slight differences in the form of transmission spectra due to various calculation expressions. {\color{red} The FE calculations show that the transmission coefficient reaches a dip within the BG range, which stands for a strong attenuation of the wave propagation; however, the attenuation intensity is larger than zero in the passing bands, implying that the output signal can be obviously detected.} Particularly, the frequency ranges of the second BG for ${{\overline{F}}_{N}}=0$ are (316~Hz, 383~Hz) and (307~Hz, 385~Hz) for the theoretical results and FE simulations, respectively. For ${{\overline{F}}_{N}}=1$, the corresponding results based on the theoretical model and FE simulation are (248~Hz, 302~Hz) and (240~Hz, 299~Hz), respectively. We note, however, that the difference in the central frequencies between the theoretical and FE predictions does not exceed 2\%.

\begin{figure}[htbp]
	\centering	
	\setlength{\abovecaptionskip}{5pt}
	\includegraphics[width=0.41\textwidth]{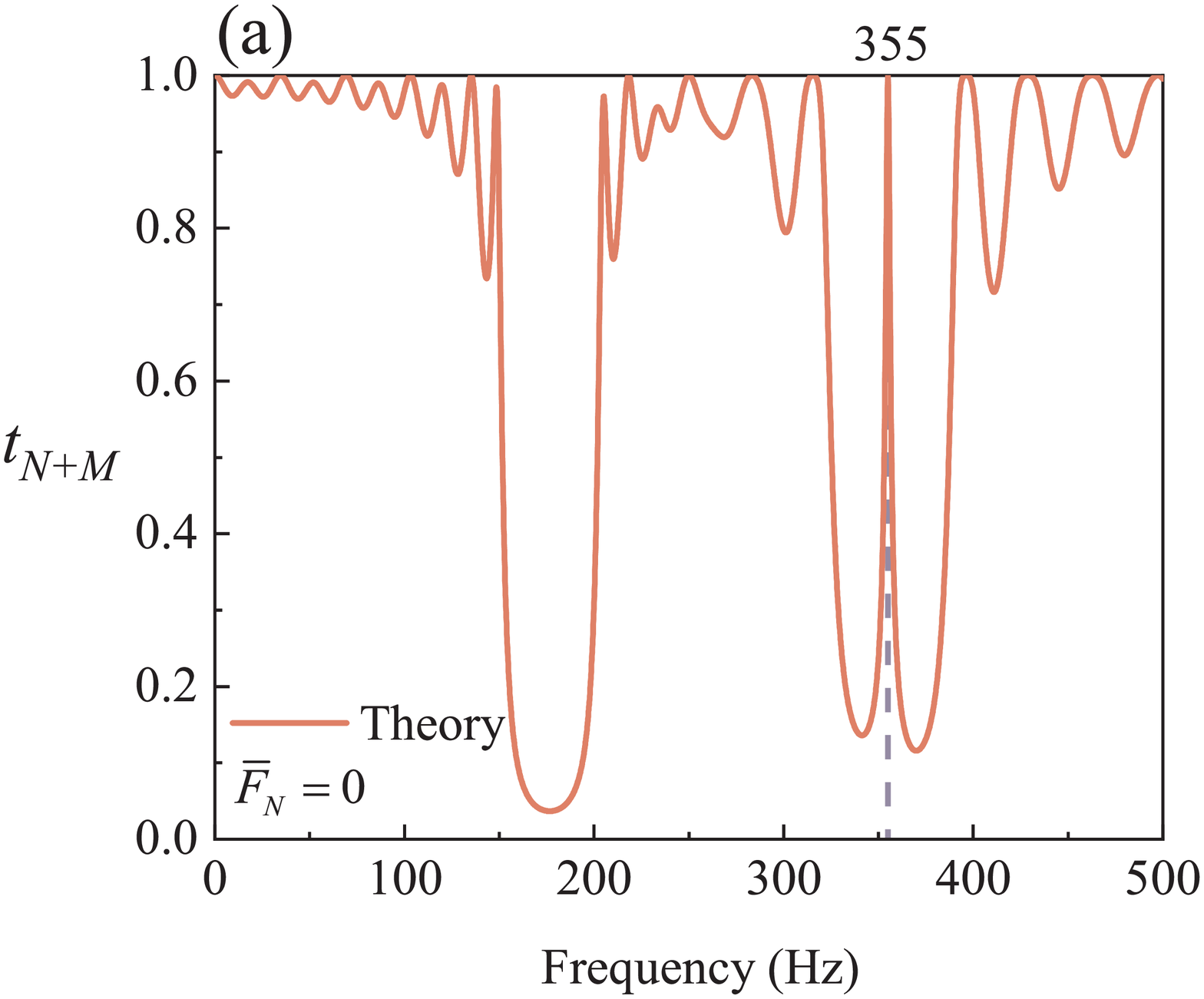}
    \hspace{0.005\textwidth}
	\includegraphics[width=0.41\textwidth]{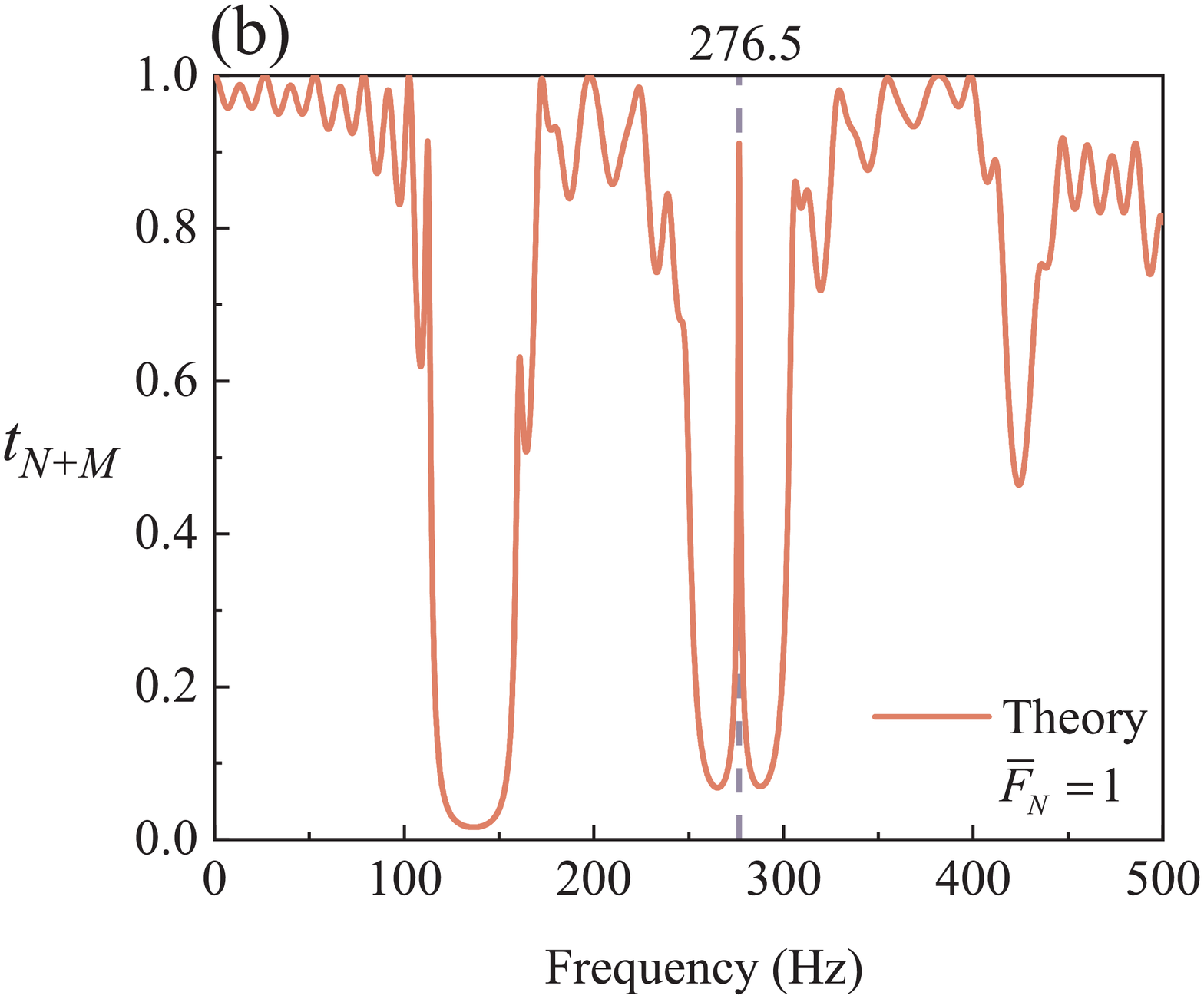}
	\includegraphics[width=0.41\textwidth]{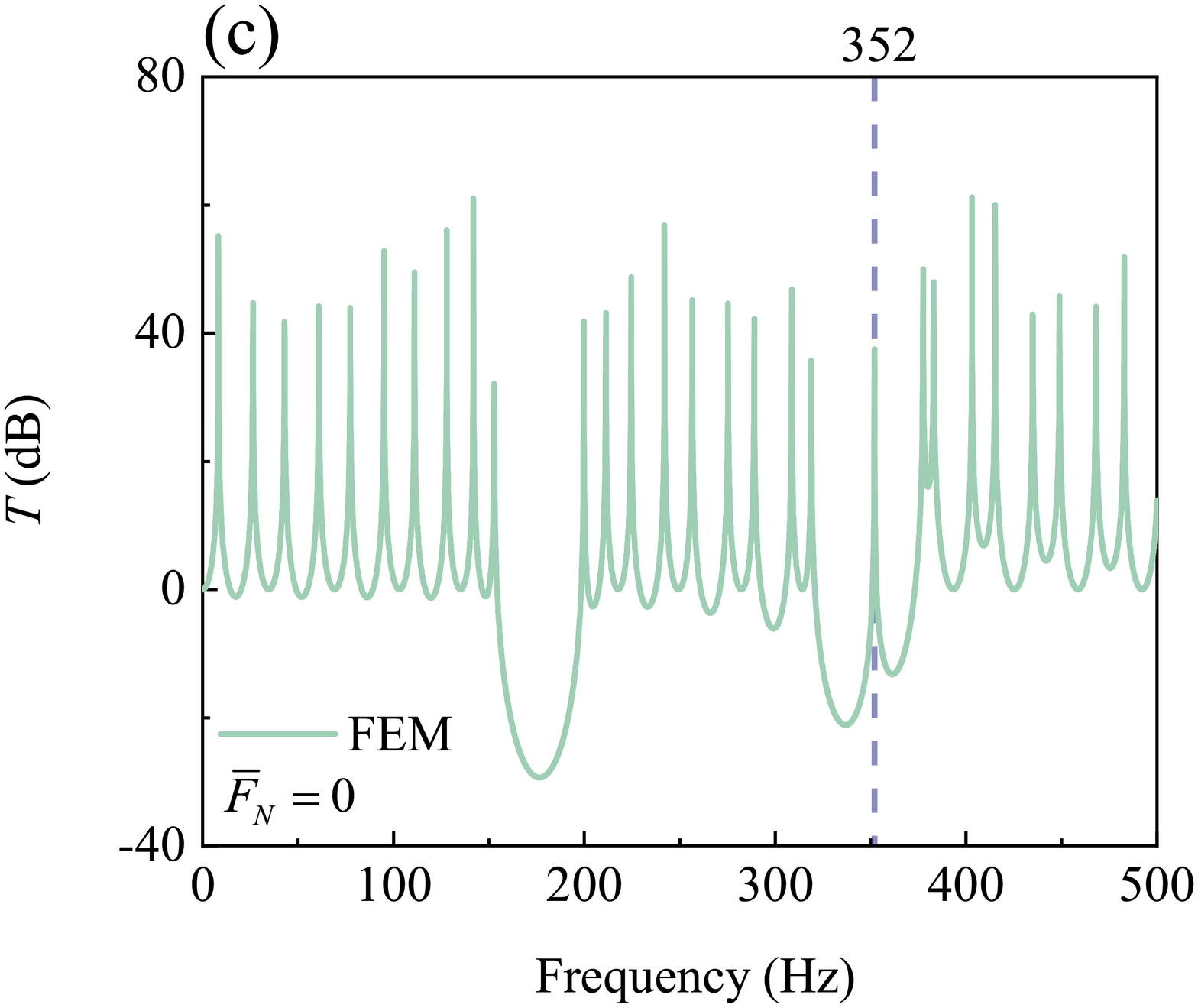}
    \hspace{0.005\textwidth}
	\includegraphics[width=0.415\textwidth]{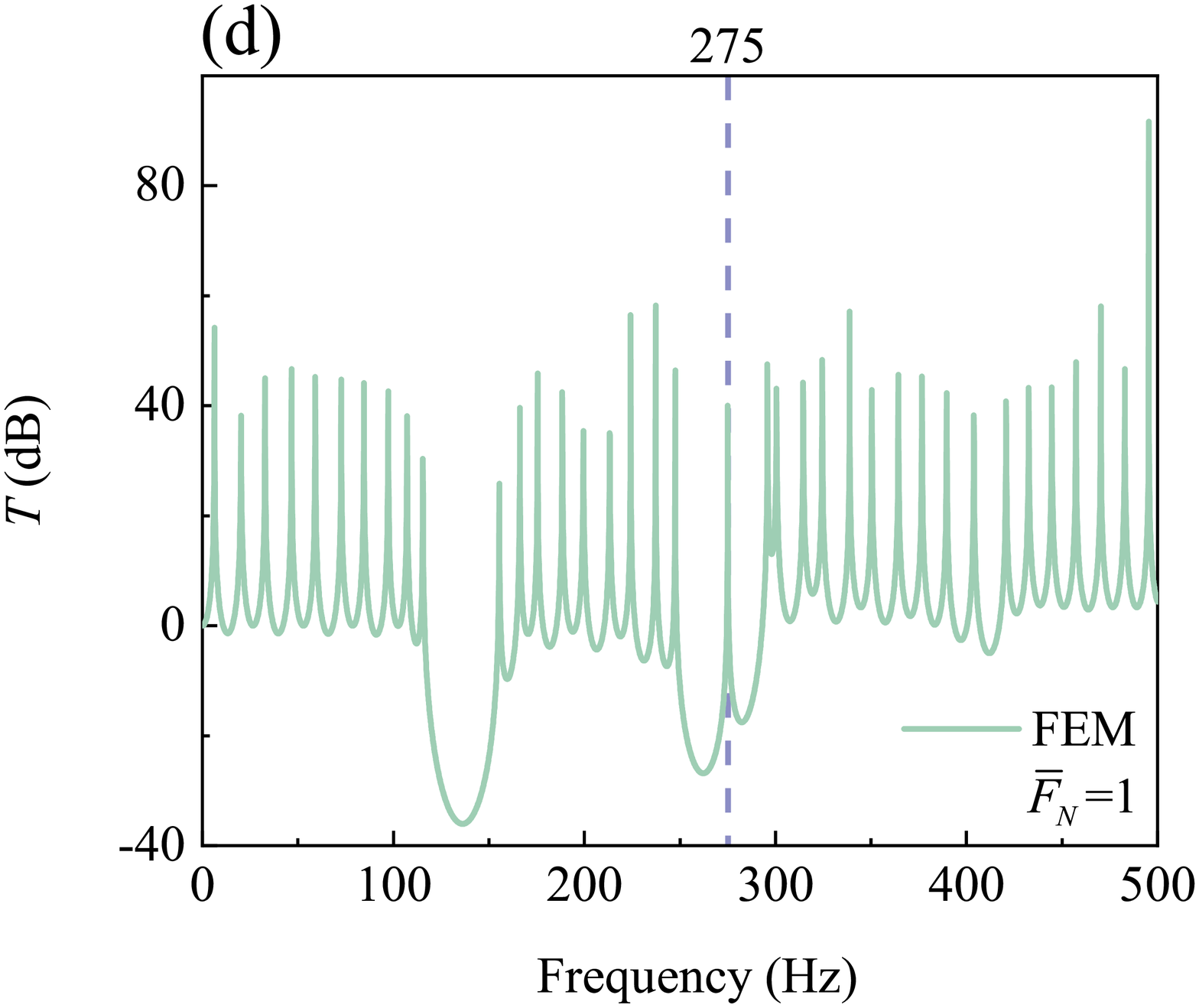}
	\caption{The transmission spectra of a \emph{mixed} finite neo-Hookean waveguide consisting of 5 S1 unit cells and 5 S2 unit cells calculated by the theoretical solutions (a, b) and the FE method (c, d): (a, c) ${{\overline{F}}_{N}}=0$; (b, d) ${{\overline{F}}_{N}}=1$. The transmission peak frequency in the second BG is labelled in the corresponding subfigure.}
	\label{Fig7}
\end{figure}

Recall that the topological phase transition exists and the topological properties of the second BG are different for S1- and S2-configurations (see Sec.~\ref{Sec5-2-1}). Thus, a topological interface state should appear in this overlapped BG at the interface delimiting the S1- and S2-configurations. Motivated by this prediction, we design a \emph{mixed} finite-size neo-Hookean waveguide consisting of 5 S1-type and 5 S2-type unit cells. We utilize the theoretical formula \eqref{49} to calculate the corresponding transmission spectra of the mixed-type PCC waveguide. Fig.~\ref{Fig7} shows the theoretical ((a) and (b)) and FE simulation ((c) and (d)) results for the \emph{mixed} soft PCC waveguide. The results are illustrated for the unloaded PCC (Figs.~\ref{Fig7}(a) and (c)), and for the PCC subjected to the axial force ${{\overline{F}}_{N}}=1$ (Figs.~\ref{Fig7}(b) and (d)). {\color{red}We observe from Fig.~\ref{Fig7} that the sharp transmission peaks emerge in the overlapped BG frequency range for both unloaded and loaded cases.} In the \emph{unloaded} mixed PCC waveguide, the peak frequencies of theoretical prediction and FE simulation are 355~Hz and 352~Hz, respectively. In the \emph{loaded} case with ${{\overline{F}}_{N}}=1$, the corresponding peak frequencies shift down to $276.5$ Hz (theory) and $275$ Hz (simulation). Thus, the application of axial force tunes the position of the transmission peak. We note that the peak frequencies predicted by the FE simulations and theoretical model are almost identical, which further validates the effectiveness of our theoretical assumption.

\begin{figure}[htbp]
	\centering
	\setlength{\abovecaptionskip}{5pt}	
	\includegraphics[width=0.45\textwidth]{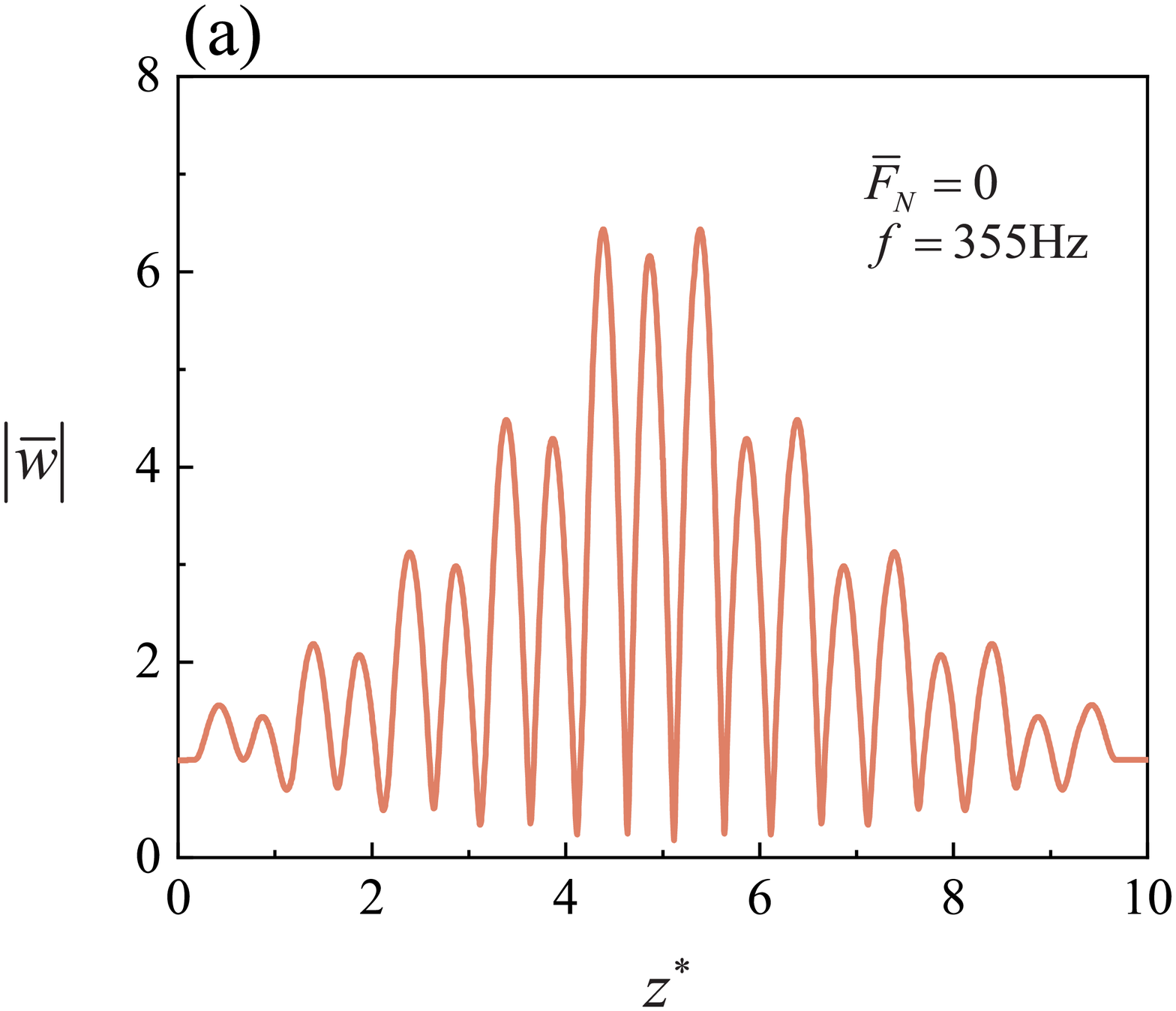}
    \hspace{0.005\textwidth}
	\includegraphics[width=0.46\textwidth]{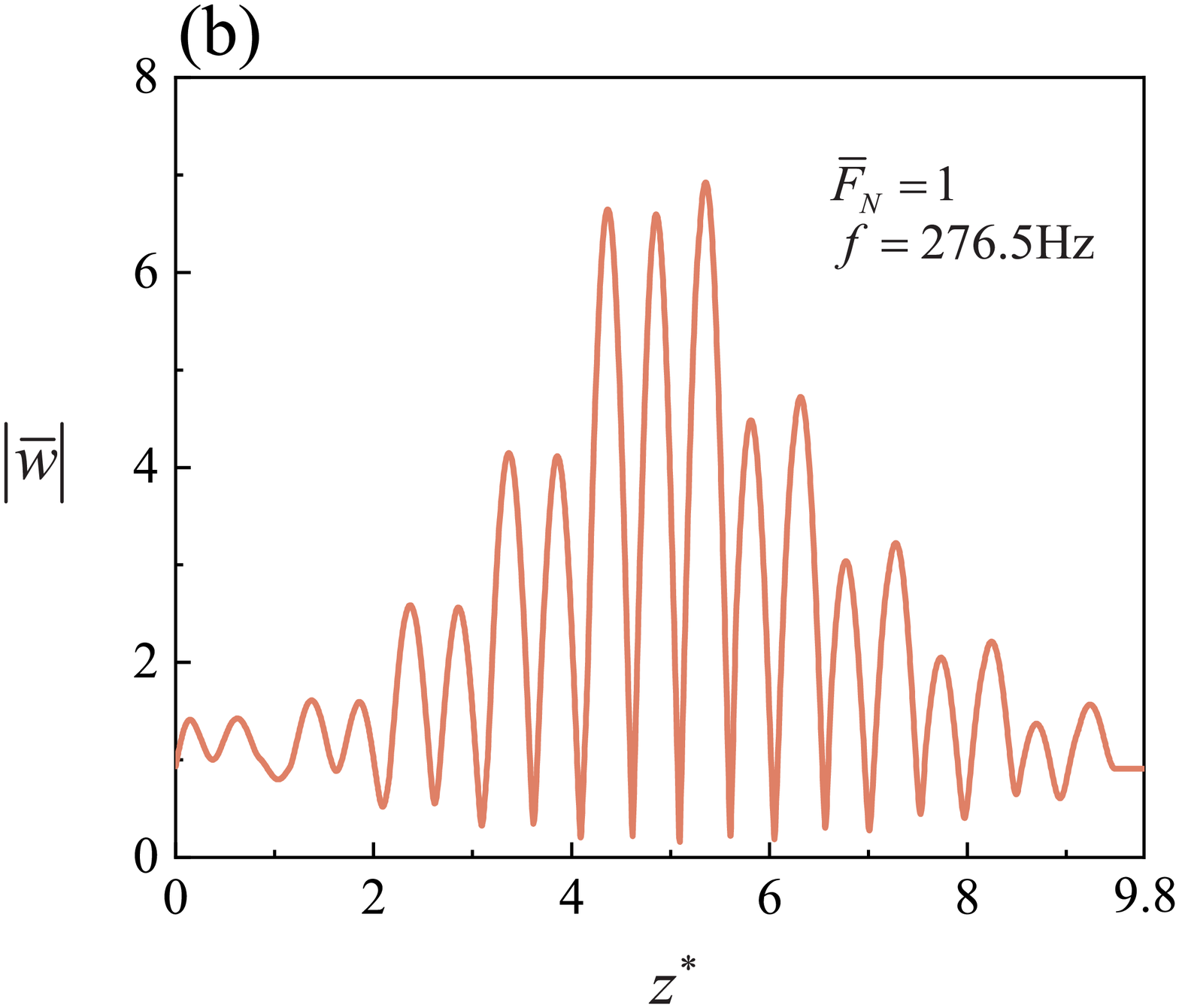}
	\includegraphics[width=0.9\textwidth]{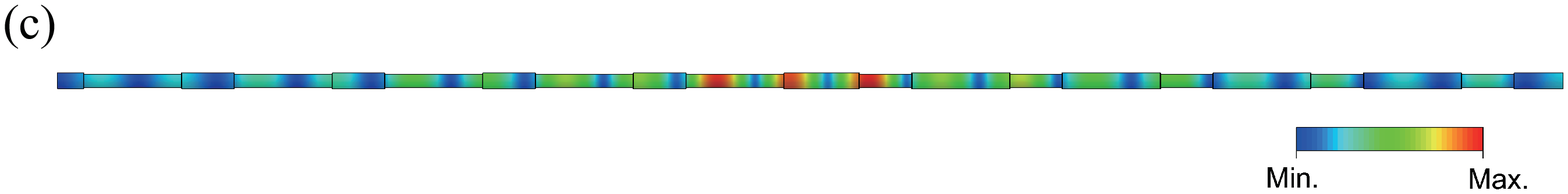}
	\includegraphics[width=0.9\textwidth]{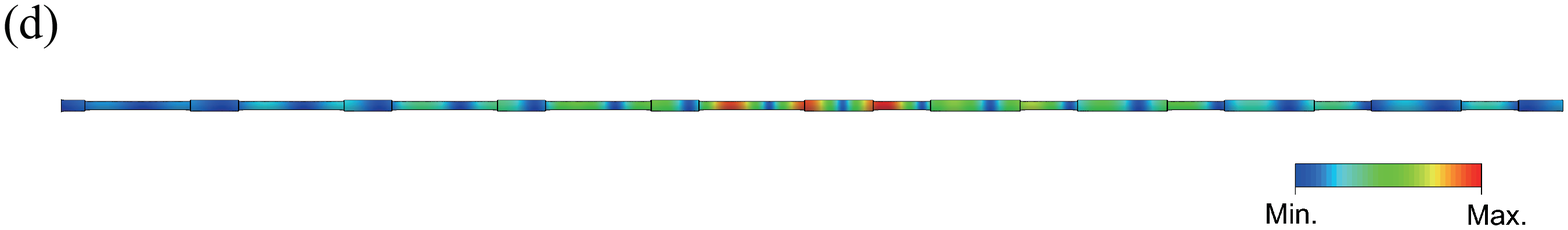}
	\caption{Theoretical calculations of the displacement field absolute value $|\overline{w}(z)|$ of a \emph{mixed} finite neo-Hookean waveguide (consisting of 5 S1 unit cells and 5 S2 unit cells) as a function of the normalized axial coordinate ${{z}^{*}=z/l}$ at 355~Hz (${{\overline{F}}_{N}}=0$) (a) and 276.5~Hz (${{\overline{F}}_{N}}=1$) (b). FE simulations of the displacement mode of the same \emph{mixed} finite waveguide at 352~Hz (${{\overline{F}}_{N}}=0$) (c) and 275~Hz (${{\overline{F}}_{N}}=1$) (d).}
	\label{Fig8}
\end{figure}

Figure~\ref{Fig8} depicts the spatial distributions of the displacement modes for the \emph{mixed} finite-size neo-Hookean waveguide at the transmission peak frequencies corresponding to the unloaded and loaded (${{\overline{F}}_{N}}=1$) states. Figs.~\ref{Fig8}(a) and (b) show theoretical results for the absolute value $|\overline{w}(z)|$ of displacement distributions at the peak frequencies 355~Hz (${{\overline{F}}_{N}}=0$) and 276.5~Hz (${{\overline{F}}_{N}}=1$). The normalized axial coordinate is defined as ${{z}^{*}=z/l}$, where $l$ is the length of the deformed unit cell of S1-configuration. Through the FE simulations, Figs.~\ref{Fig8}(c) and (d) display the displacement mode shapes at the FE peak frequencies 352~Hz (${{\overline{F}}_{N}}=0$) and 275~Hz (${{\overline{F}}_{N}}=1$). We observe that the displacement field is localized at the interface between the two PCC elements, and it decays dramatically towards the ends of the \emph{mixed} waveguide. This is an evident sign of the interface state as a result of the topological conflict of the distinct states. In particular, the displacement amplitude at the interface is 6 times more than the input signal (see Figs.~\ref{Fig8}(a) and (b)).

It is worth noting that the topological interface state is different from the concept of resonant mode \citep{li2017diatomic, chen2019tunable}. The resonant mode is greatly affected by the boundary conditions, local resonance and excitation location, whereas the topological interface state method --  based on the topological property conflict -- provides a robust mechanism against the wave propagation direction or boundary conditions \citep{yin2018band}. To validate that the occurrence of transmission peaks is ascribed to the topological interface state, we have reversed the input and output ends, and then calculated the transmission spectra as well as the displacement distributions for the unloaded and loaded (${{\overline{F}}_{N}}=1$) cases. The results (omitted here) indicate that the topological interface state is still observed at the same transmission peak frequencies as those in Fig.~\ref{Fig7}.

%


\subsubsection{Topological phase diagram} \label{Sec5-2-3}

\begin{figure}[htbp]
	\centering
	\setlength{\abovecaptionskip}{5pt}	
	\includegraphics[width=0.6\textwidth]{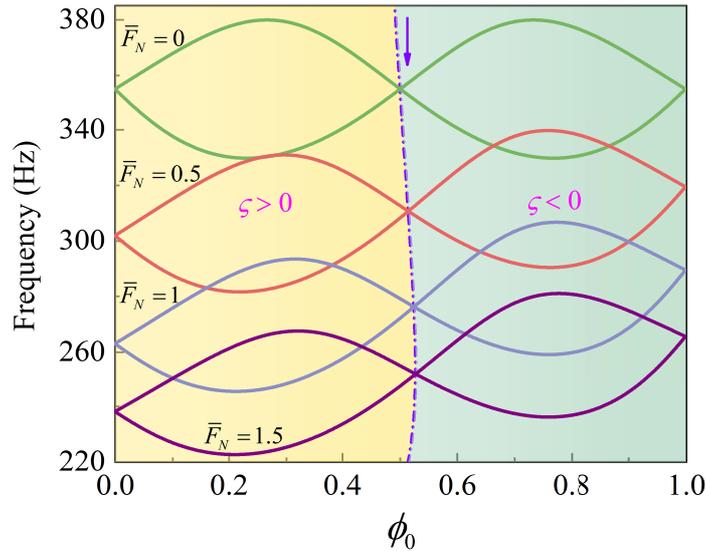}
	\caption{Topological phase diagram of a neo-Hookean PCC. The four groups of solid curves represent the two edge-state frequencies of the second BG as functions of the initial length fraction ${\phi }_{0}$ for four different axial forces. The topological phase curve (dash-dotted line) connects the topological transition points corresponding to different axial forces, and the arrow denotes the direction in which the axial force increases. The yellow and green filled regions indicate the BG signs with $\varsigma>0$ and $\varsigma<0$ respectively, which are labelled at two sides of the topological phase curve.}
	\label{Fig9}
\end{figure}

Next, to analyze the effect of axial force on topological interface states, we examine the \emph{topological phase diagram}. Fig.~\ref{Fig9} shows the two edge-state frequencies of the second BG as functions of the geometric parameter ${\phi }_{0}$ of the neo-Hookean PCC for different levels of the applied axial force. The variations of the second BG frequencies with ${\phi }_{0}$ are calculated by setting $\overline q=0$ in Eq.~\eqref{34}. Fig.~\ref{Fig9} illustrates that an increase in the axial force results in a lower frequency of the topological transition point (or band crossing point). However, there is only a slight change in ${\phi }_{0}$ where the band crossing happens. In particular, for ${{\overline{F}}_{N}}=0$, 0.5, 1 and 1.5, four topological transition points occur at ${\phi }_{0}=0.5$, 0.514, 0.524 and 0.527, with their topological transition frequencies being 355~Hz, 310.6~Hz, 275.9~Hz and 251.5~Hz, respectively. Note that for the neo-Hookean PCC subjected to a fixed positive tensile loading, the BG central frequency increases with an increase in ${\phi }_{0}$. For example, for PCC under the action of ${{\overline{F}}_{N}}=1$, the BG central frequency increases from 272.5~Hz at ${\phi }_{0}=0.35$ to 280.4~Hz at ${\phi }_{0}=0.65$. This is due to the fact that for a smaller ${\phi }_{0}$, the thinner sub-cylinder 2 occupies more in the unit cell. In the neo-Hookean PCC with smaller ${\phi }_{0}$ the unit cell elongates more, thus increasing the overall geometric size of the structure (recall Fig.~\ref{Fig3}(b)). Moreover, according to \citet{ding2006elasticity}, the natural frequency of cylindrical structure decreases with an increase in the slenderness or length-to-radius ratio. Therefore, the BG central frequency shifts down towards a lower frequency for PCC with a smaller ${\phi }_{0}$. We note, however, that this trend is reversed if a compressive axial force is applied (${{\overline{F}}_{N}}<0$); namely, the second BG central frequency decreases with an increase in ${\phi }_{0}$.

Following \citet{xiao2014surface}, the topological transition point can be obtained analytically in \ref{AppeC}. The dash-dotted line shown in Fig.~\ref{Fig9} is referred to as the \emph{topological phase curve} for different axial force levels ranging from $-0.3$ to 5, which is determined from the formulae in \ref{AppeC}. We see from Fig.~\ref{Fig9} that all the topological transition points formed by the close of the second BG are connected by the topological phase curve, which provides the frequencies of transmission peaks in Figs.~\ref{Fig7}(a) and (b) for axial forces ${{\overline{F}}_{N}}=0$ and 1. Moreover, the topological phase curve divides the topological phase diagram into two regions with different topological properties, the BG signs of which are indicated in Fig.~\ref{Fig9}. According to the topological phase diagram, we can design conveniently tunable topological interface states in a soft PCC. For example, we can construct a topological waveguide composed of two types of PCCs with ${\phi }_{0}=0.35$ and 0.65. As the axial force ${{\overline{F}}_{N}}$ increases from 0 to 2, the frequency of topological interface state is tuned from 355~Hz to 236.3~Hz. It should be emphasized that if the BG has no common frequency range for the chosen geometric size and axial force, there will be no topological interface state.

Thus, the band structure and topological interface state can be actively tuned towards a lower frequency by applying the axial force in the neo-Hookean waveguide. Its PCC elements should be properly selected to have different topological properties, so that the existence of low-frequency tunable topological interface states is guaranteed.
  
\subsection{Influence of the strain-stiffening effect} \label{Sec5-3}

\begin{figure}[htbp]
	\centering
	\setlength{\abovecaptionskip}{5pt}	
	\includegraphics[width=0.242\textwidth]{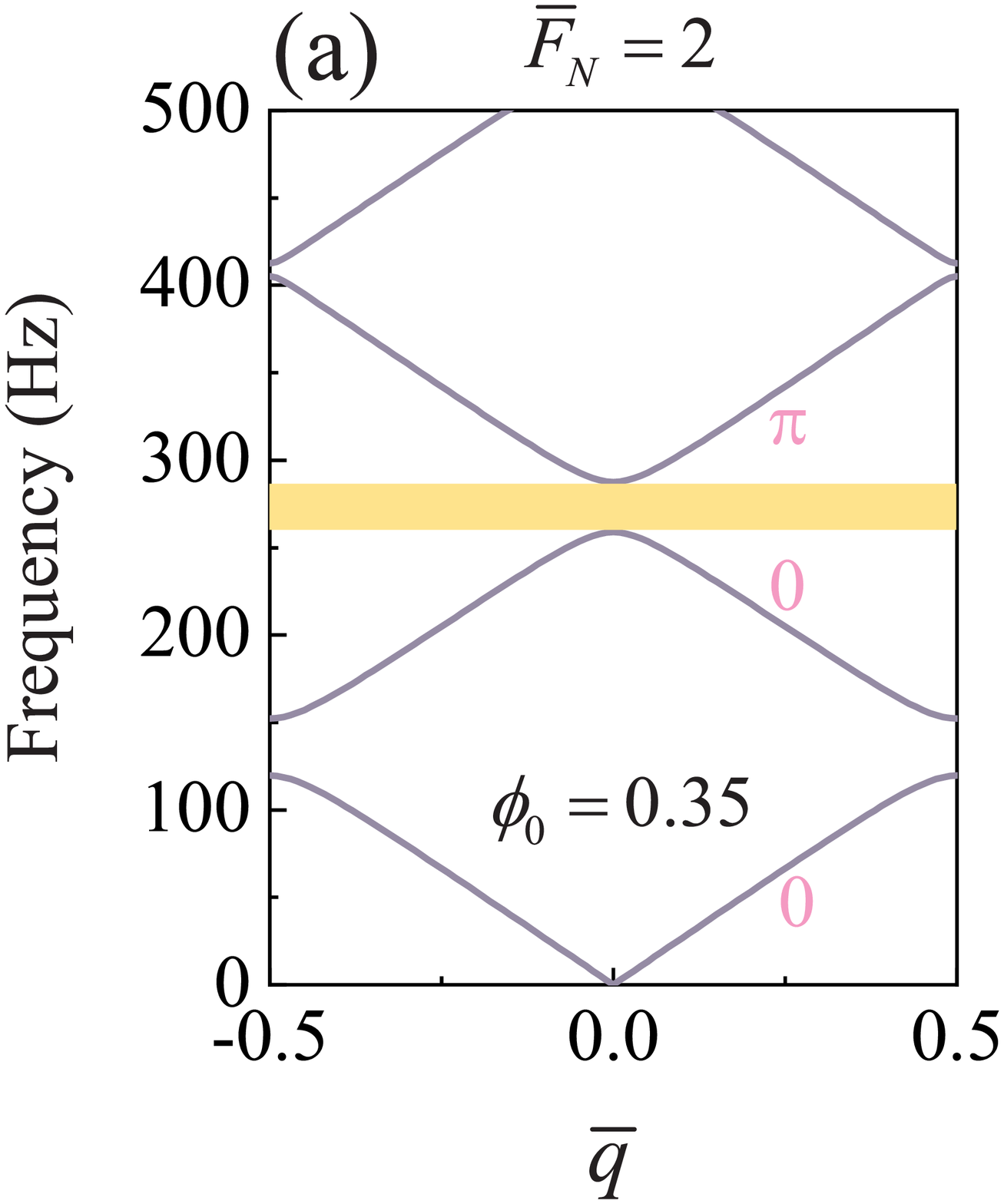}
	\includegraphics[width=0.242\textwidth]{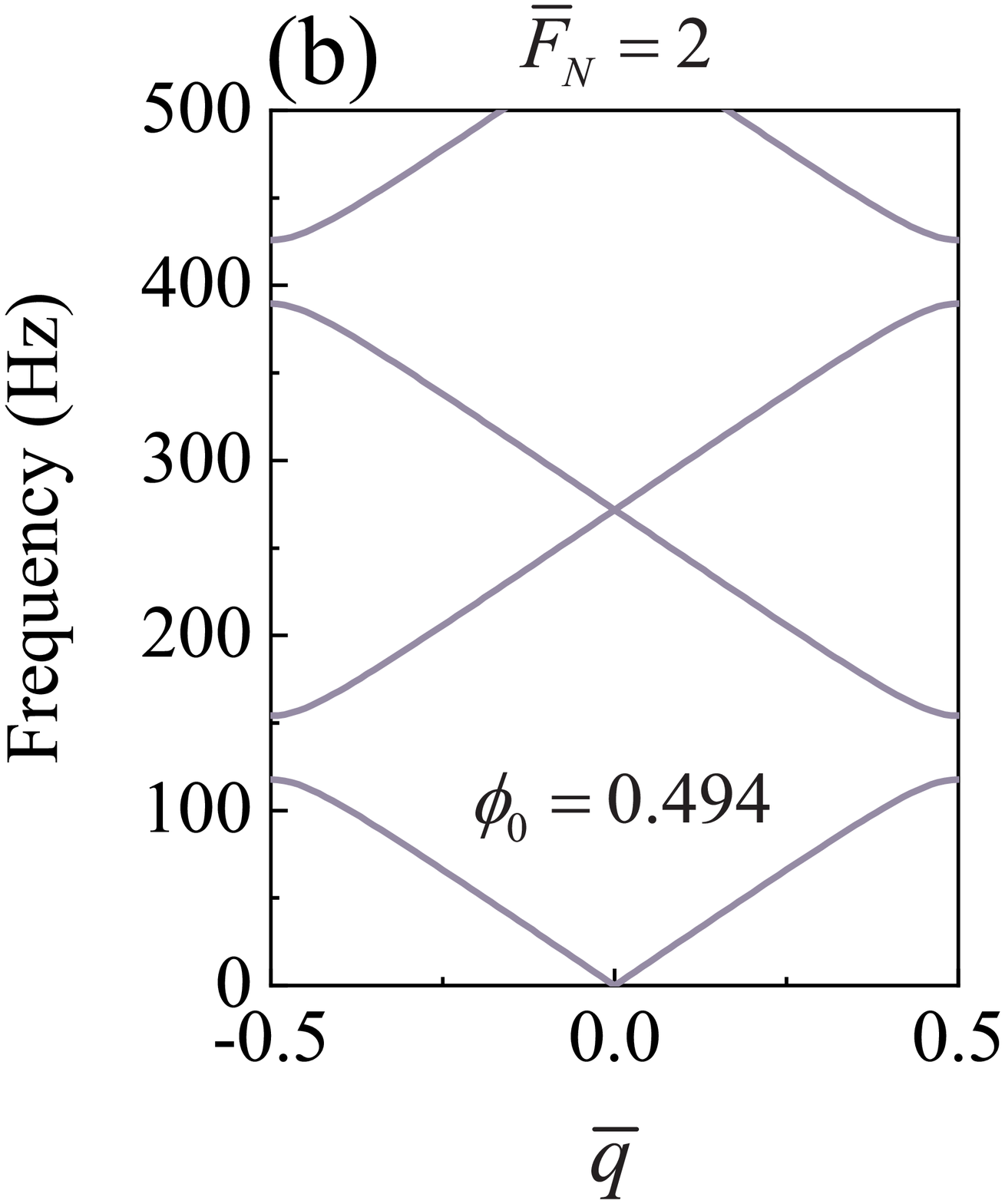}
	\includegraphics[width=0.242\textwidth]{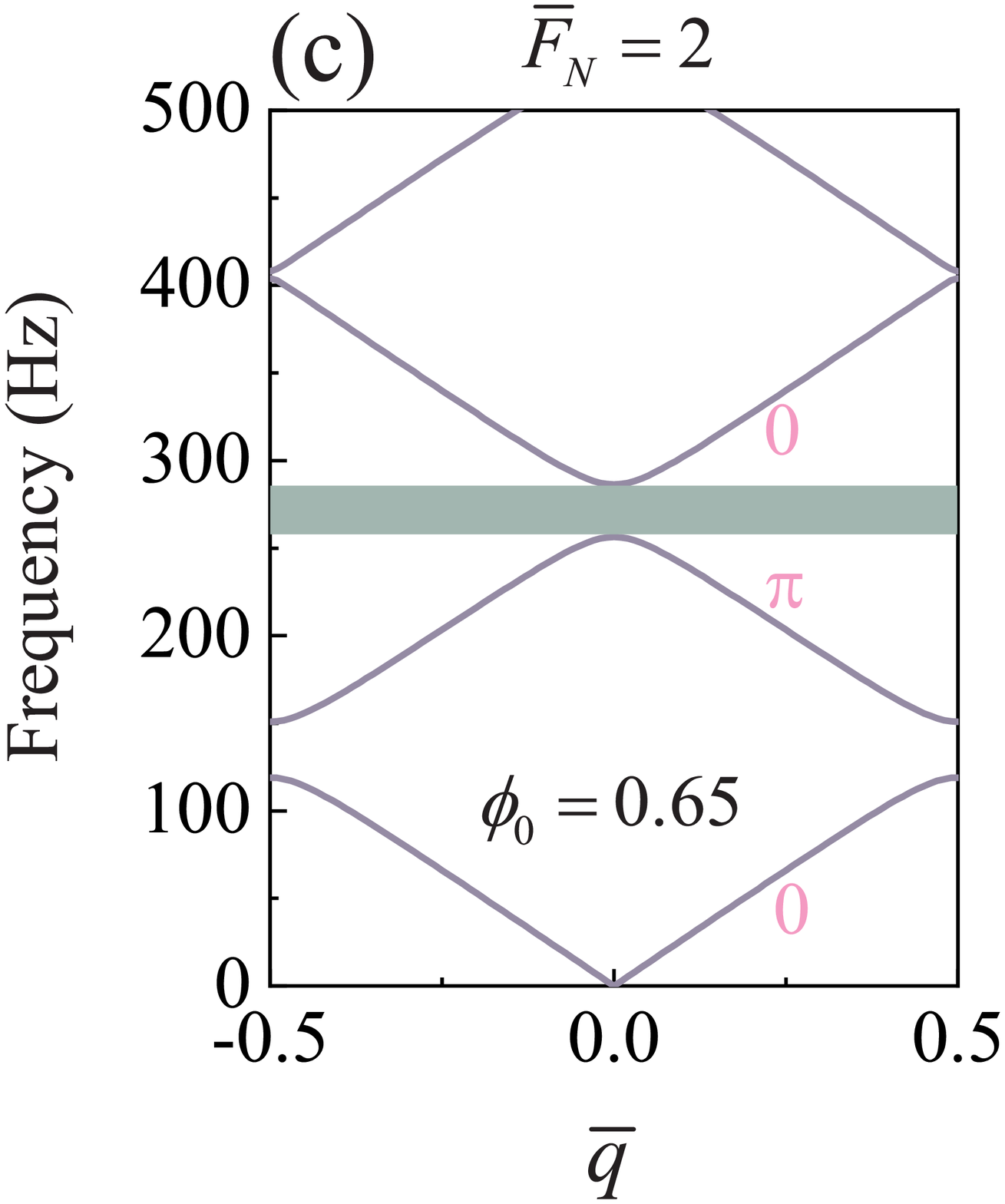}
	\includegraphics[width=0.242\textwidth]{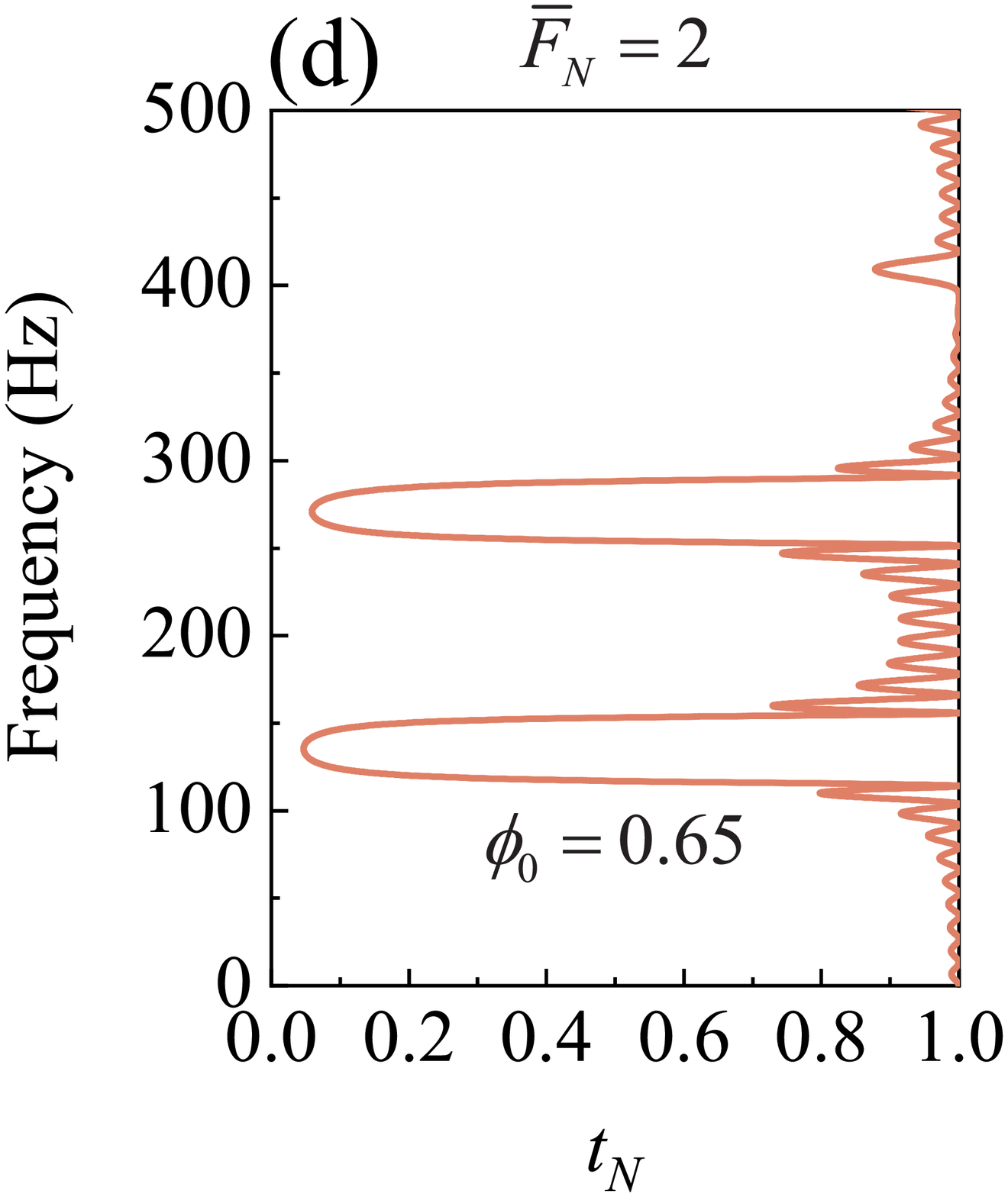}
	\caption{(a)-(c) The band structures of longitudinal waves in the Gent PCC ($J_{m}=20$) for the axial force ${{\overline{F}}_{N}}=2$ and three initial length fractions ${\phi }_{0}$. The Zak phase is marked in magenta on the corresponding bulk band. The yellow and green stripes denote the second BG signs with $\varsigma>0$ and $\varsigma<0$, respectively. (d) The transmission spectrum of a finite Gent PCC consisting of 10 identical S2 unit cells (${\phi }_{0}=0.65$) for ${{\overline{F}}_{N}}=2$.}
	\label{Fig10}
\end{figure}

Here, we consider a PCC made out of nonlinear material with a strong strain-stiffening behavior. In particular, we employ the Gent material model and analyze the stiffening effect on the tunable topological interface states. When the axial force is not large enough (e.g., ${{\overline{F}}_{N}}\le 1.25$), the band structure and transmission behavior of the Gent model resemble those of the neo-Hookean model (recall Figs.~\ref{Fig4} and \ref{Fig6}) since the hyperelastic material has not reached the stiffening stage (see Fig.~\ref{Fig3}) for the given locking parameter $J_{m}=20$. Therefore, in Fig.~\ref{Fig10}, we display the topological transition of band structures and the transmission spectrum for a large axial force ${{\overline{F}}_{N}}=2$. In Figs.~\ref{Fig10}(a) and (c), the calculated Zak phase 0 or ${\pi}$ is indicated on the related bulk band and the second BGs are marked by yellow ($\varsigma  > 0$) and green ($\varsigma < 0$) stripes with different topological properties. Similar to the previous observations in the neo-Hookean PCC (recall Fig.~\ref{Fig4}), the band inversion can be obtained for the Gent PCC. Here, the second BG closes as ${\phi }_{0}$ increases from 0.35 to 0.494, and it reopens with a further increase in ${\phi }_{0}$. Other illustrations are not repeated for brevity. Nevertheless, in view of the triggered strain-stiffening effect, the position of the second BG for the Gent PCC is higher than that of the neo-Hookean PCC for ${{\overline{F}}_{N}}>1.25$. Particularly, for the S1-configuration PCC subjected to the axial force ${{\overline{F}}_{N}}=2$, the frequency limits of the second BG vary from (225~Hz, 254~Hz) for the neo-Hookean model to (256~Hz, 286~Hz) for the Gent model.

\begin{figure}[htbp]
	\centering
	\setlength{\abovecaptionskip}{5pt}	
	\includegraphics[width=0.4\textwidth]{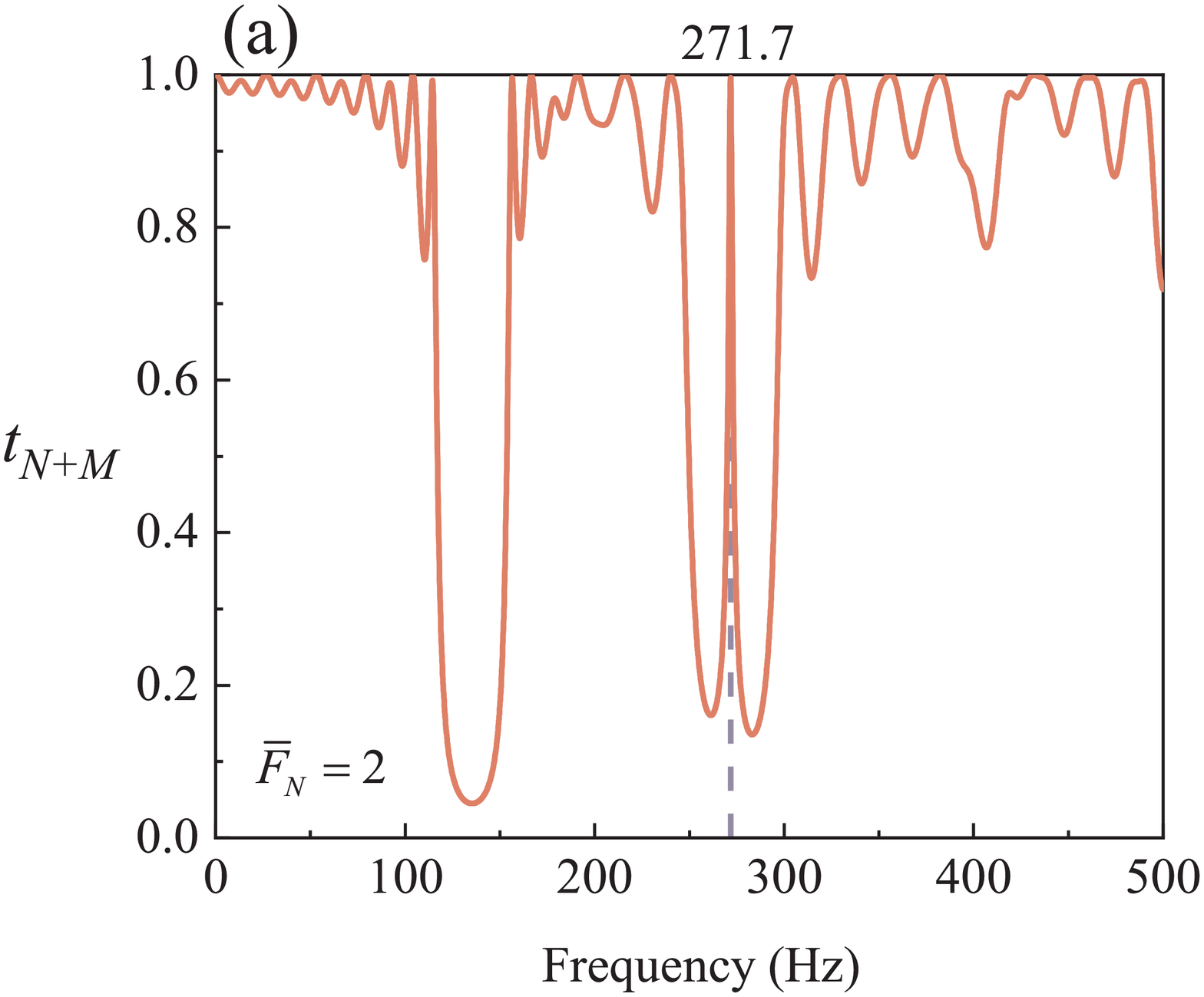}
	\hspace{0.005\textwidth}
	\includegraphics[width=0.415\textwidth]{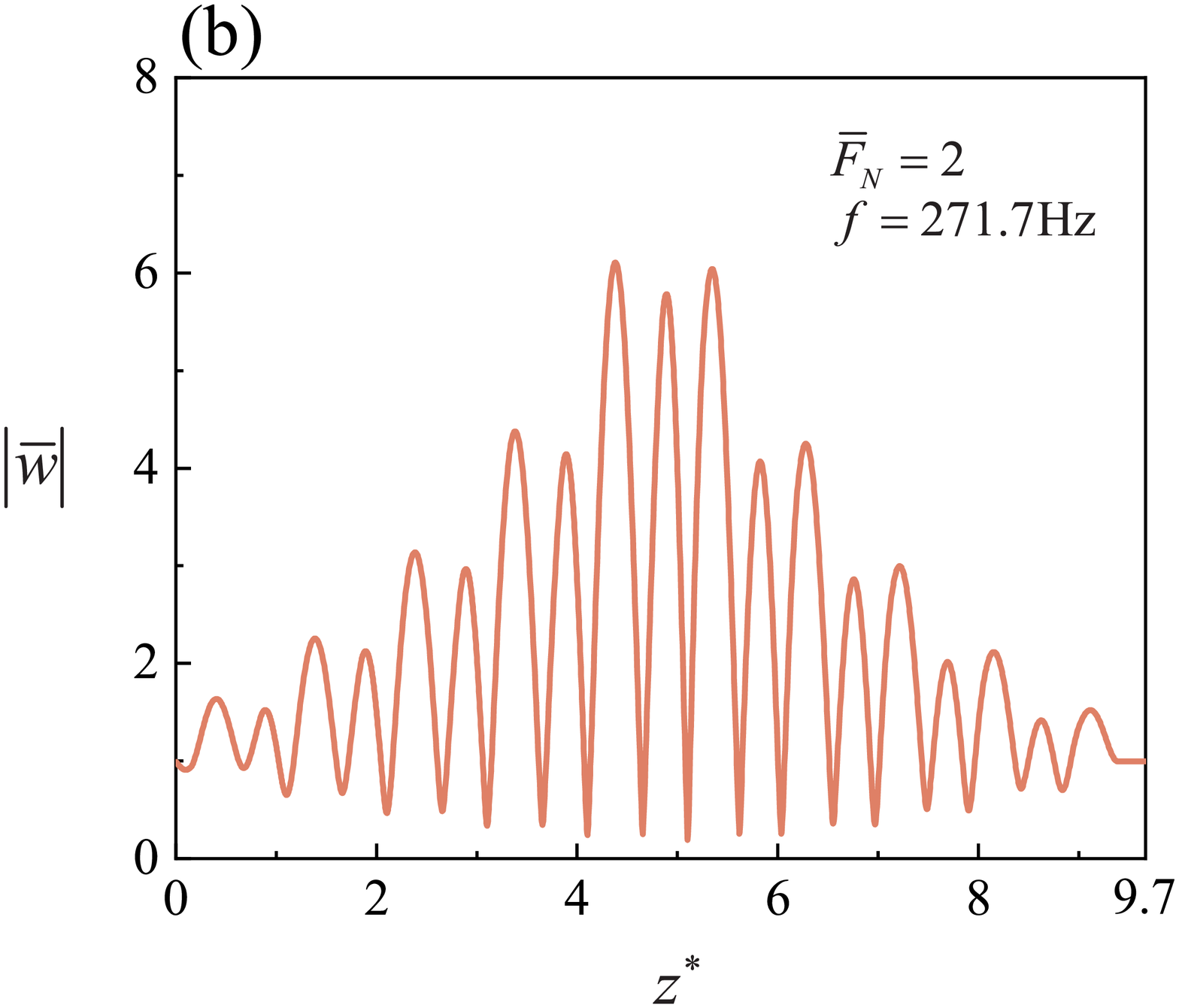}
	\caption{(a)The transmssion spectrum of a \emph{mixed} finite waveguide consisting of 5 S1 unit cells and 5 S2 unit cells at axial force ${{\overline{F}}_{N}}=2$ for the Gent model ($J_{m}=20$). (b) The absolute value $|\overline{w}(z)|$ of displacement field of the \emph{mixed} finite waveguide as a function of the normalized axial coordinate ${{z}^{*}=z/l}$ at the peak frequency 271.7~Hz.}
	\label{Fig11}
\end{figure}

Next, we examine the wave characteristics of a \emph{mixed} Gent PCC waveguide subjected to the axial force ${{\overline{F}}_{N}}=2$. Fig.~\ref{Fig11}(a) demonstrates the transmission spectrum of the waveguide consisting of 5 S1-type unit cells (${\phi }_{0}=0.35$) and 5 S2-type unit cells (${\phi }_{0}=0.65$). The transmission spectrum is calculated with the help of Eq.~\eqref{49}. The transmission peak emerges in the second overlapped BG, with the peak frequency 271.7~Hz  for this case (see Fig.~\ref{Fig11}(a)). The corresponding displacement field distribution at 271.7~Hz is shown in Fig.~\ref{Fig11}(b). It is confirmed that the displacement is mainly confined in vicinity of the interface (with the  amplitude nearly 6 times over the input signal) and attenuates rapidly towards the ends of the hyperelastic waveguide. 

\begin{figure}[htbp]
	\centering
	\setlength{\abovecaptionskip}{5pt}	
	\includegraphics[width=0.6\textwidth]{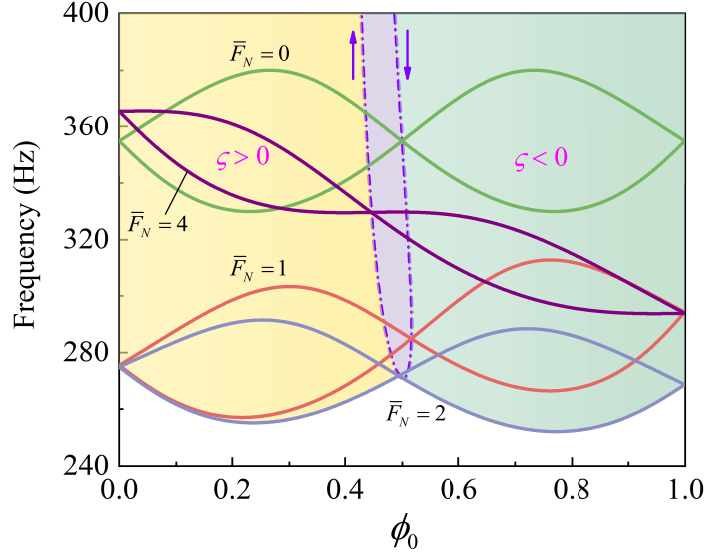}
	\caption{Topological phase diagram of a Gent PCC ($J_m=20$). The four groups of solid curves represent the two edge-state frequencies of the second BG as functions of ${\phi }_{0}$ for four different axial forces. The topological transition points are located on the topological phase curve (dash-dotted curve), and the arrow denotes the direction in which the axial force increases. The yellow and green filled regions indicate the BG signs with $\varsigma>0$ and $\varsigma<0$ respectively, which are labelled at two sides of the topological phase curve. The violet region marks the morphologies for which the topological property switches depending on the axial force.}
	\label{Fig12}
\end{figure}

To further explore the influence of strain-stiffening effect on the topological interface state, we plot the corresponding topological phase diagram in Fig.~\ref{Fig12}. The results are depicted for the Gent PCC ($J_{m}=20$) with four topological transition variations at ${{\overline{F}}_{N}}=0$, 1, 2 and 4. The \emph{topological phase curve} calculated according to the theoretical formulae (see \ref{AppeC}) is also included in Fig.~\ref{Fig12}. The topological phase curve and the band inversion curves for the Gent PCC are almost the same as those for the corresponding neo-Hookean PCC (compare with Fig.~\ref{Fig9}) for the range of loadings not reaching the stiffening stage (i.e., when ${{\overline{F}}_{N}}\le 1.25$). In particular, the frequency of the topological transition point continuously decreases with an increase in axial force level in the range of ${{\overline{F}}_{N}}\le 1.25$. However, a further increase in the axial force leads to different variation trend in the  topological transition point frequency: after decreasing to the lowest value 270.8~Hz at around ${{\overline{F}}_{N}}=1.8$, the frequency starts to increase conversely and rapidly (see Fig.~\ref{Fig12}). This is a unique feature of the nonlinear PCC with a strong strain-stiffening effect.

Remarkably, the tensile force --  if large enough -- can cause a reverse trend in the second BG central frequency versus ${\phi }_{0}$ in Gent PCC. For example, in the Gent PCC subjected to ${{\overline{F}}_{N}}=2$, the BG central frequency  decreases with an increase in ${\phi }_{0}$. This reverse trend is even more prominent for the Gent PCC subjected to ${{\overline{F}}_{N}}=4$ (see the corresponding curves in Fig.~\ref{Fig12}). Note that this BG central frequency trend reversion is not observed in the neo-Hookean PCC (see Fig.~\ref{Fig9}). This is again an important manifestation of the strain-stiffening effect that begins to prevail over the deformation-induced geometric change upon achieving a certain level of the applied tensile force.

Interestingly, we find from Fig.~\ref{Fig12} that the length fraction ${\phi }_{0}$ corresponding to the topological transition point has a more obvious variation with the increasing axial force applied to the Gent PCC, compared with the neo-Hookean case (see Fig.~\ref{Fig9}). This phenomenon combined with the frequency non-monotonous change results in a peculiar state for morphologies enclosed by the topological phase curve. In this morphology domain (denoted by the violet region in Fig.~\ref{Fig12}), the BG sign can switch depending on the applied axial force. For example, for ${{\overline{F}}_{N}}=1$, this special area is on the left of the topological phase curve with $\varsigma>0$, while for ${{\overline{F}}_{N}}=4$, the violet area turns to be on the right of the topological phase curve with $\varsigma<0$. Therefore, for ${{\overline{F}}_{N}}\le 1.8$, $\varsigma>0$; but $\varsigma<0$ for ${{\overline{F}}_{N}} > 1.8$.

In addition, when the axial force exceeds ${{\overline{F}}_{N}}=4$, the BG frequency ranges for PCC elements (whose unit-cell length fractions $\phi_0$ take values from different sides of the topological transition point) do not overlap (see Fig.~\ref{Fig12}), and hence the topological interface state does not exist for any combination of the geometric parameters. For example, for Gent PCC subjected to the axial force ${{\overline{F}}_{N}}=4.1$, the topological transition point is $\phi_0=0.444$ and the corresponding frequency is 333.8~Hz. In vicinity of this topological transition point, the BG frequency limits for $\phi_0=0.44$ are (333.8~Hz, 334.4~Hz), while those for $\phi_0=0.45$ become (332.9~Hz, 333.8~Hz); thus, there is no overlapped BG frequency.

Furthermore, for a \emph{mixed} finite Gent PCC waveguide composed of 5 S1-type unit cells (${\phi }_{0}=0.35$) and 5 S2-type unit cells (${\phi }_{0}=0.65$), the topological interface state exists until the axial force reaches the level  ${{\overline{F}}_{N}}\approx3.8$. When the applied axial force exceeds this level, the frequency limits of the second BG have no overlapped part, and although the topological properties of the two PCC elements are different, the topological interface state cannot be activated.

\section{Conclusions}\label{section6}

We studied a class of 1D soft PCs possessing the topologically protected interface states. The large-deformation ability of soft waveguides combined with the material stiffening effect is exploited to tune the topologically protected states. In particular, we illustrate this concept based on the example of the 1D waveguide composed of two types of soft PCCs with different topological characteristics. Here, the topological interface state for longitudinal waves is tunable by application of an external axial force. First, we utilized the nonlinear elasticity theory combined with the assumption of uniform deformations to determine the nonlinear static response of PCCs under the action of an axial force. Next, the dispersion relation for small-amplitude longitudinal waves, the unit-cell mode shape as well as the displacement field distribution and signal transmission coefficient of the finite-size cylindrical waveguide were derived analytically. Finally, the theoretical predictions along with the numerical calculations were analyzed to elucidate how the external loading and the material stiffening affect the frequency tunability and the existence of topological interface states. Our main observations are summarized below:

\begin{enumerate}[(1)]
	\item The BG inversion process (i.e., the BG open, close and reopen process) accompanied by the topological phase transition can be realized by altering the initial PCC geometric parameter and be tuned by adjusting the axial force.
	
	\item For the neo-Hookean waveguides, the axial tensile force lowers monotonically the frequency of topological interface states owing to the generated elongation in the whole system.
	
    \item For the Gent waveguides, the frequency of topological interface states varies with the axial force in a non-monotonous way; this is a result of the competition between the geometric change and the material strain-stiffening effect.	
	
	\item In reference to the topological phase diagram, the tunable position and existence condition of topological interface states are clearly demonstrated when changing the axial force in a properly pre-designed system.
\end{enumerate}

Our results -- based on the example of 1D soft PCC waveguides -- indicate the possibility to realize on-demand tunability of the topological interface states. The present study provides guidelines for further design of actively tunable topological wave devices operating at the low-frequency range. These systems may find a wide range of potential applications such as tunable energy harvesters, low-pass filters and high-sensitivity biomedical detectors.

{\color{red} It should be emphasized that the tunable topological interface or edge states in 2D metamaterial systems could be achieved by means of the electromechanical biasing fields \citep{li2018observation, zhou2020voltage}, which is an interesting topic to be addressed for further applications in the future.}

\section*{Acknowledgements}

This work was supported by the Government of Ireland Postdoctoral Fellowship from the Irish Research Council (No. GOIPD/2019/65), the National Natural Science Foundation of China (Nos. 11872329 and 11621062) and the China Scholarship Council (CSC). Partial supports from the Fundamental Research Funds for the Central Universities, PR China (No. 2016XZZX001-05) and the Shenzhen Scientific and Technological Fund for R\&D, PR China (No. JCYJ20170816172316775) are also acknowledged.
  
\appendix


\section{Calculation of the Zak phase} \label{AppeA}


In this appendix, we will employ the method developed by \citet{xiao2015geometric} to calculate the Zak phase for the $j$th bulk band of the phononic cylinder, which is defined in Eq.~\eqref{53}. Specifically, we select $P$ points to equally divide the first Brillouin zone from $q=-\pi/l$ to $q=\pi/l$. In the limit of $P \to \infty$, we have $\Delta {q} = {q_{i + 1}} - {q_i} \to 0$, which leads to ${{\partial }_{q}}{{W}_{j,q}}=({{W}_{j,{q}+\Delta q}}-{{W}_{j,q}})/\Delta q$. Thus, Eq.~\eqref{53} for the Zak phase can be equivalently expressed as 
\begin{equation} \label{A1}
\theta _{j}^{\textrm{Zak}}=\int_{-\pi /l}^{\pi /l}{\left[ \dfrac{\text{i}}{\Delta q}\left( \int\limits_{\text{ unit cell}}{\dfrac{1}{2\rho c^2}\textrm{d}\mathbf{r}\operatorname{dz}W_{j,q}^{*}{{W}_{j,q+\Delta q}}}-1 \right) \right]}\textrm{d}q,
\end{equation}
where the physical meaning of the related quantities has been defined in Eq.~\eqref{53} and the periodic in-cell part $W_{j,q}$ of the Bloch displacement eigenfunction is normalized with the orthogonal relationship $\int_{\text{unit cell}} {\textrm{d}\mathbf{r} \operatorname{dz} ({1}/{2\rho c^2}) \left|W_{j,q}\left( z,\mathbf{r} \right) \right|^2}=1$.

By discretizing $q$ and noting the relation $\ln \left( {x - 1 + 1} \right) \to x - 1$ in the limit of ${x \to 1}$, Eq.~\eqref{A1} with $\Delta q \to 0$ can be rewritten in a discretized form as 
\begin{equation} \label{A2}
\theta _{j}^{\textrm{Zak}}=\sum\limits_{i = 1}^P \left[ \dfrac{\text{i}}{\Delta q_i} \ln \left(\int\limits_{\text{ unit cell}}{\dfrac{1}{2\rho c^2}\textrm{d}\mathbf{r}\operatorname{dz}W_{j,q_i}^{*}{{W}_{j,q_i+\Delta q_i}}} \right) \left( {{q_{i + 1}} - {q_i}} \right)\right],
\end{equation}
which, after some simplifications, yields
\begin{equation} \label{A3}
\theta _{j}^{\textrm{Zak}}=-\text{Im}\sum\limits_{i = 1}^N \ln \left[\int\limits_{\text{ unit cell}}{\dfrac{1}{2\rho c^2}\textrm{d}\mathbf{r}\operatorname{dz}W_{j,q_i}^{*}{{W}_{j,q_{i+1}}}} \right].
\end{equation}

Consequently, after obtaining the Bloch displacement distribution ${{{w}}_{j,q}}$ of the deformed unit cell in Subsec~\ref{Sec4-3}, we can exploit the relation ${{W}_{j,q}} = {{{w}}_{j,q}} {{{e}}^{-\text{i}qz}}$ and Eq.~\eqref{A3} to  calculate the Zak phase numerically.


\section{FE simulations} \label{AppeB}


To validate our theoretical model, numerical simulations are conducted by using Abaqus, an FE analysis and solver software. The simulations to calculate the transmission spectra are implemented for the finite PCC structure, including its geometric parameters and material properties of silicon rubber (Zhermarck Elite Double 32) \citep{galich2017elastic}. Here, we adopt the neo-Hookean hyperelastic model and establish an axisymmetric structural model with a fine mesh of 8-node hybrid elements (i.e., element type CAX8H in Abaqus).

In order to understand the longitudinal wave propagation behaviors in a pre-deformed structure, the static analysis (i.e., pre-stretching the structure) and frequency domain analysis (i.e., wave propagation in the structure) are performed consecutively in Abaqus:

\textbf{Step 1 (Static analysis)}: The boundary conditions in accordance with the theoretical model are applied to both sides of the structure to simulate its extension procedure. The resultant displacement and stress fields in the deformed structure are recorded and saved.

\textbf{Step 2 (Frequency domain analysis)}: The displacement and stress fields calculated in the static analysis are imported to the structural model as the initial deformed state. For calculating the transmission spectra, a sinusoidal axial-displacement excitation over a frequency range of interest is imposed to one input side of the finite-size PCC structure and its average displacement amplitude is calculated as the input signal ${{A}_{\text{input}}}$. Additionally, the average displacement amplitude at output side is collected as the output signal ${{A}_{\text{output}}}$. Thus, the attenuation intensity $T\left( \text{dB} \right)$ is defined as $T\left( \text{dB} \right)=20\log \left( {{A}_{\text{output}}}/{{A}_{\text{input}}} \right)$, which describes the elastic wave transmission behavior. 


\section{Conditions for the band crossing}\label{AppeC}


Following the argument of \citet{xiao2014surface} for 1D photonic crystals, this appendix will provide the conditions for two bands to cross for a soft PCC subjected to an axial force.

The dispersion relation of incremental longitudinal waves in a deformed soft PCC is given in Eq.~\eqref{34} with ${{k}^{\left( p \right)}}=\omega/c^{\left( p \right)}$ $(p=1,2)$. If $\sin \left( {{k^{\left( 1 \right)}}{l^{\left( 1 \right)}}} \right)$ and $\sin \left( {{k^{\left( 2 \right)}}{l^{\left( 2 \right)}}} \right)$ in Eq.~\eqref{34} vanish simultaneously, ${{k^{\left( 1 \right)}}{l^{\left( 1 \right)}}}=\gamma m_1 \pi$ and ${{k^{\left( 2 \right)}}{l^{\left( 2 \right)}}} =\gamma m_2 \pi$ hold with $m_1$, $m_2$ and $\gamma$ being positive integers, which results in the frequency at the band crossing point as
\begin{equation} \label{C1}
{\omega _c} = \gamma {m_1}\pi {c^{\left( 1 \right)}}/{l^{\left( 1 \right)}} = \gamma {m_2}\pi {c^{\left( 2 \right)}}/{l^{\left( 2 \right)}}.
\end{equation}
Now we define a dimensionless parameter $\alpha$ as
\begin{equation} \label{C2}
\alpha  \equiv \frac{{{l^{\left( 1 \right)}}{c^{\left( 2 \right)}}}}{{{l^{\left( 2 \right)}}{c^{\left( 1 \right)}}}}.
\end{equation}
It can be proved \citep{xiao2014surface} from Eqs.~\eqref{C1} and \eqref{C2} that if $\alpha=m_1/m_2$ is a \emph{rational number}, the pass bands $\gamma(m_1+m_2)$ and $\gamma(m_1+m_2)+1$ cross each other at the frequency $\omega_c$ given in Eq.~\eqref{C1}, which also yields
\begin{equation} \label{C3}
{\omega _c} = \frac{{\gamma \left( {{m_1} + {m_2}} \right)\pi }}{{{l^{\left( 1 \right)}}/{c^{\left( 1 \right)}} + {l^{\left( 2 \right)}}/{c^{\left( 2 \right)}}}}.
\end{equation}

At $\omega _c$, we have $\cos \left( {{k^{\left( 1 \right)}}{l^{\left( 1 \right)}}} \right) = (-1)^{\gamma m_1}$ and $\cos \left( {{k^{\left( 2 \right)}}{l^{\left( 2 \right)}}} \right)  = (-1)^{\gamma m_2}$. Thus the dispersion relation \eqref{34} becomes
\begin{equation} \label{C4}
\cos \left( {ql} \right) = {\left( { - 1} \right)^{\gamma \left( {{m_1} + {m_2}} \right)}}.
\end{equation}
Therefore, the $\gamma(m_1+m_2)$-th BG will close either at $q=0$ when $\gamma(m_1+m_2)$ is even, or at $q=\pm \pi/l$ for the odd $\gamma(m_1+m_2)$. It can be easily verified \citep{xiao2014surface} that for those frequencies in vicinity of $\omega _c$ at the crossing points, the two bands have linear dispersion such that a Dirac cone is formed (see Figs.~\ref{Fig4}(b) and \ref{Fig4}(e) for example).

In this work, we focus on the second BG. In order to make the second BG close, we have $\gamma=m_1=m_2=1$ and hence $\alpha=1$ is a rational number. We can take two steps to obtain the position of the band crossing point (or topological transition point). First, for a given axial force $F_N$, the mechanical biasing field state and the parameters $c^{(p)}$ are determined. Second, using the relation $\alpha=1$ and the condition that the length of the undeformed unit cell is kept fixed (i.e., $L=L^{(1)}+L^{(2)}$), we can obtain the corresponding geometric sizes $L^{(1)}$ and $L^{(2)}$ (or initial length fraction $\phi_0=L^{(1)}/L$) and then calculate the frequency $\omega_c$ of topological transition points from Eq.~\eqref{C3}.


\section*{References}

\bibliographystyle{elsarticle-harv.bst}
\nocite{*}
\bibliography{1D_topo_Sep_10.bib}







\end{document}